\documentclass[preprint2]{aastex}
\usepackage{graphicx}
\usepackage{natbib}
\usepackage{enumerate}
\usepackage{amssymb}
\usepackage{amsmath}

\shorttitle{}
\shortauthors{}

\begin{document}
\title{The Response of Metal Rich Gas to X-Ray Irradiation from a Massive Black Hole at High Redshift: Proof of Concept}


\author{A. Aykutalp\altaffilmark{1,2}, J. H. Wise\altaffilmark{3}, R. Meijerink\altaffilmark{1}, and M. Spaans\altaffilmark{1}}
\altaffiltext{1}{Kapteyn Astronomical Institute, University of Groningen, PO Box 800, 9700 AV Groningen, The Netherlands}
\altaffiltext{2}{Scuola Normale Superiore, Piazza dei Cavalieri 7, I-56126, Pisa, Italy}
\altaffiltext{3}{Center for Relativistic Astrophysics, Georgia Institute of Technology, 837 State Street, Atlanta, GA 30332}
\email{aycin.aykutalp@sns.it}
\email{jwise@physics.gatech.edu}
\email{meijerink@astro.rug.nl}
\email{spaans@astro.rug.nl}

\begin{abstract}Observational studies show that there is a strong link between the formation and evolution of galaxies and the growth of supermassive black holes (SMBH) at their centers. However, the underlying physics of this observed relation is poorly understood. In order to study the effects of X-ray radiation on the surroundings of the black hole, we implement X-ray Dominated Region (XDR) physics into Enzo and use the radiation transport module $\it{Moray}$ to calculate the radiative transfer for a polychromatic spectrum. In this work, we investigate the effects of X-ray irradiation, produced by a central massive black hole (MBH) with a mass of M = $5 \times 10^4$ M$_{\odot}$, on ambient gas with solar and zero metallicity. We find that in the solar metallicity case, due to high opacity of the metals, the energy deposition rate in the central region ($\le$ 20 pc) is high and hence the temperatures in the center are on the order of 10$^{5-7}$ K. Moreover, due to the cooling ability and high intrinsic opacity of solar metallicity gas, column densities of 10$^{24}$ cm$^{-2}$ are reached at a radius of 20 pc from the MBH. These column densities are about 3 orders of magnitudes higher than in the zero metallicity case. Furthermore, in the zero metallicity case an X-ray induced H \textsc{ii} region is formed already after 5.8 Myr. This causes a significant outflow of gas ($\sim 8 \times 10^6$ M$_{\odot}$) from the central region, with the gas reaching outflow velocities up to $\sim$ 100 km s$^{-1}$. At later times, $\sim 23$ Myr after we insert the MBH, we find that the solar metallicity case also develops an X-ray induced H  \textsc{ii} region, but delayed by $\sim$ 17 Myr.

\end{abstract}
%% example has been keyed in interApJ style. See the instructions to authors




\section{INTRODUCTION}

Most galaxies today are thought to host supermassive black holes (SMBHs) in their centers \citep{1995ARA&A..33..581K, 1998Natur.395A..14R, 2001AIPC..586..363K}. Dynamical studies of central black holes in nearby galaxies reveal that there is a relation between the central black hole mass and the bulge mass ($\rm{M}_{BH}/\rm{M}_B \sim 10^{-3}$, \cite{1998AJ....115.2285M}) or the velocity dispersion ($\rm{M}_{BH} \propto \sigma^p$, p $\sim 4$, \cite{2000ApJ...539L...9F, 2000ApJ...539L..13G}) of the stellar component of the host galaxy. More recent studies have shown that the value of p varies between 4-6 depending on the type of the galaxy  \citep{2011MNRAS.412.2211G, 2011Natur.480..215M}.  These relations suggest that there is a strong link between the formation and evolution of galaxies, and the growth of SMBHs at their centers.

The black hole masses  of quasars observed at high redshifts ($z>6$) are on the order of 10$^9$ M$_\odot$ \citep{2003AJ....125.1649F, 2006AJ....132..117F, 2007ApJ...669...32K}. These SMBHs are thought to form (in less than 1 billion years) through the accretion of gas onto seed black holes of masses between 10$^2$ - 10$^5$ M$_\odot$. The origin of these seed black holes  \citep{2003ApJ...596...34B, 2006ApJ...652..902S} is still an open question. However, the existence of quasars with inferred black hole masses in excess of 10$^9$ M$_\odot$ implies that black hole growth should be dominated by high accretion rates ($\geq$ 0.1 Eddington) during their quasi-stellar object (QSO) phase.

The amount of gas available for the SMBHs to accrete depends on the mergers or interactions with galaxies, which drive large quantities of gas into galaxy centers \citep{1991ApJ...370L..65B}, star formation and feedback processes. Observational \citep{2004MNRAS.353.1035M} and theoretical \citep{2006ApJS..163....1H} studies indicate that there is an evolution in the relation between the black hole mass and the host stellar mass such that at high redshifts (z $>6$), the ratio between the black hole mass and the host stellar mass is on the order of 0.02-0.1, which is more than one order of magnitude larger than the ratio measured in local galaxies \citep{2006ApJ...649..616P, 2009astro2010S.217M}.

The growth of SMBHs during a quasar phase can be traced by the luminosity function of QSOs as a function of redshift \citep{1982MNRAS.200..115S}. The peak activity of luminous QSOs occurs at z $\sim 2-3$, where the majority of the most massive galaxies were also forming most of their stars. On the other hand, the peak activity of lower luminosity active galactic nuclei (AGN) occurs at z $\sim 0.7-1$. X-ray selected AGN serve as a population to robustly determine the luminosity function and evolution of unobscured AGN. However, the obscured fraction of AGN depends on X-ray luminosity as well as redshift. Therefore, this obscured fraction might affect the peak determination \citep{2008AN....329..122C}. The average black hole mass increases with increasing redshift \citep{2004MNRAS.353.1035M}, meaning that there are fewer low-mass black holes at high redshifts, and the average accretion rate decreases towards lower redshift. This observed anti-hierarchical behaviour of AGN evolution indicates that SMBHs and their host galaxies co-evolve.

The underlying physics of the observed relation between the formation and evolution of galaxies and their central SMBHs is still unknown. Several efforts have been made to explain the origin of this relation by using semi-analytical modelling \citep{2000MNRAS.311..576K, 2003ApJ...595..614W} and N-body SPH simulations \citep{2005ApJ...620L..79S,2005MNRAS.361..776S,2005Natur.433..604D}. Numerical simulations have shown that the gravitational tidal torques excited during major mergers lead to rapid inflows of gas into the centers of galaxies \citep{1991ApJ...370L..65B, 1996ApJ...471..115B} which can be the mechanism to trigger quasar activity and starbursts in galaxies.  Observations of low-redshift quasars show a connection between galaxy mergers and quasar activity \citep{1984AJ.....89..958H, 1996ApJ...457..557B}. In the local universe, ultraluminous infrared galaxies (ULIRGs) have bolometric luminosities similar to bright quasars and are often in mergers \citep{1986ApJ...305L..45S, 1996ARA&A..34..749S}, and there is observational evidence that ULIRGs form the birthplaces of QSOs \citep{1988ApJ...325...74S, 2011ApJ...741L..38V}.

\cite{2005Natur.433..604D} argue that a merger generates a burst of star formation and leads to strong inflows that feed gas to the SMBH and thereby power the quasar. Hence, the energy released by the quasar expels enough gas to inhibit further star formation and quenches the growth of the black hole.  Also, \cite{2010MNRAS.407.1529H} find that nuclear star formation is more tightly coupled to AGN activity than the global star formation rate of a galaxy.

AGN, in which interstellar material is still rapidly accreting onto SMBHs, are key to the study of the Magorrian relation. The accretion of gas onto the central black hole yields a luminous source of X-ray, UV, and optical photons. Here, UV and X-ray radiation influence the chemistry of the accreting and star-forming gas and affect the thermodynamics of the ISM. Gas clouds in the inner regions of galaxies are exposed to radiation originating from the active nucleus, newly formed stars or both. The hot O and B stars that are formed in the starburst regions are radiating in the far-ultraviolet (FUV) wavelength range ($6<\rm{E}<13.6$ eV), whereas hard X-rays ($1<\rm{E}< 100$ keV) are emanating from the nucleus when gas is falling in. When assuming that the energetics are dominated by either FUV or X-rays and effects by shocks can be neglected, the thermal balance and chemical structure are determined by the radiation field and result in Photon-dominated Regions (PDRs, \cite{1985ApJ...291..722T}) or XDRs \citep{1996ApJ...466..561M}. AGN are thought to be partially obscured by dusty interstellar matter (ISM). X-rays (1-100 keV) have smaller absorption cross sections than UV photons and, therefore, can penetrate large columns ($N> 10^{22}$ cm$^{-2}$). Hence, they provide a probe of the innermost regions of AGNs. Furthermore, the main heating mechanism in X-ray dominated regions (XDRs) is photoionization (Coulomb heating with thermal electrons), whereas in PDRs it is photoelectric heating. Hence, the heating efficiency in XDRs and PDRs is $\sim 30\%$ and 0.1-1$\%$, respectively. We refer the interested reader to \cite{2005A&A...436..397M} for a full introduction to the distinctions between PDRs and XDRs.

\cite{2011ApJ...730...48P} have post-processed the 3-D hydrodynamical models of an AGN torus  \citep{2009ApJ...702...63W} by using the chemical network of \cite{2005A&A...436..397M}. They estimated the X-ray flux emanating from the black hole and found that X-ray irradiation affects the thermodynamics of the AGN torus up to $~100$ pc. They have calculated that the temperature of AGN gas, exposed to X-rays is a factor $\sim 5$ higher than gas in a starburst of equal bolometric power. This has strong effects for star forming clouds, since the Jeans mass (M$\rm_{J}$) scales with the temperature of the ambient gas as M$\rm{_J} \propto$ T$^{3/2}$. Therefore, this might inhibit star formation in the central galactic regions or change the initial mass function [IMF, \cite{2010A&A...522A..24H,2011A&A...536A..41H}]. Moreover, \cite{2009ApJ...702...63W} studied the molecular gas disks around AGN and have suggested that XDR physics may change the distribution of H$_2$ around an AGN. Recent work by \cite{2011ApJ...738...54K} focused on galaxy formation with feedback-regulated star formation and black hole growth. In their simulations they have taken into account the radiative feedback from the central MBH and found that radiative feedback from the MBH locally suppresses star formation and self-regulates its growth. However, they use a monochromatic spectrum for X-ray photons and have a limited chemical network. 

In order to study the effects of X-ray irradiation from an accreting black hole in the central region of a galaxy, we perform 3-D cosmological hydrodynamic simulations by implementing the XDR/PDR chemical network of \cite{2005A&A...436..397M} into Enzo. This work is the first part of a trilogy where we investigate the effects of X-ray irradiation by an AGN on ambient gas with a metallicity of zero and solar. We will study the Magorrian relation and derive observational diagnostics to find the X-ray fingerprints in the high redshift universe in two subsequent papers. Here, as a proof of concept, we show that X-ray physics has been successfully included into Enzo. We further show that treating radiative feedback from the central MBH with a self-consistent XDR/PDR network and a polychromatic energy distribution is crucial to assess the relation between AGN and their host galaxies. In this study, we  concentrate on the effects of X-rays on the multi-phase ISM near the central black hole. This will help us to better understand the conditions of star formation near an AGN and hence the interplay between stellar feedback and radiative feedback from the central BH.

This paper is structured as follows. In Section 2 we detail our implementation of the XDR code into Enzo and the treatment of the polychromatic spectrum. We further discuss the relevant cooling and heating processes in XDRs. In Section 3 we describe the set-up of the simulations. In Section 4 we present the results and implications of our simulations. Finally, in Section 5 we discuss our results. We present some of the outputs of our chemical network in the Appendix.

\section{XDR IMPLEMENTATION}
\subsection{Black Hole Radiation}

We set up our intrinsic specific X-ray flux (F$_i$) to have a spectral shape of the form 

\begin{equation}\label{eq1}
F_{\rm i}(\rm E) = F_0 \Big(\frac{E}{1 keV}\Big)^{-\alpha}\exp(-E/E_c) \mbox{ erg s$^{-1}$ cm$^{-2}$ eV$^{-1}$},
\end{equation}
where F$_0$ is the constant that determines the total X-ray flux in the spectrum, E ($\geq1$ keV) is the energy, $\alpha = 0.9$ is the characteristic spectral index of the power law component of Seyfert 1 galaxies \citep{1990Natur.344..132P, 1995ApJ...438..672M, 1995ApJ...438L..63Z}, and E$_c = 100$  keV is the cut-off energy \citep{1995ApJ...438..672M}. In order to determine F$_0$, one needs an independent estimate of the bolometric X-ray flux F$_{X}$. We estimate the central F$_{X}$ by assuming that only 10$\%$ of the total luminosity (L$_{bol}$) is emitted in X-rays \citep{2010A&A...513A...7S},

\begin{equation}
F_{\rm{X}} = \int_{\geq 1 \rm{keV}} F_i(E)dE = 0.1\times L_{bol} / 4\pi r^2 =  0.1 F_{bol},
\end{equation} 
where r is distance from the central black hole. We assume that the MBH has a radiative efficiency of $\epsilon = 0.1$ and thus has a luminosity $\rm{L}=\epsilon \times \rm{L}_{\rm{edd}} = 1.2 \times 10^{37} (\epsilon/0.1)$ M/M$_{\odot}$ erg s$^{-1}$.

We set M$_{BH}  = 5 \times 10^4$ M$_{\odot}$, which gives a bolometric flux F$_{bol} = 4 \times 10^3$ erg s$^{-1}$ cm$^{-2}$ at our finest resolution of 3.6 pc. The UV part of the AGN spectrum is produced by the usual multi-color \cite{1973A&A....24..337S} accretion disk, with a viscosity parameter of $\alpha =0.1$. Here, $\alpha$ determines the efficiency of the transportation of angular momentum via a turbulent viscosity that is proportional to the local pressure in the disk. Our choice of $\alpha = 0.1$ is consistent with observations \citep{1998AcA....48..677S, 1999AcA....49..391S, 2001A&A...373..251D} and theoretical simulations of global accretion disks \citep{2010MNRAS.408..752P}.

\indent In our XDR models only X-rays with E $\geq 1$ keV are considered. Therefore, we take the integral of equation \ref{eq1} over this energy range and equate it  to the estimated bolometric flux to find F$_0 \approx 0.8$ erg s$^{-1}$ cm$^{-2}$ eV$^{-1}$. Furthermore, with increasing distance from the AGN, the X-ray flux does not solely decrease by 1/$\rm{r}^2$ but also by attenuation along the line of sight. This attenuation is determined by the opacity, which is a function of energy and position

\begin{equation}
\tau(\rm{E},\bold{r}) = \sigma_{pa}(\rm{E})N_H(\bold{r}),
\end{equation}
where $\sigma_{pa}$(E) is the photoelectric absorption cross section per hydrogen nucleus, and N$_H(\bold{r})$ is the total column density of hydrogen between the central black hole and the position $\bold{r}$. Taking this opacity effect into account we calculate the total X-ray flux for a given column density as

\begin{equation}
F_X =  \int_{\geq 1 \rm{keV}} F_0(\frac{\rm{E}}{1 \rm{keV}})^{-\alpha} e^{-E/E_c} e^{-\tau(\rm{E},\bold{r})} dE
\end{equation}
\subsection{Radiative Feedback}

We use the Enzo radiation transport module
$\it{Moray}$ \citep{Wise11_Moray} to calculate the transfer of X-ray radiation field
produced by the black hole.  $\it{Moray}$ solves the radiative transfer equation
with ray tracing that is adaptive in spatial and angular coordinates. The ray normal
directions are determined by HEALPix \citep[Hierarchical Equal Area
isoLatitude Pixelation;][]{HEALPix}.  The rays originate from point
sources and are progressively split when the angular sampling becomes
too coarse, which can happen when the ray either propagates to larger
radii or encounters a fine resolution AMR grid.  

The rays are traced through the grid in a typical fashion
\citep[e.g.,][]{Abel99_RT}, in which we calculate the next cell
boundary crossing.  In each ray segment crossing a single cell, we
solve the radiative transfer equation for a ray,
\begin{equation}
\frac{1}{c} \frac{\partial P}{\partial t} + \frac{\partial P}{\partial
  r} = -\kappa P,
\end{equation}
where $P$ is the monochromatic photon flux associated with the ray and
$\kappa = \kappa(\bold{r},t)$ is the absorption coefficient.  In the static case, its
solution has a simple exponential analytic solution and the photon
flux of the ray is reduced by
\begin{equation}
  \label{eqn:dP}
  \mathrm{d}P = P \times (1 - e^{-\tau}),
\end{equation}
where $\tau$ is the optical depth (see above) across the
ray segment.  Because $\it{Moray}$ equates the photoionization rate to the
absorption rate, it is photon conserving \citep{Abel99_RT, 2006NewA...11..374M}.
Thus, the photoionization and photoheating rates are, respectively,
\begin{equation}
  \label{eqn:kph}
  k_{\rm ph} = \frac{dP}{n_{\rm abs} V_{\rm cell} dt_P}
\end{equation}
and
\begin{equation}
  \label{eqn:gamma}
  \Gamma_{\rm ph} = k_{\rm ph} (E_{\rm ph} - E_i)
\end{equation}
where $V_{\rm cell}$ is the cell volume, $E_{\rm ph}$ is the photon
energy and $E_i$ is the ionization energy of the absorbing material.
In each cell, the resulting $k_{\rm ph}$ and $\Gamma_{\rm ph}$ values
are the sum from all rays that cross that cell, which then are
inputs into the chemistry rate equations and energy equation.

\subsection{Treatment of a General Input Spectrum}

In the original release of $\it{Moray}$ \citep{Wise11_Moray}, radiation is discretized into 
energy bins.  Since then, a more general approach of
\citet{Shapiro04} and \cite{2006NewA...11..374M} has been
implemented that can consider arbitrary (polychromatic) spectral shapes. In our work, 
we use a \textit{polychromatic energy distribution}. In this method, the radiative transfer 
equation is numerically solved before the simulation, giving a relative ionizing photon flux $I_\nu$ as a function of
neutral hydrogen column density $N_H$.  The relative ionizing photon flux
\begin{equation}
  I_\nu(N_H) = \int_{\nu_{\rm th}}^{\infty}
  \frac{L_\nu \sigma_\nu \exp(-\sigma_\nu N_H)}
       {h \nu} \;
       \mathrm{d}\nu
\end{equation}
for H \textsc{i}, He \textsc{i}, and He \textsc{ii} is computed and
stored for 300 column densities, equally log-spaced over the range $N_H =
10^{12-25} \rm{cm}^{-2}$.  $\nu_{\rm th}$ is the ionization energy
threshold for each absorbing species.  Also, the average photon energy
\begin{equation}
  \langle E \rangle = 
  \frac{\int_{\nu_{\rm th}}^{\infty} h \nu I_\nu(N_H) \; \textrm{d}\nu}
       {\int_{\nu_{\rm th}}^{\infty} I_\nu(N_H) \; \textrm{d}\nu}
\end{equation}
is pre-calculated as a function of hydrogen column density $N_H$.  The
attenuation across each segment is thus d$P = I_\nu(N_H+dN_H) -
I_\nu(N_H)$, which is determined by interpolating from
pre-calculated tables.  These values of d$P$ and $\langle E \rangle$
are used in Equations (\ref{eqn:kph}) and (\ref{eqn:gamma}), resulting
in a solution that retains the full spectral shape until the source
radiation is fully attenuated.

\subsection{XDR physics}

The dominant parameter that drives the chemical and thermal structure of the gas in XDRs is the ratio of the X-ray energy deposition rate to gas density, H$_X$/n. This occurs because the molecular destruction and heating rates per unit volume due to X-ray induced ionizations are proportional to nH$_X$, whereas the molecular formation and cooling rates are generally proportional to n$^2$ times a rate coefficient \citep{1996ApJ...466..561M}.

The main heating mechanism in XDRs is photoionization. The rate at which photoionizations happen depends on the cross-section, which is a function of the frequency of the radiation and the properties of the atom. The absorption cross section of X-rays roughly scales with energy as $\sim 1/ E^3$, which allows X-rays to penetrate deep into interstellar clouds. A photon with an energy of 1 keV penetrates a hydrogen column of about 10$^{22}$ cm$^{-2}$ for solar metallicity.

UV radiation is absorbed by outer-shell electrons, whereas X-ray photons are absorbed by inner-shell electrons. When an electron from an inner shell (e.g., the K-shell)  is ejected it leaves the atom behind in an excited state with an inner shell vacancy and more energy than is required to remove the least bound electron. The inner shell vacancy can be filled by an outer-shell electron (e.g., an L-shell electron). The excess energy then can be shed in one of two ways: by emitting photons (fluorescence) or by ejecting outer electrons, which is referred to as the Auger effect.

When inner-shell ionization is followed by Auger ionization, two high energy electrons are produced: the first from the primary ionization with an energy equal to the photon energy minus the binding energy of the ejected electron, and the second from Auger ionization which has an energy equal to the difference in binding energies of electrons in the inner and outer shell minus the binding energy of the outer shell electron. This can be hundreds or thousands of eV. In an ionized gas, these suprathermal electrons undergo frequent elastic collisions with thermal electrons and their kinetic energy is converted into heat.

In a predominantly neutral gas, these suprathermal electrons can collisionally excite and/or ionize ions before interacting with free electrons. These secondary ionizations are more important for H, H$_2$ and He than the primary ionizations, which is a consequence of the large primary photo-electron energies \citep{2005A&A...436..397M}. The importance of secondary ionization is determined by the ionization fraction x = n(H$^+$)/n(H) since the probability that a fast electron will share its energy with a free electron, compared to its probability of striking an atom or molecule, is proportional to this fraction. The ratio of secondary to primary ionizations is $\sim$ 26, depending on the chemical composition of the gas \citep{1999ApJS..125..237D}. If the ionization fraction of the gas is $x>0.9$, then most of the kinetic energy goes into heat through Coulomb interactions with the ambient thermal electrons. If $x< 0.9$, then approximately 40$\%$ of the primary photo-electron energy goes into secondary ionization and excitation (H I excitation, mainly Ly$\alpha$) whether the gas is atomic \citep{1985ApJ...298..268S,2010MNRAS.404.1869F} or molecular \citep{1991ApJ...377..158V}. 

The heating efficiency of a molecular gas is larger than that of a mostly neutral atomic gas. When H$_2$ is ionized by a fast electron and recombines dissociatively, about 10.9 eV of the ionization energy can go into kinetic energy. On the other hand, an H$_2^+$ ion can react on with a hydrogen molecule to produce H$_3^+$, which may then either recombine dissociatively or react with other species. Due to the strong exothermic nature of the recombination reaction, \cite{1973ApJ...179L.147G} argue that for every H$_3^+$ ion formed about 8 eV goes into gas heating.

In general, ro-vibrational H$_2$ cooling is important in XDRs due to the high gas temperatures (T$_g>$ 1000 K). The ro-vibrational bands of molecular hydrogen can be excited by the nonthermal electrons produced through X-ray ionization and thermal collisions with e$^-$, H and H$_2$. Collisional de-excitation of vibrationally excited H$_2$ can be an important heating source when non-thermal electrons dominate the excitation (at low H$_X$/n). These excited H$_2$ molecules can also enhance chemical reactions with an activation barrier \citep{2005A&A...436..397M}.

XDRs are exposed to X-rays as well as FUV photons. Internally generated FUV photons are produced when energetic nonthermal electrons collide with atomic and molecular hydrogen. These collisions result in the emission of Ly$\alpha$ and Lyman-Werner band photons, and significantly affect the chemistry of the X-ray irradiated gas clouds. If the electron fraction of the gas is x $\leq$ 0.01, then about 40$\%$ of the energy deposited by X-rays results in the production of these internal FUV photons. Thus, the dense interiors of X-ray irradiated clouds can chemically resemble photodissociation regions with significant abundances of neutral oxygen as well as singly ionized iron, silicon and carbon \citep{1996ApJ...466..561M}. 

Most of the FUV photons that are locally produced by the nonthermal electrons will be absorbed by dust grains. The resulting temperature of the dust grains is proportional to the locally absorbed X-ray energy per hydrogen atom (H$_X$). Therefore, if the dust abundance is high, then there is less energy per dust particle and the average dust temperature T$_d$ drops and is given by

\begin{equation}
T_d = 1.5\times10^4(H_X /  x_d)^{0.2} \mbox{ K},
\end{equation}
where $x_d=1.9 \times 10^{-8}$ is the grain abundance (for solar metallicity) and $H_X$ is in erg s$^{-1}$ (Yan 1997).  The minimum grain size is set to a$_{min}=10 \rm{\AA}$. In XDRs, the dust temperature for the same impinging flux by energy is lower than in PDRs \citep{2005A&A...436..397M}. Also, in XDRs the gas heating efficiency is 10-50$\%$ whereas it is 0.1-1$\%$ in PDRs. Furthermore, the chemistry in XDRs is less stratified, with C$^+$, C, CO co-existing over large columns, than in PDRs.

At high temperatures (T $>$ 5000 K) the gas cooling is dominated by collisional excitation of Ly$\alpha$, and forbidden transitions of [O I] ($\lambda \lambda$ 6300, 6363 $\mu$m), [C I] ($\lambda \lambda$ 9823, 9850 $\mu$m), [Fe II]($\lambda \lambda$ 1.26, 1.64 $\mu$m), and [Si II] ($\lambda \lambda$ 6716, 6731$\mu$m). 

The cooling below a few 1000 K is dominated by the fine-structure lines of [OI] 63 $\mu$m, [SiII] 35 $\mu$m, [CII] 158 $\mu$m, [CI] 269 and 609 $\mu$m. Rotational and vibrational transitions of H$_2$, H$_2$O, and CO may also be important when these molecules are abundant. Cooling lines may become optically thick and such column density dependent transfer effects are included using a non-local escape probability method \citep{2005A&A...440..559P}. At high densities, and if the grain temperature is less than the gas temperature, gas-grain collisions can be a source of gas cooling and grain heating. For further details, we refer the interested reader to \citet{2005A&A...436..397M}.

\subsection{Modifications to the 2005 XDR code}

The XDR code as described in \cite{2005A&A...436..397M} has been modified by including all the heavy elements up to iron with abundances $> 10^{-6}$. As a result, the elements treated in the chemistry of the code were extended with Ne, Na, Ar and K. In the original XDR code, the doubly ionized states of C, N, O, S, and Fe were included. Here, we treat the singly and doubly ionized states of all elements, also He$^{2+}$. Most of the additional reactions were adopted from \cite{2011ApJ...736..143A}, who give an elaborate description on the X-ray ionization of heavy elements. We followed their method to determine the secondary ionization rates of He$^{2+}$, which were scaled to those for hydrogen, by using the peak electron impact cross sections as obtained from Tarawa \& Kato (1987). The primary ionization rates were obtained from \cite{1995A&AS..109..125V}, where we assumed that no Auger ionization is possible, implying that only one electron is released per absorbed X-ray photon.

\section{Simulations set-up}

In this work, we use the cosmological adaptive mesh refinement (AMR) code Enzo-2.0 \footnote{http://enzo-project.org}  (Bryan $\&$ Norman 1997; O'Shea et al. 2004) that is modified to include XDR physics. We perform simulations in a three-dimensional periodic box with a side length of 3 h$^{-1}$ Mpc, initialized at $z=99$. The size of the root grid is 128$^3$ with three nested subgrids, each refined by a factor of two. The
finest grid has an effective resolution of 1024$^3$ with a side length of 375 h$^{-1}$ kpc. Refinement is restricted to the finest grid and occurs during the simulations whenever the baryonic matter, or dark matter density, is greater than the mean density by a factor of four. The maximum
level of refinement that is reached in the finest grid is 10, allowing us to have a resolution of 3.6 physical pc. 
We use Wilkinson Microwave Anisotropy Probe seven-year cosmological parameters (Komatsu et al. 2009), which have the following values: $\Omega _{\Lambda}$ = 0.734, $\Omega _{m} = 0.266$,  $ \Omega _b$ = 0.0449, $\sigma_8$ = 0.81, and $h=0.701$. Here, $\Omega _{\Lambda}$ is the vacuum energy, $\Omega _{m}$ is the matter density, $\Omega _b$ is the baryon density, $\sigma_8$ is the variance of random mass fluctuations in a sphere of radius 8 $h^{-1}$ Mpc, and $h$ is the Hubble parameter in units of 100 km s$^{-1}$ Mpc$^{-1}$.

In order to see the impact of X-ray radiation from the central MBH on the surrounding gas, we have performed two simulations. In these simulations we use the primordial chemical network of Enzo until z = 15, where the most massive halo with a mass of $2 \times 10^8$ M$_{\odot}$ forms. At z=15, we insert a MBH with a mass of $5 \times 10^4$ M$_{\odot}$, in order to be roughly consistent with the Magorrian relation,  and consider its radiative feedback afterwards. In one of the runs, we start using the XDR tables compiled for solar metallicity gas (hereafter referred to XDR$_{\rm{S}}$), and in the other one, we keep using the primordial chemical network of Enzo (hereafter referred to XDR$_{\rm{Z}}$). We compare a non-zero and zero metallicity run that both enjoy the full Enzo-XDR treatment. In both simulations, we employ the radiative transfer module, \textit{Moray}, which uses a polychromatic energy distribution. Hence, the only difference between the two runs is the metallicity of the ambient gas that is exposed to X-ray radiation from the accreting MBH.

Our XDR chemical network consists of more than 170 species. We have constructed tables of XDR solutions for species abundances and temperatures for solar metallicity and for a wide range of X-ray flux F$_{\rm{X}} = 10^{-1.25}-10^{5.5}$ erg cm$^{-2}$ $\rm{s}^{-1}$, density n = 10-10$^6$ cm$^{-3}$ and column density  N$_{\rm{H}} = 10^{20}$-10$^{24}$ cm$^{-2}$. This large parameter space enables us to model the ISM properties close to an AGN properly. Here, the XDR model assumes that the gas is instantaneously in local thermal equilibrium (LTE). In XDRs, the dominant heating mechanism is Coulomb heating and is transferred through collisions between electrons and gas. The slowest process of these two is the collisions. The particle collision time, which is the time scale on which heating occurs as electrons, generated by photoionization, thermalize with the ambient medium through elastic collisions and is given by $1/(\sigma n dv)$ in seconds. Here, n is the number density in cm$^{-3}$, $\sigma$ is the cross-section in cm $^{-2}$, and dv is the velocity in cm$3$ s$^{-1}$. This corresponds to t $<3$ yr  for densities in excess of $10^2$ cm$^{-3}$. Furthermore, at solar metallicity the cooling time $kT/n\Lambda$ is $<30$ yr for densities in excess of $10^2$ cm$^{-3}$. The time step of our simulations dt $\gg$ 30 yrs (on the order of  600 yrs). Hence,  the LTE assumption holds.
 
We use Enzo's 9 species (H, H$^+$, H$^-$, He, He$^+$, He$^{2+}$, H$_2$, H$_2^-$, and e$^-$) non-equilibrium chemical network for zero metallicity case \citep{1997NewA....2..181A,1997NewA....2..209A}. In the XDR$_{\rm{S}}$ case, we run both the XDR routine and the chemical network of Enzo and compare the computed temperatures to determine if the cell is X-ray dominated. We take the highest value of the two found temperatures and continue to iterate for the next step. By taking the maximum we basically divide the simulation box into XDR and non-XDR zones. Of course, we might overestimate the  temperature in the grid when X-ray heating and non-X-ray heating are comparable, but this pertains to a very small part of the grid given the deep penetration of X-rays into dense gas, where shocks are of modest importance.

We choose solar metallicity for our XDR tables since this is a good approximation for regions around SMBHs, even inside z $\sim$ 6 AGNs \citep{2007AJ....134.1150J}. Moreover, attenuation of X-ray photons in a non-zero metallicity ISM has never been done before for a high redshift accreting MBH, and here we show that we can treat the impact of X-rays on such non-zero metallicity ambient gas. By choosing solar metallicity, we seek to highlight the full range of effects that metals introduce in X-ray exposed gas. 
 
Furthermore, in our tables, we also store the abundances of all species that Enzo's chemical network is using, attached to the corresponding temperature values.  For a given unattenuated flux, we create a table of attenuated fluxes as a function of column density. Combined we thus have an XDR grid of models in n, F$_{\rm{X}}$, and N$_{\rm{H}}$ (varying metallicity will be presented in paper II) that uses $\it{Moray}$ to compute the full (chemical, thermal and hydrodynamic) response of X-ray exposed gas at non-zero metallicity. In the code, we calculate the column densities between the cell and MBH and incident flux with $\it{Moray}$, take the corresponding flux values from the pre-calculated tables and apply the radial dependence. By using the flux, density, and  column density computed in Enzo, we find the corresponding temperature values and the species abundances from the pre-computed XDR grid tables for each cell and feed these back into Enzo. We stop the simulations after $\sim$ 42 Myr at z=13.54 This is long enough for ambient gas in a radius of a few hundred pc to be affected by X-rays emanating from the $5 \times10^4$ M$_\odot$ MBH and to find (if at all) an equilibrium again.

In this work, we did not take into account the effects of momentum transfer from the ionizing radiation field. Addition of radiation pressure to cosmological hydrodynamic simulations is not very well studied and hence we did not want to introduce another unknown parameter to our simulations. However, recently \cite{2012MNRAS.427..311W} have performed the first cosmological simulations with radiation pressure that is calculated by solving the RT equation, and have shown that the main mechanism for blowing away the gas from the central region of a galaxy is SNe. But one thing to keep in mind is that, in their simulations they reached maximum densities of $\sim$ 100 cm$^{-3}$ and did not take into account dust. In his work, \cite{2011ApJ...732..100D}  has made a systematic study on the effects of radiation pressure in a static, dusty H \textsc{ii} region and has shown that radiation pressure becomes important at high densities (n$\geq 10^3$ cm$^{-3}$.) Hence, here we only concentrate on the X-ray effects and leave the investigation of the interplay between radiation pressure and X-ray irradiation for a follow up paper.

For the analysis of our cosmological simulations we use $\it{yt}$, a cross-platform analysis toolkit written in Python \citep{2011ApJS..192....9T}. 

\section{Results $\&$ Implications}

In the simulations that we perform, as a proof of concept, we see significant differences between the solar metallicity and zero metallicity cases. Below we explain the physical processes that play a role in causing the differences in the ISM properties of the modelled halos.

\begin{figure*}
\includegraphics[angle=0,width=16cm]{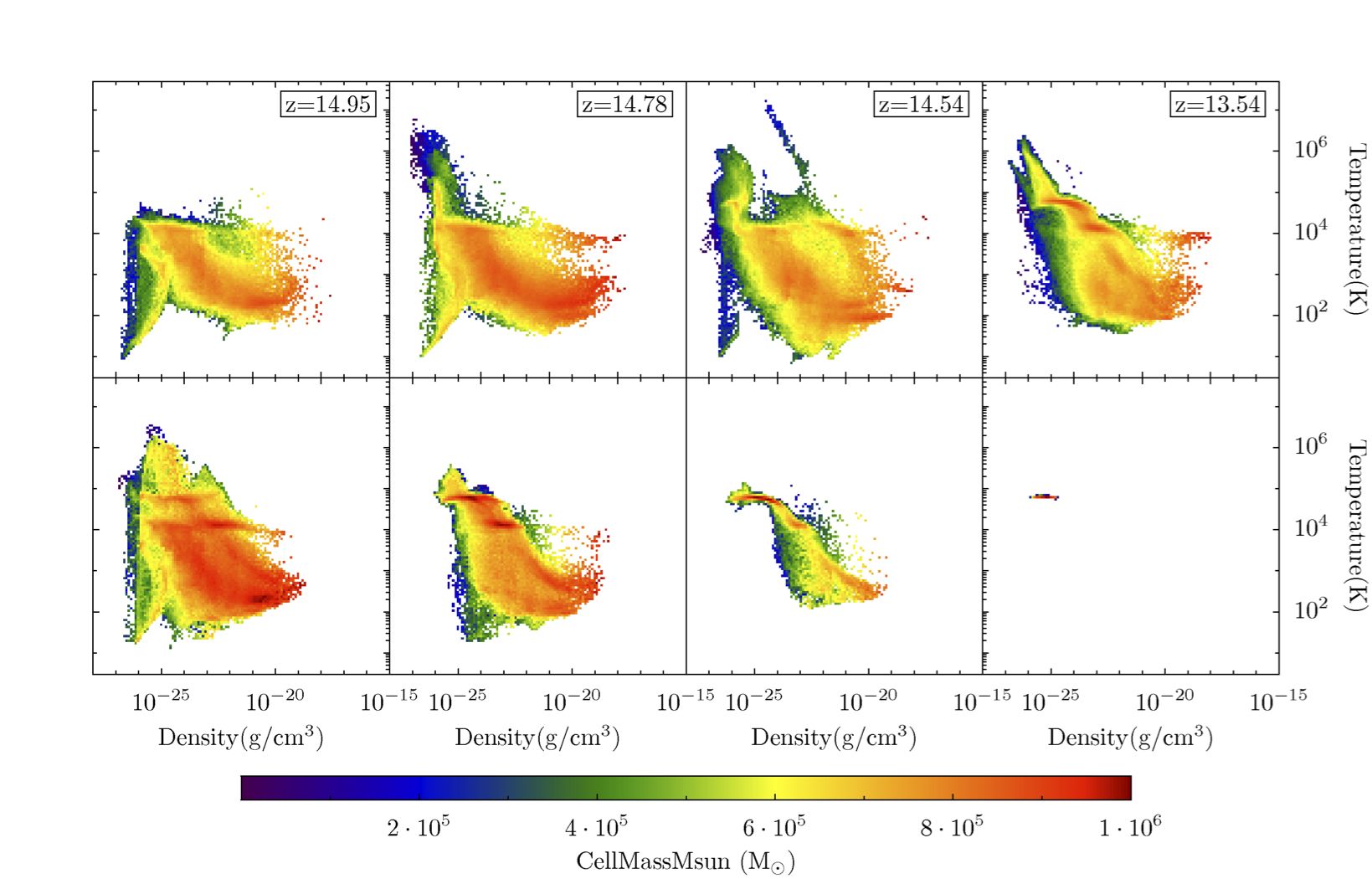}
\caption {Density-temperature phase diagrams within a sphere of 500 pc diameter for the XDR$_{\rm{S}}$  (top) and XDR$_{\rm{Z}}$ (bottom) cases at z = 14.95, 14.78, 14.54 and 13.54, from left to right. Note the missing mass already at z = 14.78 in the XDR$_{\rm{Z}}$ case.}\label{fig:phase1}
\end{figure*}

\begin{figure*}
\includegraphics[angle=0,width=8cm]{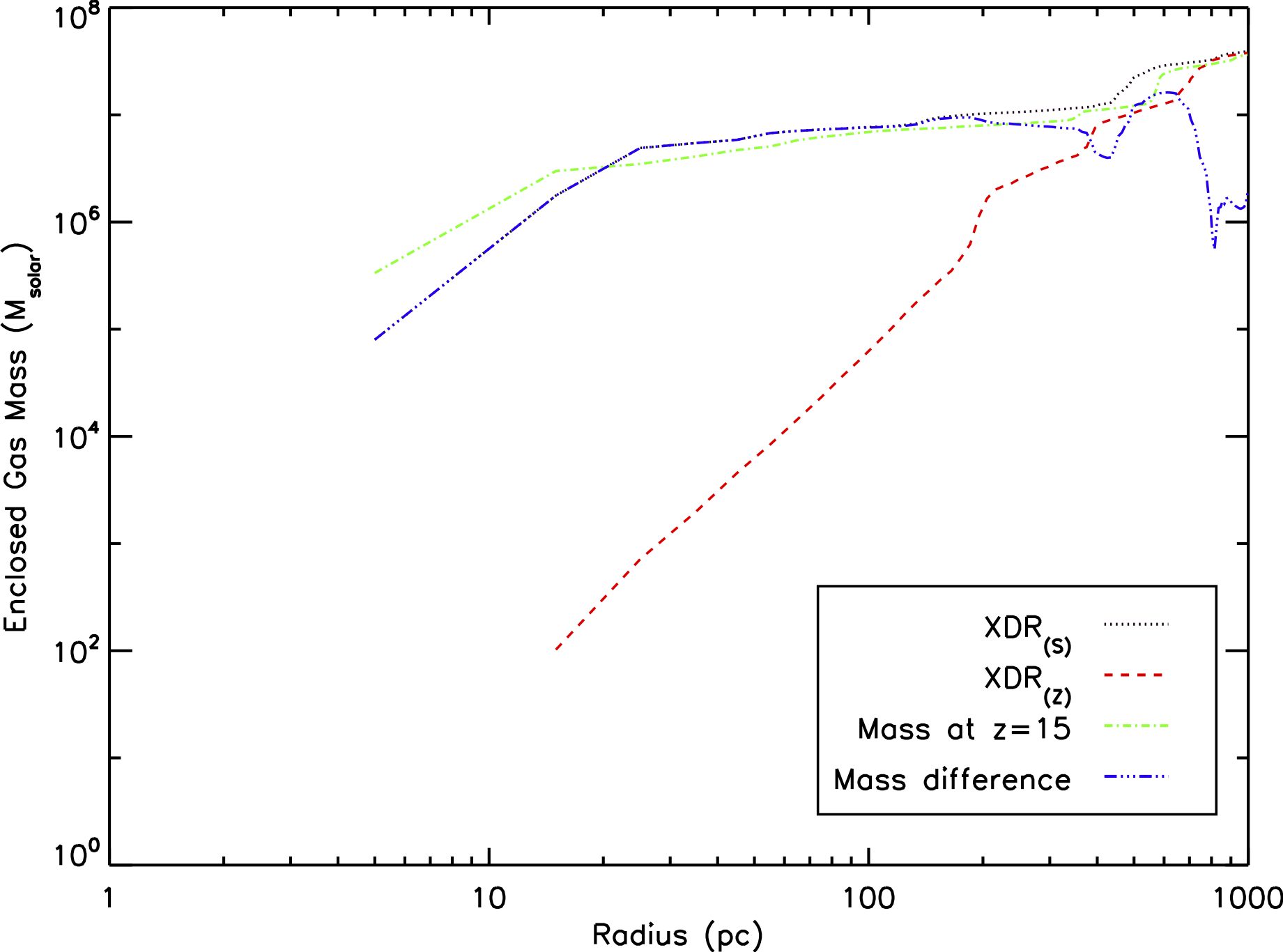}
\vspace{0.3cm}
\includegraphics[angle=0,width=8cm]{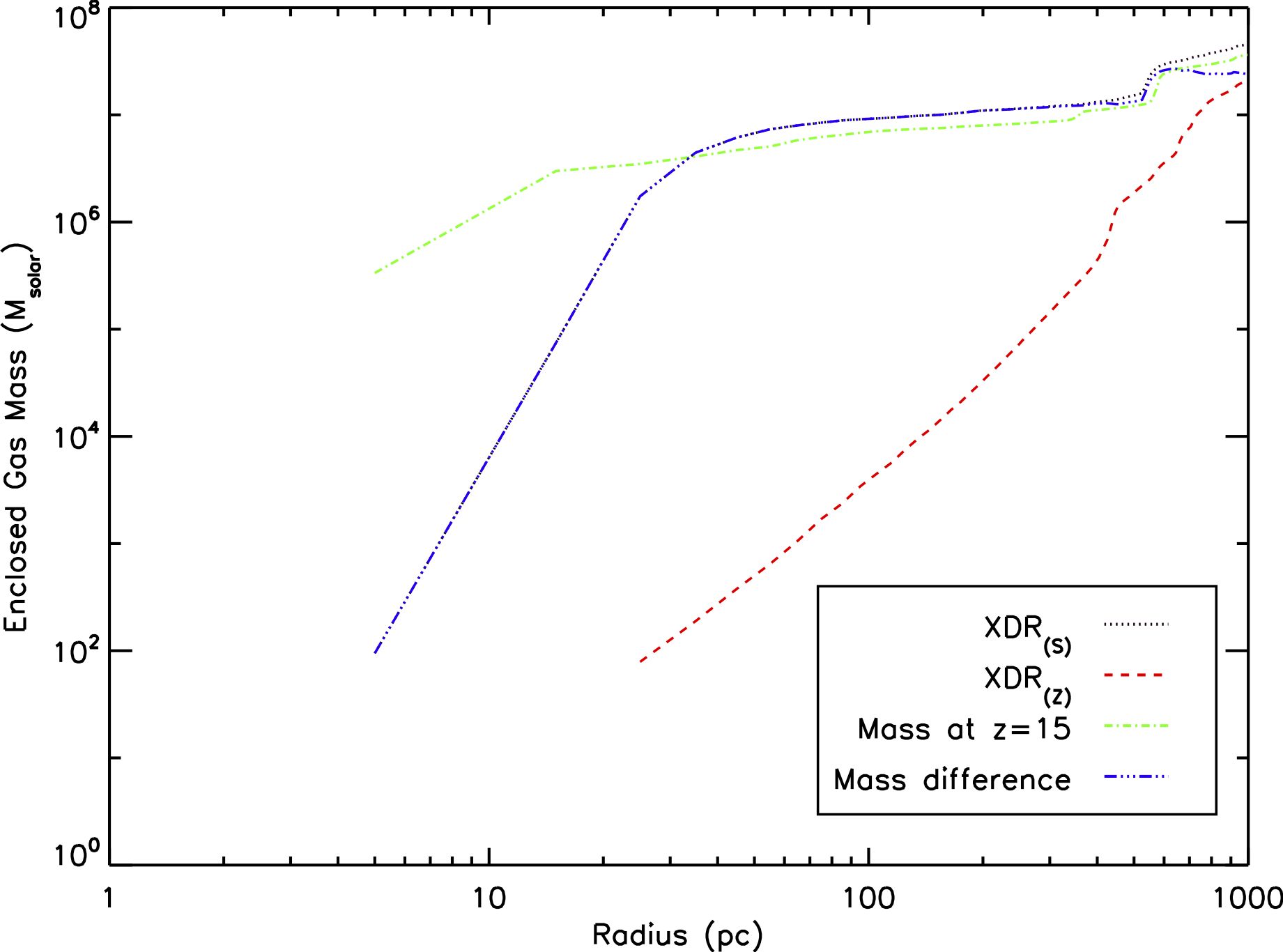}
\caption {Enclosed gas mass of the XDR$_{\rm{S}}$ (black dotted line) run, the XDR$_{\rm{Z}}$ (red dashed line) run, and the gas mass difference between the two runs (blue dotted dashed line) at z = 14.78 (left) and 14.54 (right) as a function of radius. The green dotted-dashed line is the enclosed gas mass of the halo at z = 15.}\label{fig:menc} 
\end{figure*}

Shortly after we insert a MBH, we already note considerable differences between the two runs in the central region around the MBH. In Figure \ref{fig:phase1}, we plot the density-temperature histograms, within a sphere of 500 pc diameter, for the XDR$_{\rm{S}}$ (top row) and XDR$_{\rm{Z}}$ (bottom row) runs at redshifts z = 14.95, 14.78, 14.54 and 13.54 from left to right. At  $\rm{z}=14.78$, only 5.8 Myr after we turn on the central MBH, phase diagrams show that there is less gas present in the XDR$_{\rm{Z}}$ case than in the XDR$_{\rm{S}}$ case. This difference in the gas mass, becomes even more pronounced after 12 Myr we turn on the MBH ($\rm{z} = 14.54$). To estimate the amount of gas that is missing in the zero metallicity case we compare the enclosed gas mass of both simulations as a function of radius at redshift $\rm{z} = 14.78$ (left) and $\rm{z}=14.54$ (right) in Figure \ref{fig:menc}. We calculate the difference in enclosed gas mass (blue dot-dashed line) in the inner 200 pc to be $8\times10^6$ M$_\odot$  between the XDR$_{\rm{S}}$ and XDR$_{\rm{Z}}$ cases, after 5.8 Myr we turn on the MBH ($\rm{z} = 14.78$).

\begin{figure*}[!htb]
\includegraphics[angle=0,width=15cm]{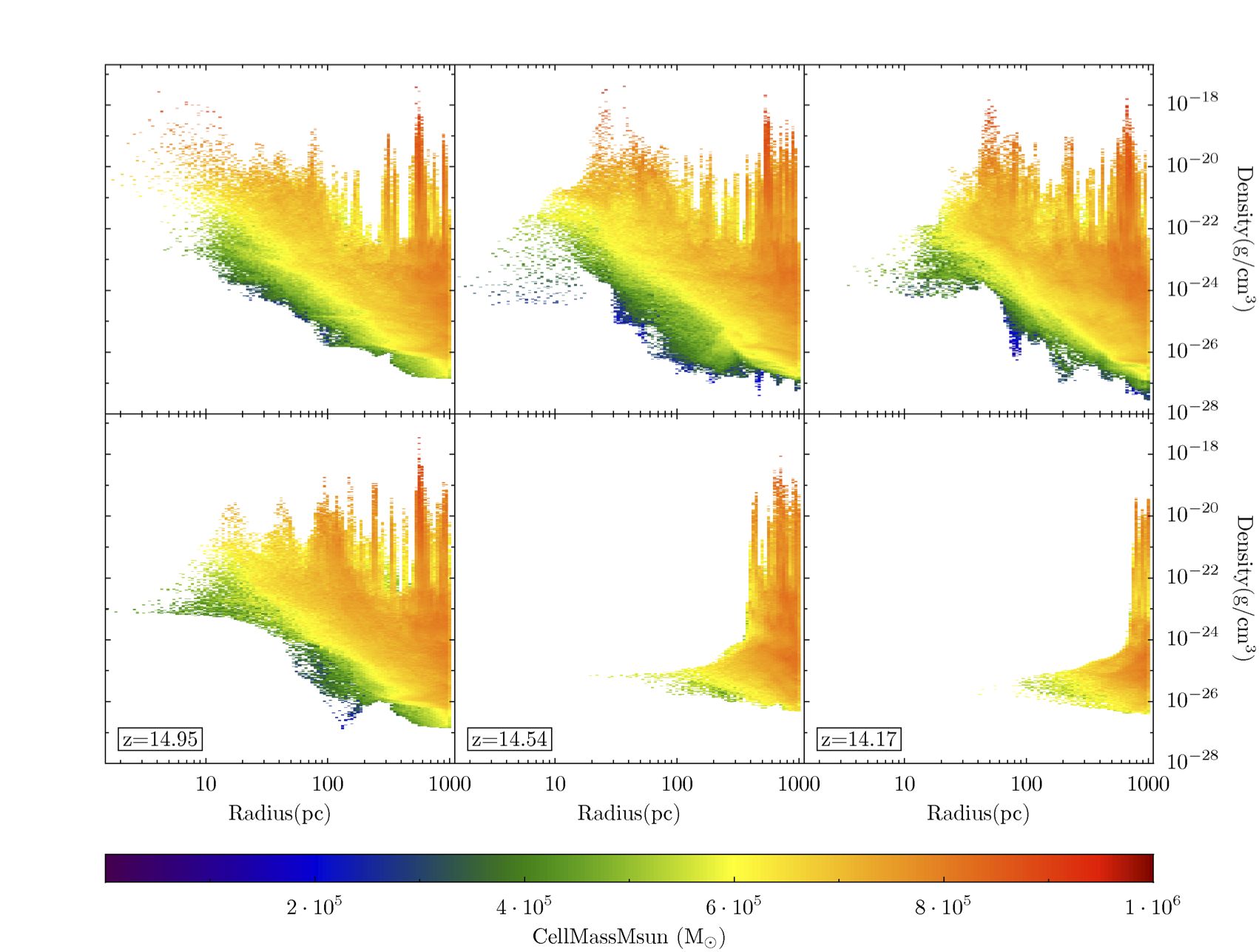}
\caption {2D profiles of density versus radius for the XDR$_{\rm{S}}$ (top) and XDR$_{\rm{Z}}$ (bottom) at z = 14.95 (left), 14.54 (middle), and 14.17(right).}\label{fig:rad1}
\end{figure*}

\begin{figure*}[!htb]
\includegraphics[angle=0,width=15cm]{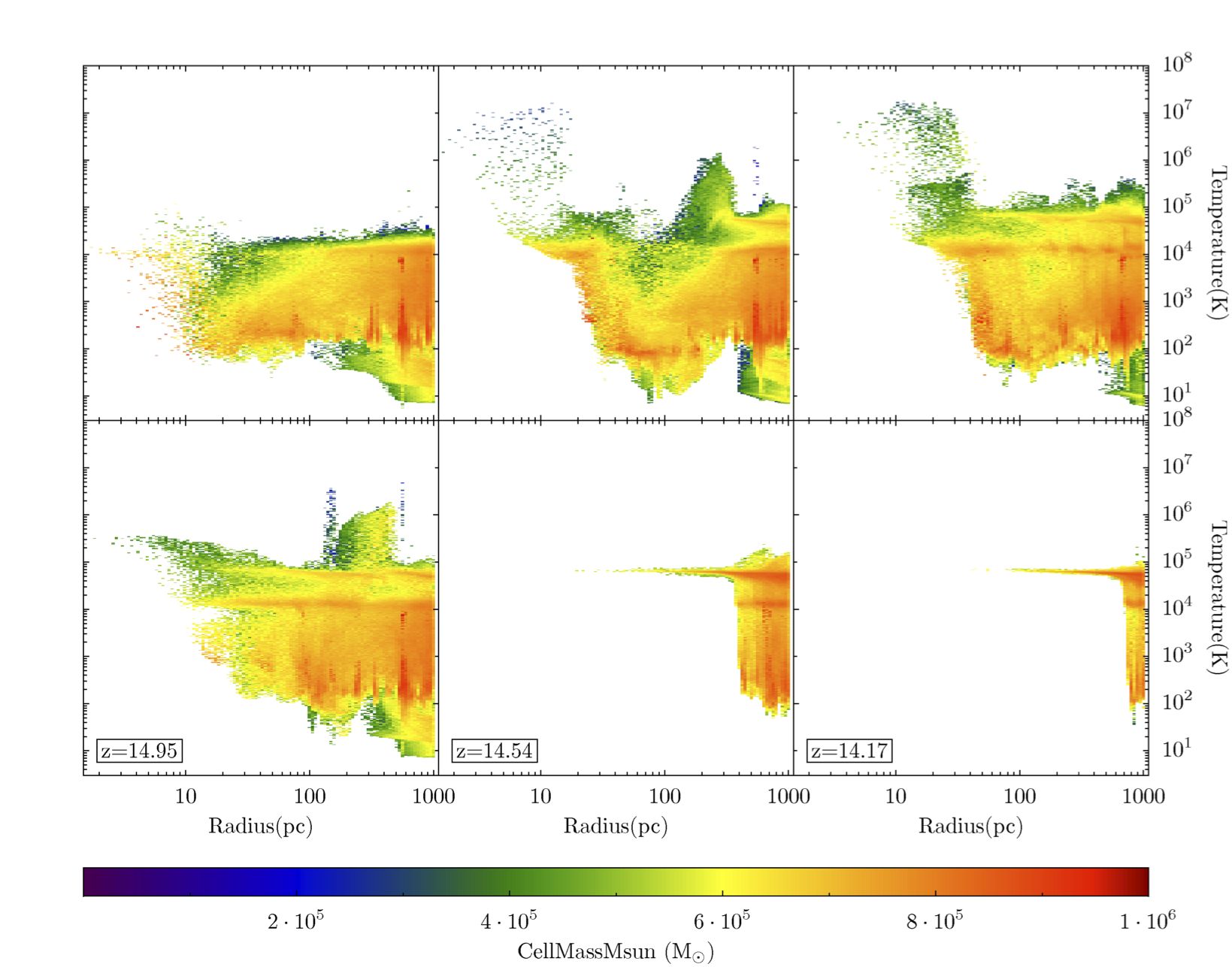}
\caption {2D profiles of temperature versus radius for the XDR$_{\rm{S}}$ (top) and XDR$_{\rm{Z}}$ (bottom) at z = 14.95 (left), 14.54 (middle) and 14.17 (right).}\label{fig:rad2}
\end{figure*}

Furthermore, when we look at the 2D radial density and temperature profiles (see Figures \ref{fig:rad1} and \ref{fig:rad2}) of the XDR$_{\rm{S}}$ and XDR$_{\rm{Z}}$ cases we see significant differences in the inner 100 pc while at larger radius (R $>500$ pc) the differences become less pronounced. (see section \ref{sec:TD}). 

\begin{figure*}[!htb]
\includegraphics[angle=0,height=11cm]{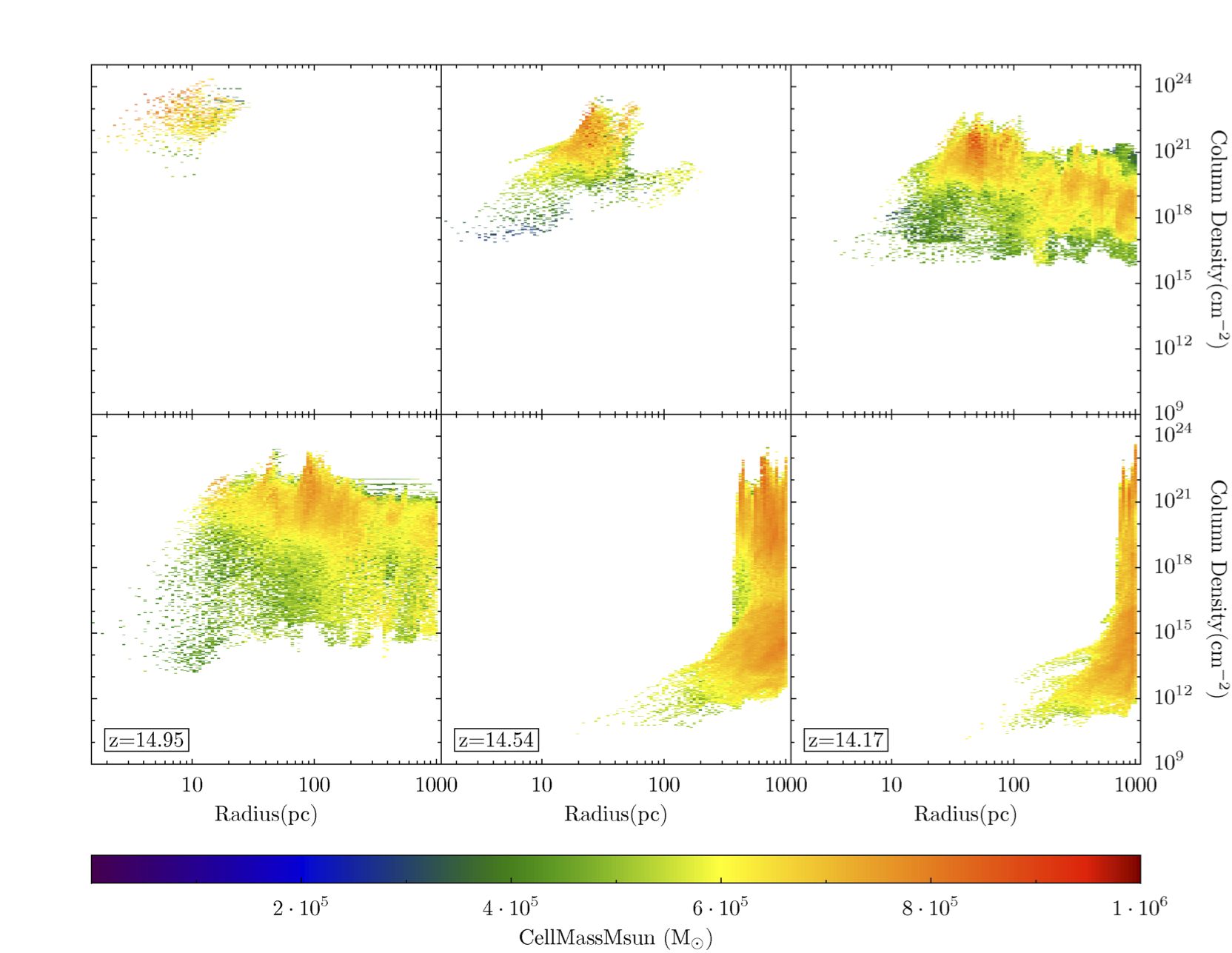}
\caption {2D profiles of column density versus radius for the XDR$_{\rm{S}}$ (top) and XDR$_{\rm{Z}}$ (bottom) runs at z = 14.95 (left), 14.54 (middle) and 14.17 (right).} \label{fig:col}
\end{figure*}

The missing mass in the inner 200 pc of the XDR$_{\rm{Z}}$ run can be explained as follows. In the XDR$_{\rm{S}}$ case because of the metals, X-rays produced by the central MBH see a high opacity, about a factor of 3-30 higher than for zero metallicity gas. This is because metals like C, N, O, Si, and Fe have large cross sections, for inner shell absorptions (see Figure 5.3 in \cite{2006PhDT........11M} for E $>1$ keV and \citet{1983ApJ...270..119M}), which lead to an ``opacity wall'', if column densities exceed 10$^{22}$ cm$^{-2}$. Thus, the energy deposition rate into the medium close to MBH is high and the X-ray power is dissipated only locally. On top of that, the cooling ability of solar metallicity gas is high. Therefore, the cooling time is short beyond the opacity wall. Combined, large column densities of 10$^{24}$ cm$^{-2}$ are reached in the central 20 pc as metal enriched gas falls in and builds up sufficient column to shield itself from irradiation (see Figure \ref{fig:col}). However, in the XDR$_{\rm{Z}}$ case, in the absence of metals, the maximum column density of the bulk of the gas in the inner 20 pc ($\sim 10^{20}$ cm$^{-2}$) is 3 orders of magnitude less than in the XDR$_{\rm{S}}$ case at $\rm{z}=14.95$ and the X-rays penetrate to larger columns. Thus, the energy deposition rate into the medium close to the MBH is significantly less and X-ray power is dissipated globally. This difference in column densities is shown in Figure \ref{fig:col}, where we plot the 2D column density-radius profiles for both cases at redshift z = 14.95, 14.54 and 14.17. In this plot it is easily seen that the gas column, viewed from the MBH, builds up much faster in the XDR$_{\rm{S}}$ case due to the efficient accumulation of solar metallicity gas. Here, in the XDR$_{\rm{S}}$ case there are no column densities shown at radius larger than 30 and 110 pc at z= 14.95 and 14.54, respectively. This is due to the fact that we do not track the column density once the ray is terminated after it is almost completely absorbed.

Therefore, in the XDR$_{\rm{Z}}$ case, the impinging X-ray flux onto the relatively low density gas that sits at larger radii is much higher. Hence, this forms an $\it{X-ray}$ $\it{induced}$ H\textsc{ii} $\it {region}$ in the  XDR$_{\rm{Z}}$ case already after only 5.8 Myr we turn on the central MBH.

\begin{figure*}[!htb]
\includegraphics[angle=0,height=11cm]{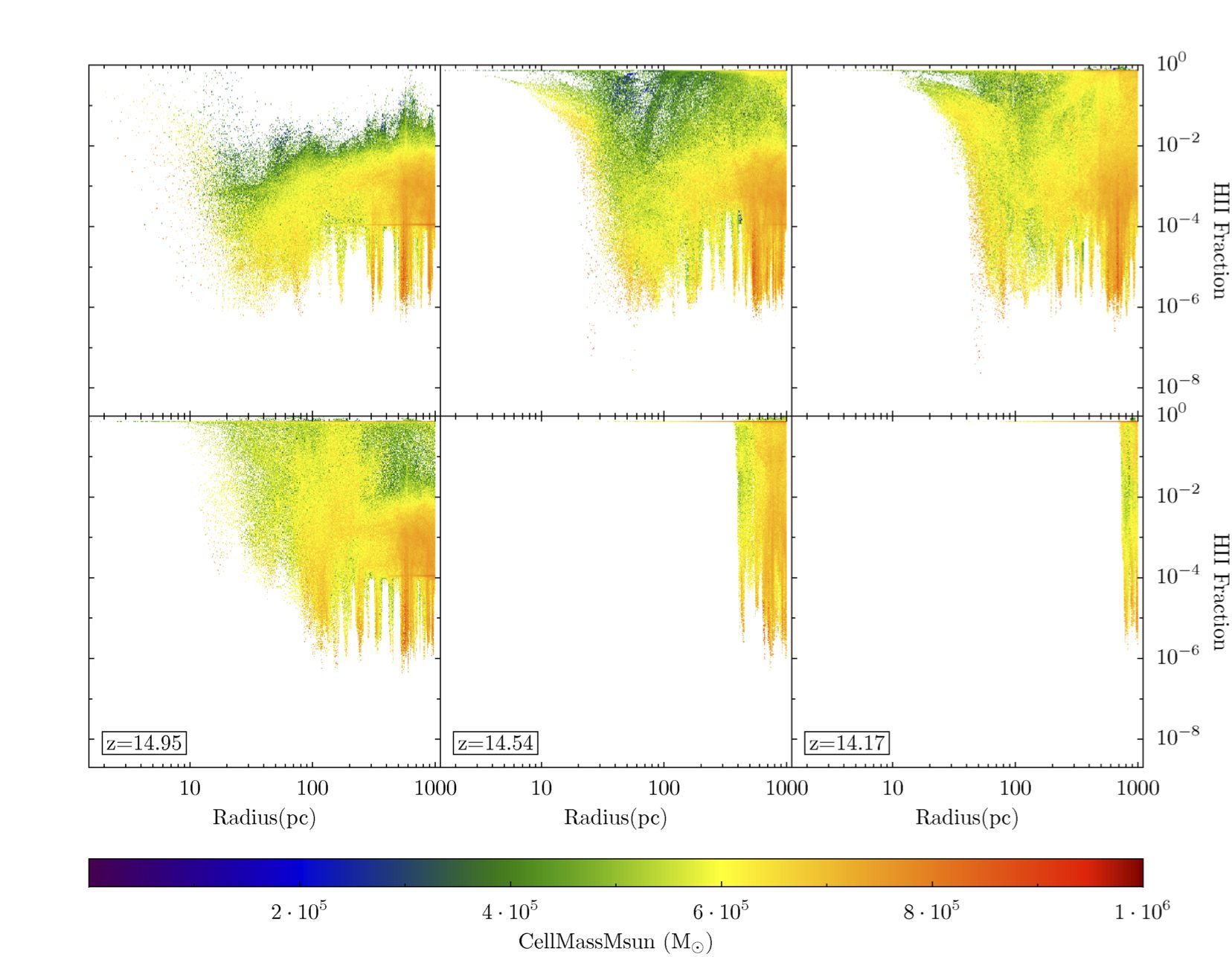}
\caption {2D profiles of H{\sc I} ionization fraction versus radius for the XDR$_{\rm{S}}$ (top) and XDR$_{\rm{Z}}$ (bottom) runs at z = 14.95 (left),14.54(right) and 14.17 (right).} \label{fig:HI}
\end{figure*}

The H \textsc{ii} ionization front will sweep through the surrounding medium until the recombination rate inside the H \textsc{ii} region balances the energy output of the MBH (see Figure \ref{fig:HI} for the H{\sc I} ionization fraction). As a consequence, in an attempt to establish pressure balance, the newly formed H \textsc{ii} region will expand, thus driving large quantities of gas to larger radii. This also further lowers the column densities at small radii in the XDR$_{\rm{Z}}$ case. A back of the envelope calculation, where we compare the gas kinetic energy to the energy in X-rays, shows that the energy input from the black hole (L$_{Edd}$ $\sim10^{42}$ erg s$^{-1}$) is sufficient to drive $\sim 10^7$ M$_\odot$  gas out with an average speed of $\sim$ 100 km s$^{-1}$ in 6.5 Myr. 

\subsection{Dynamics of the H \textsc{II} regions}

\begin{figure*}[!htb]
\includegraphics[angle=0,width=12cm]{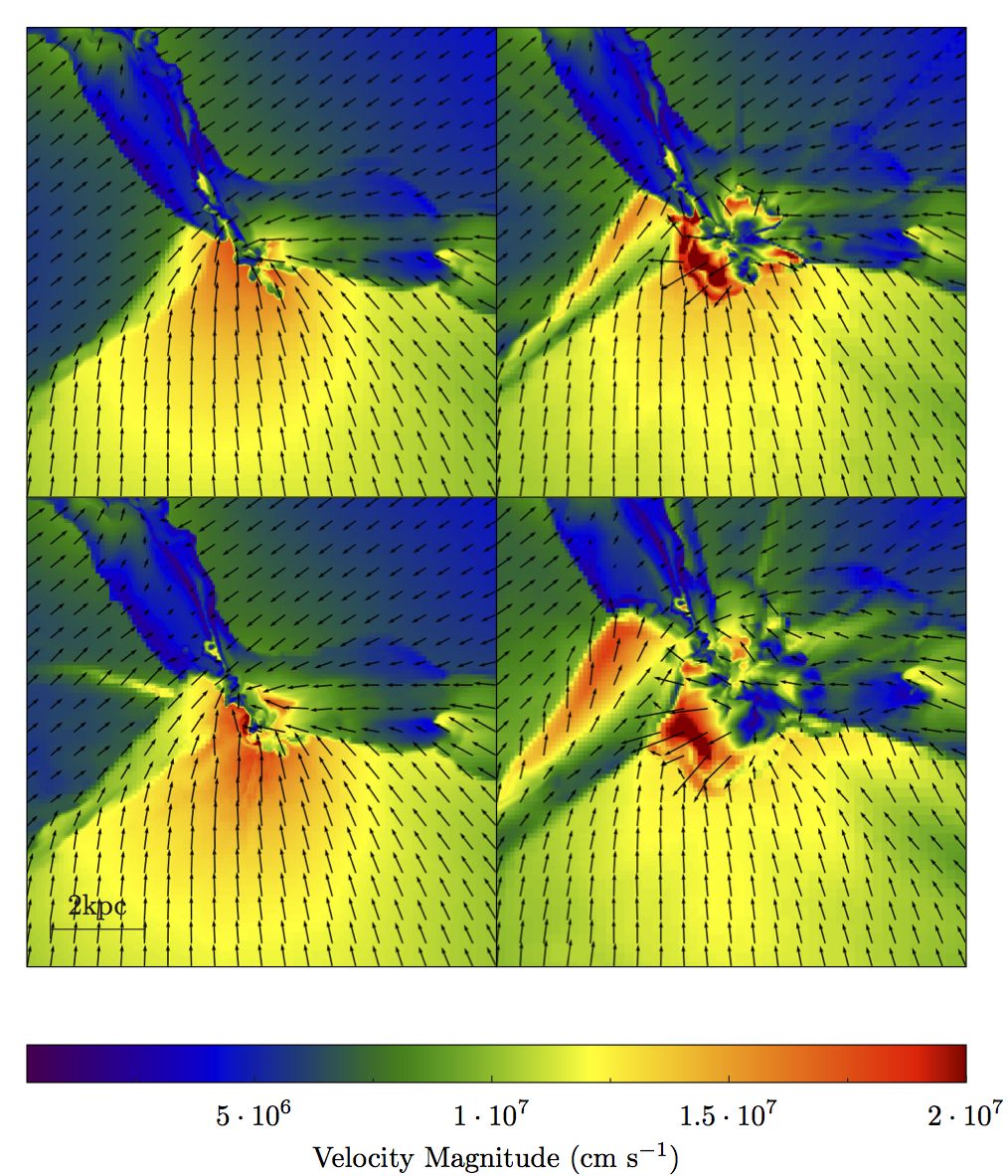}
\caption {Velocity magnitude slice along the y-axis overlaid with velocity vectors for the XDR$_{\rm{S}}$ (left column) and XDR$_{\rm{Z}}$ (right column) runs at z = 14.78 (top) and 14.54 (bottom).}\label{fig:vel}
\end{figure*}

\begin{figure*}
\includegraphics[angle=0,width=16cm]{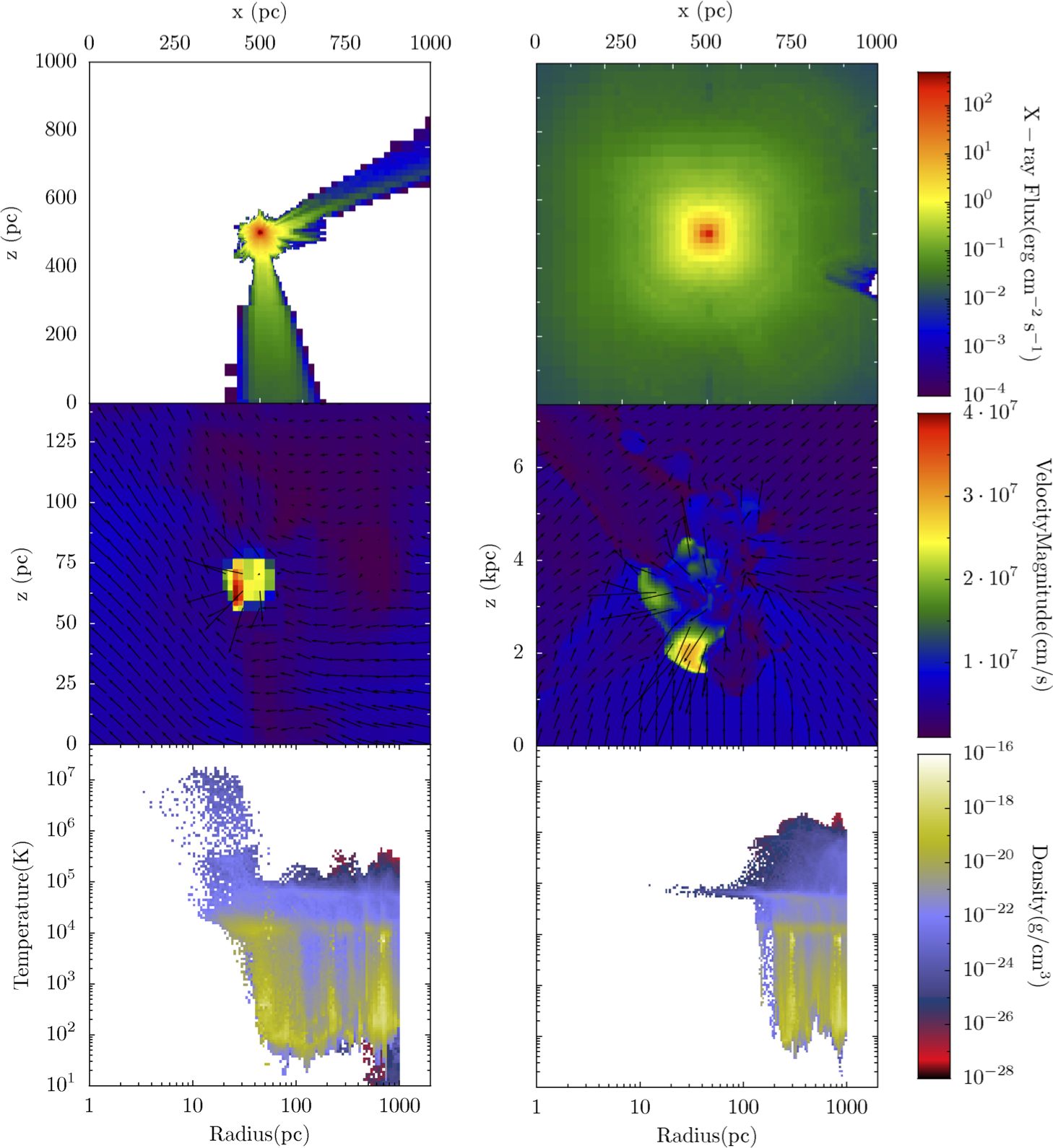}
\caption {X-ray flux (top) and velocity magnitude (middle) slices along the x-axis, and temperature vs radius profile, colour coded with density (bottom) for the XDR$_{\rm{S}}$ run at z = 14.17 (left), 13.54 (right).  The velocity plot is overlaid with velocity vectors in order to show the outflow.}\label{fig:vel2}
\end{figure*}

In Figure \ref{fig:vel}, we show the velocity magnitude at z = 14.78 (top) and 14.54 (bottom) for the XDR$_{\rm{S}}$ (left column) and XDR$_{\rm{Z}}$ (right column) runs overlaid with velocity vectors showing the inflow/outflow. Indeed, at $\rm{z}=14.78$ (after only 5.8 Myr we turn on the MBH) we see an outflow in the XDR$_{\rm{Z}}$ case, at a speed of $\sim 100$ km s$^{-1}$, which is caused by the X-ray induced H \textsc{ii} region. However, in the XDR$_{\rm{S}}$ case, due to the opacity wall, the X-rays are strongly attenuated. This leaves the gas at larger radii virtually unaffected. Due to the lack of a large scale H \textsc{ii} region in the XDR$_{\rm{S}}$ run, the bulk of the gas keeps falling towards the MBH, increasing the densities around the black hole. One thing to keep in mind here is that our smallest cell size is 3.6 pc, which means that we might be overestimating the metals' opacity effect and hence the energy deposition rate. It is possible that higher resolution simulations will show the formation of an H \textsc{ii} region in the XDR$_{\rm{S}}$ case as well, although at a much smaller spatial scale than in the XDR$_{\rm{Z}}$ run. In fact, at z = 14.17 (23 Myr after we turn on the MBH), we see that X-rays start to penetrate much further ($>$ 200 pc) in the XDR$_{\rm{S}}$ case as well and drive an H \textsc{ii} region, but with a delay of 17 Myr with respect to XDR$_{\rm{Z}}$ case, which pushes apart the central opacity wall with a speed of $\sim 200$ km s$^{-1}$. This is shown in  Figure \ref{fig:vel2} where we plot the X-ray flux (top) and velocity magnitude (middle) slices along the x-axis, and 2D profiles of temperature vs radius for the XDR$_{\rm{S}}$ run at z=14.17 (left column) and 13.54 (right column). The H \textsc{ii} region forms after 23 Myr we turn on the MBH ($\rm{z} = 14.17$), expands to a few kpc in 20 Myr at z = 13.54 (middle row) and drive the gas away from the central MBH (bottom row). 

To understand this delay, two effects are relevant. First, in the XDR$_{\rm{Z}}$ case, the initial H \textsc{ii} region radius is larger because of the lower X-ray optical depth through zero metallicity gas. Second, for a local ionizing photon rate S$_i$, the Str$\ddot{o}$mgren radius scales as R$_S \propto (S_i/n^2)^{1/3}$, the recombination time scales as t$_s \propto 1/n$, and the ionization front velocity scales as V$_I \propto S_i/nR^2$. Because X-rays are absorbed in denser gas and at smaller radii in the XDR$_{\rm{S}}$  run, the H \textsc{ii} region grows more slowly when compared to the XDR$_{\rm{Z}}$ case.

\subsection{Thermodynamics}\label{sec:TD}

At small radius (R $< 100$ pc), as a consequence of high metal opacity, the attenuated X-rays heat up the gas in the XDR$_{\rm{S}}$ case as shown in Figure \ref{fig:rad2}. Therefore, the temperature in the inner 20 pc in the XDR$_{\rm{S}}$ case increases from 10$^5$ K to almost $\sim 10^7$ K. As a consequence of this the dÄensities drop from $10^{-18}$ g cm$^{-3}$ to $10^{-24}$ g cm$^{-3}$ as shown in Figure \ref{fig:rad1}. However, in the XDR$_{\rm{Z}}$ case, the densities at $\rm{z}=14.95$ in the inner 20 pc are 5 orders of magnitude less than in the XDR$_{\rm{S}}$ case. The X-ray induced H \textsc{ii} region sweeps away most of the gas from the inner 100 pc, keeping the densities low (10$^{-26}$ g cm$^{-3}$) and temperatures around 10$^5$ K at larger radii. 

\begin{figure*}[!htb]
\includegraphics[angle=0,width=16cm]{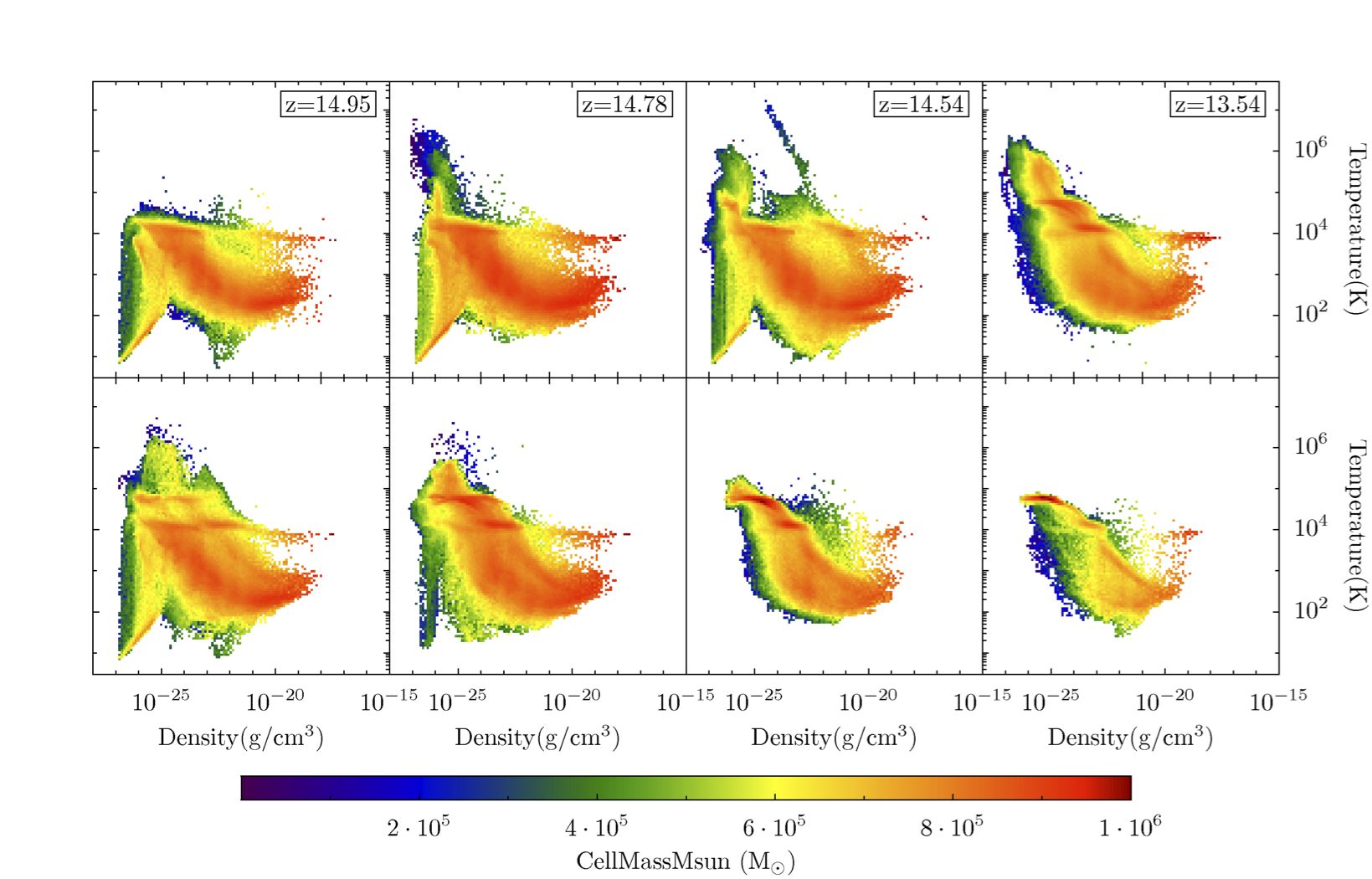}
\caption {Density-temperature phase diagram within a sphere of 1 kpc diameter for the XDR$_{\rm{S}}$ (top) and XDR$_{\rm{Z}}$ (bottom) cases at z = 14.95, 14.78, 14.54 and 13.54, from left to right.}\label{fig:phase2}
\end{figure*}

At large radius (R $>$ 500 pc), we do not see much difference in the multiphase ISM of the XDR$_{\rm{S}}$ and XDR$_{\rm{Z}}$ runs. This is shown in Figure \ref{fig:phase2} where we plot the density-temperature phase diagrams within a sphere of 1 kpc radius for the XDR$_{\rm{S}}$ (top) and XDR$_{\rm{Z}}$ (bottom) runs at redshifts z = 14.95, 14.78, 14.54 and 13.54, from left to right. In a gas with modest ionization degree and weak X-ray radiation field the H$^-$ route will drive the formation of H$_2$ and HD, due to the availability of free electrons. The presence of dust will boost the formation of H$_2$ \citep{2009A&A...496..365C}. However, an increase in H$_2$ abundance will also increase the H$_2$ ionization rate, and hence the gas heating rate, when X-rays are present. In our simulations, most of the gas has temperatures of a few thousand K, and at those thermally unstable temperatures H$_2$ is the dominant cooling and heating channel. Therefore, we do not see a fundamental difference in the density and temperature profile of the ambient gas between the two runs.

In the XDR$_{\rm{S}}$ run we see that at redshifts z = 14.78, as well as slightly later z = 14.54, high density gas ($> 10^{-20}$ g cm$^{-3}$) is more abundantly present at temperatures of $\geq$ 500 K (see Figure \ref{fig:phase2}). There are two effects that play a role here. First of all, in the XDR$_{\rm{S}}$ run the opacity is large enough to absorb any X-ray photon below 5-10 keV. Also, we reach column densities of 10$^{24}$ cm$^{-2}$ at small distances from the MBH, as shown in Figure \ref{fig:col}. This explains the lower H{\sc I} ionization rates in the XDR$_{\rm{S}}$ case as shown in Figure \ref{fig:HI}. Along these large columns, the X-ray flux is almost completely absorbed by the gas and thus the gas at larger distances is not much affected. Secondly, as mentioned before, solar metallicity gas can cool down and evolve to higher densities faster than zero metallicity gas, which was also found in the PDR studies of \cite{2011ApJ...737...63A}. During the fast collapse, adiabatic heating raises the temperature of the gas in the XDR$_{\rm{S}}$ case as $\propto n^{3/2}$ while for X-ray heating this goes as $\propto$ n. However, in the XDR$_{\rm{Z}}$ run the maximum column density, within the central 200 pc, is only 10$^{20}$ cm$^{-2}$, and the X-rays are not fully attenuated in this central region. Thus, the incident X-ray flux at large distances is high, leading to a high heating rate and H$_2$ formation rate. Furthermore, due to the long cooling times of the zero metallicity gas in the XDR$_{\rm{Z}}$ case, it takes longer to reach high densities and adiabatic heating is not significant.
	
\section{Discussion}

As a proof of concept, we investigated the importance of metals in the vicinity of an Eddington-limited X-ray emitting MBH. The high temperatures ($\rm{T}> 10^6$ K, Figure \ref{fig:rad2}) that are found in the inner regions around the MBH in the XDR$_{\rm{S}}$ case might have severe consequences. These high temperatures can quench the gas accretion onto the MBH. Consequently, the accretion power decreases and thus negative radiative feedback becomes self-limiting. Furthermore, high opacity due to metals might aid star formation in some regions by shielding the high density gas against X-ray irradiation and allow gas to cool and collapse. On the other hand, in metal-poor regions, if an X-ray induced H \textsc{ii} region forms, the latter can either quench star formation by removing gas or induce star formation by gas compression: As shown in our zero metallicity simulation, the X-ray driven H \textsc{ii} region sweeps out a significant amount of mass which means that the growth of the black hole will be reduced. Moreover, this would also quench the star formation in the inner regions around the black hole. But, the H \textsc{ii} ionization front might also induce star formation by compressing gas where otherwise stars would not form in gravitationally stable gas. Also, the redistribution of gas will cause a much flatter density profile in the central region than in the high metallicity case. 

In this work, we showed that the difference between zero and solar metallicity gas irradiated by X-rays is very large and that in order to study the Magorrian relation a self-consistent treatment of the effects of metals on X-ray physics should be included. Here, we only concentrate on the X-ray effects and did not take into account the radiation pressure. As discussed earlier, the effects of momentum transfer from the ionising radiation field might further help to sweep away the gas from the central region of the MBH when densities of n$\geq 10^3$ cm$^{-3}$ are reached. Also, in our simulations we compare the computed temperatures from the XDR routine and the chemical network of Enzo and take the highest value of the two found temperatures. By taking the maximum we basically divide the simulation box into XDR and non-XDR zones. Of course, we might overestimate the temperature in the grid when X-ray heating and non-X-ray heating are comparable, but this pertains to a very small part of the grid given the deep penetration of X-rays into dense gas, where shocks are of modest importance. In a follow-up paper, we will include a star formation recipe, a Bondi Hoyle accretion rate for the growth of the MBH and $\rm{H}_2$ self-shielding.

\section*{Appendix}

In the following, we show a few models from the grid of XDR models. These are merely intended to illustrate the dynamic range and variety in XDR physics that Enzo has been augmented with. A copy of the electronic tables will be made available upon request to the authors.

\begin{figure*}
\includegraphics[angle=0,width=7cm]{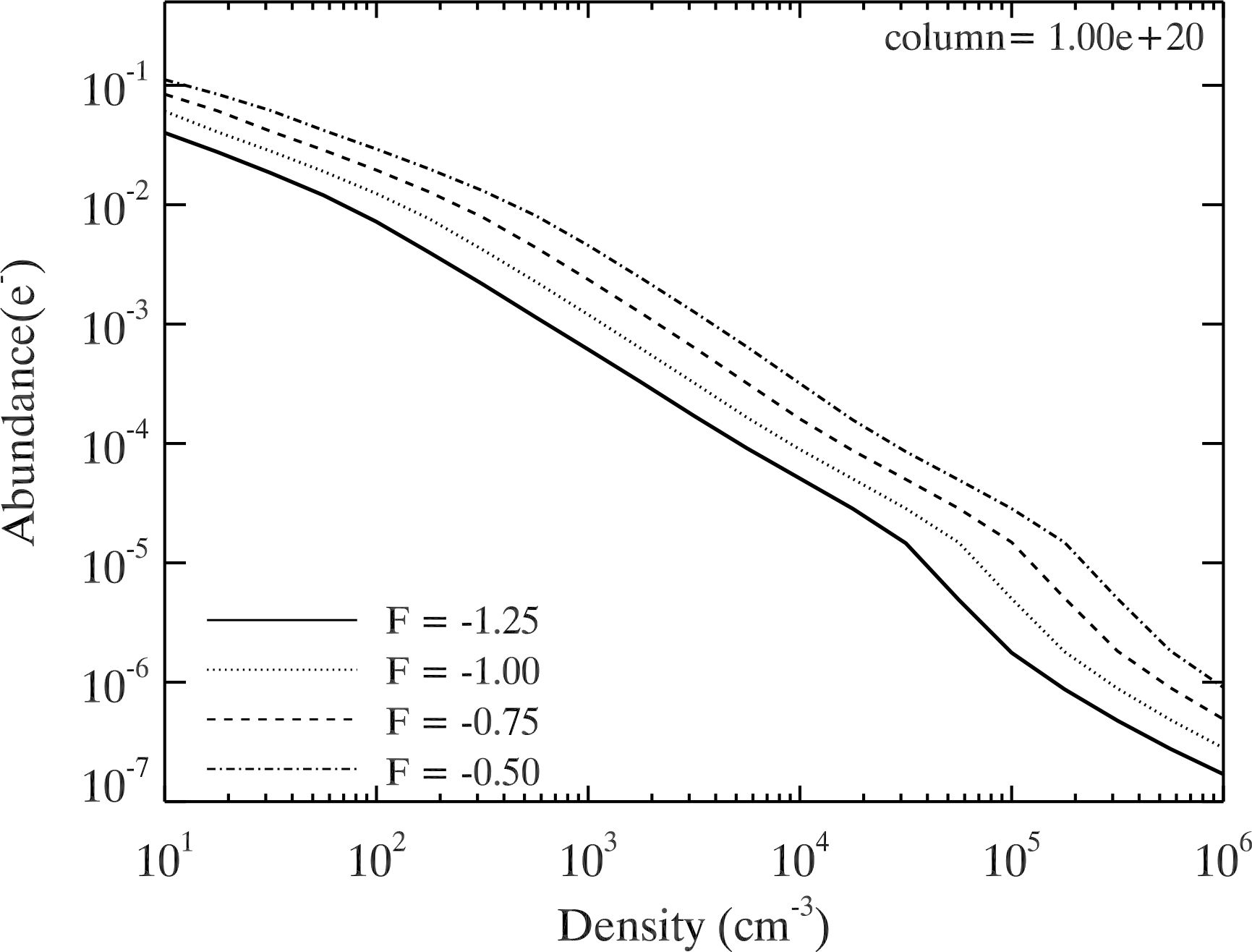}
\vspace{0.05cm}
\includegraphics[angle=0,width=7cm]{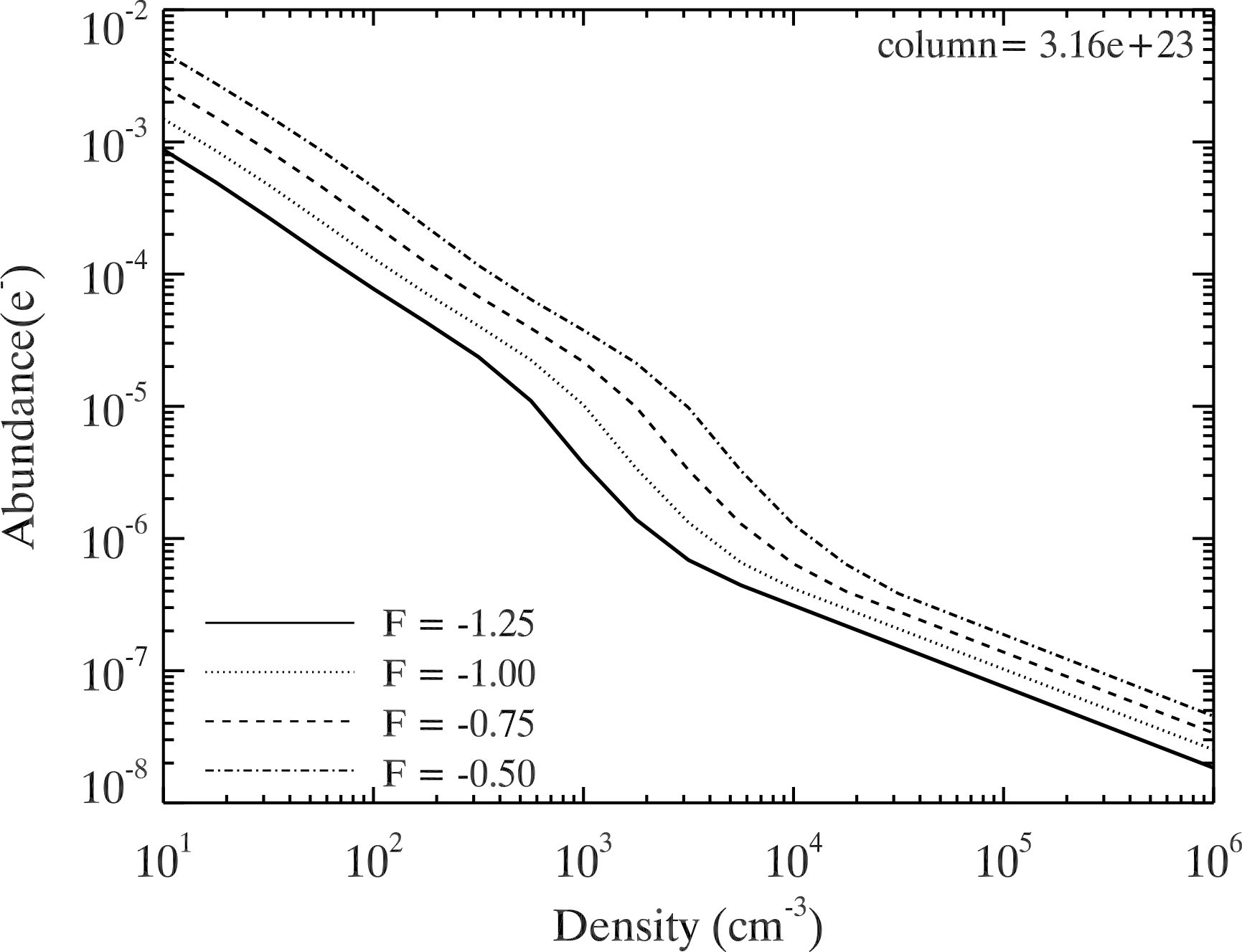}\\
\includegraphics[angle=0,width=7cm]{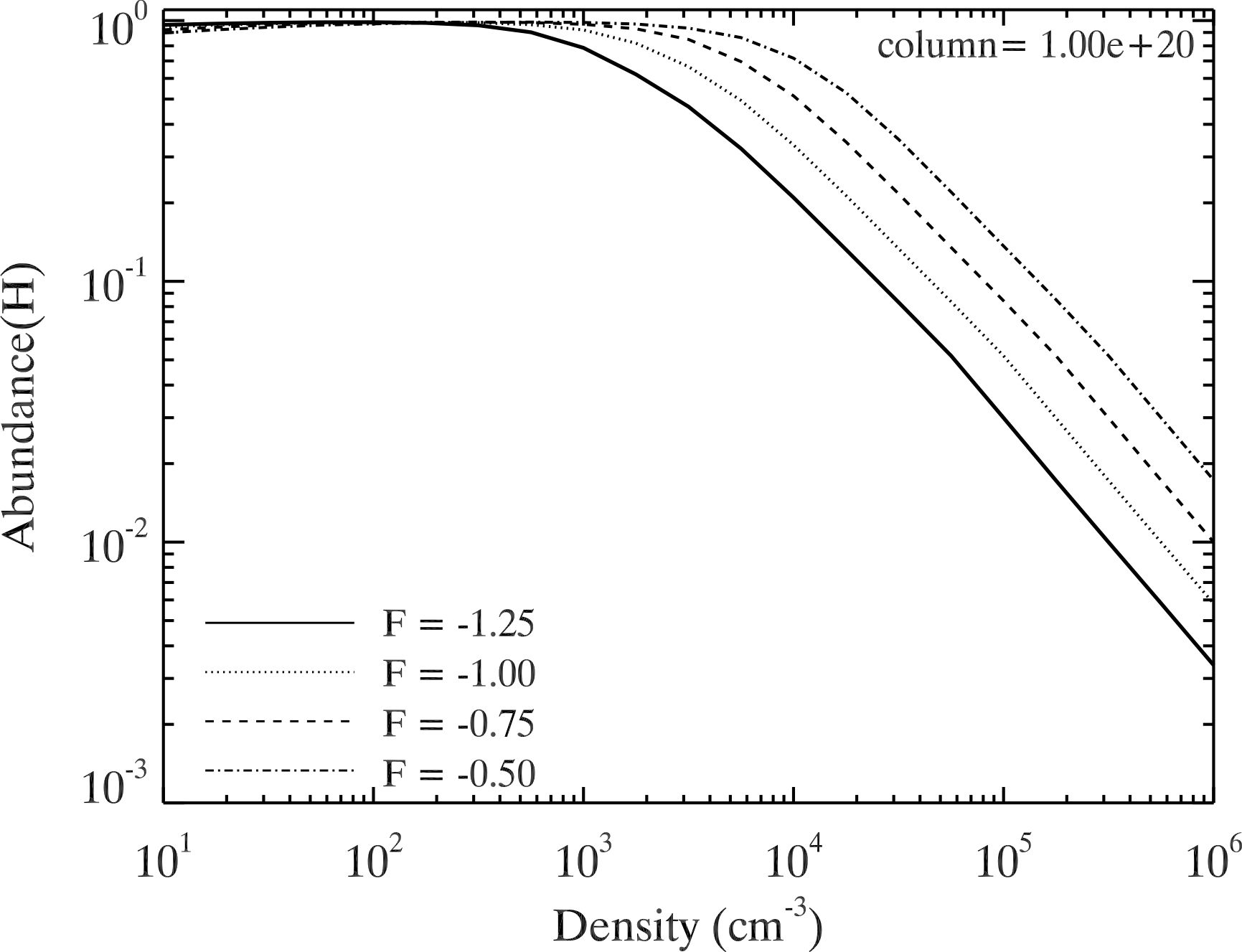}
\vspace{0.05cm}
\includegraphics[angle=0,width=7cm]{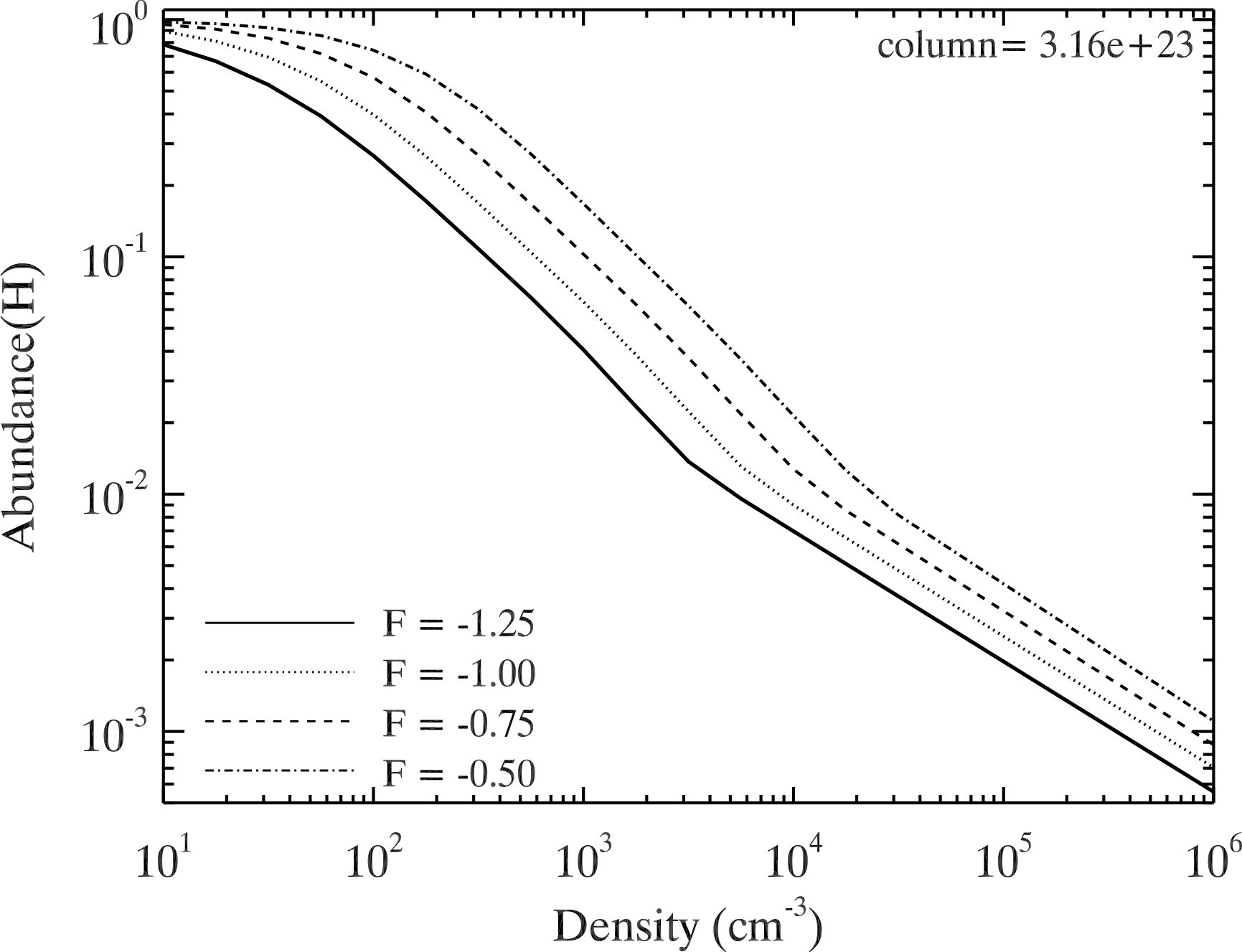}\\
\includegraphics[angle=0,width=7cm]{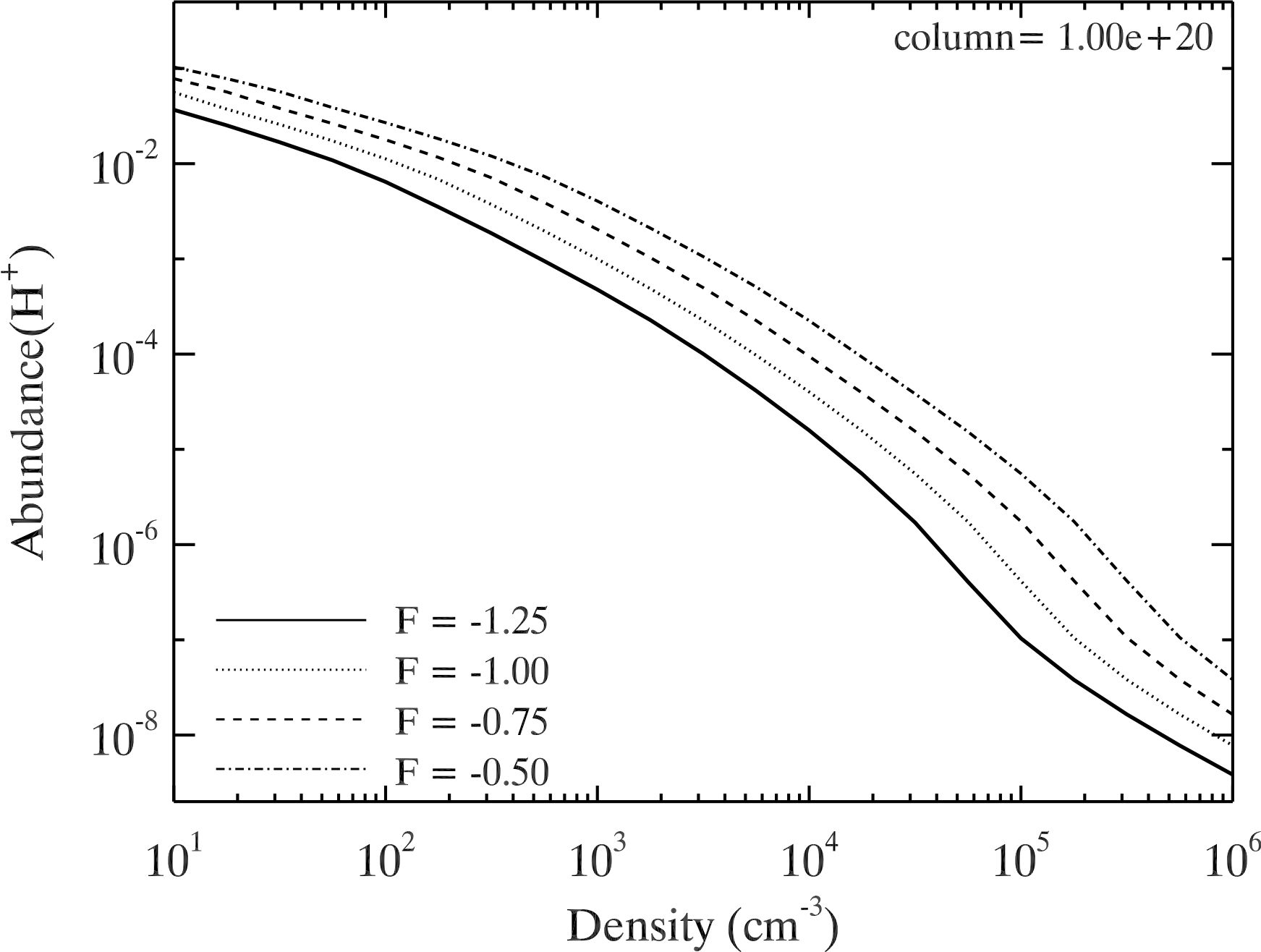}
\vspace{0.05cm}
\includegraphics[angle=0,width=7cm]{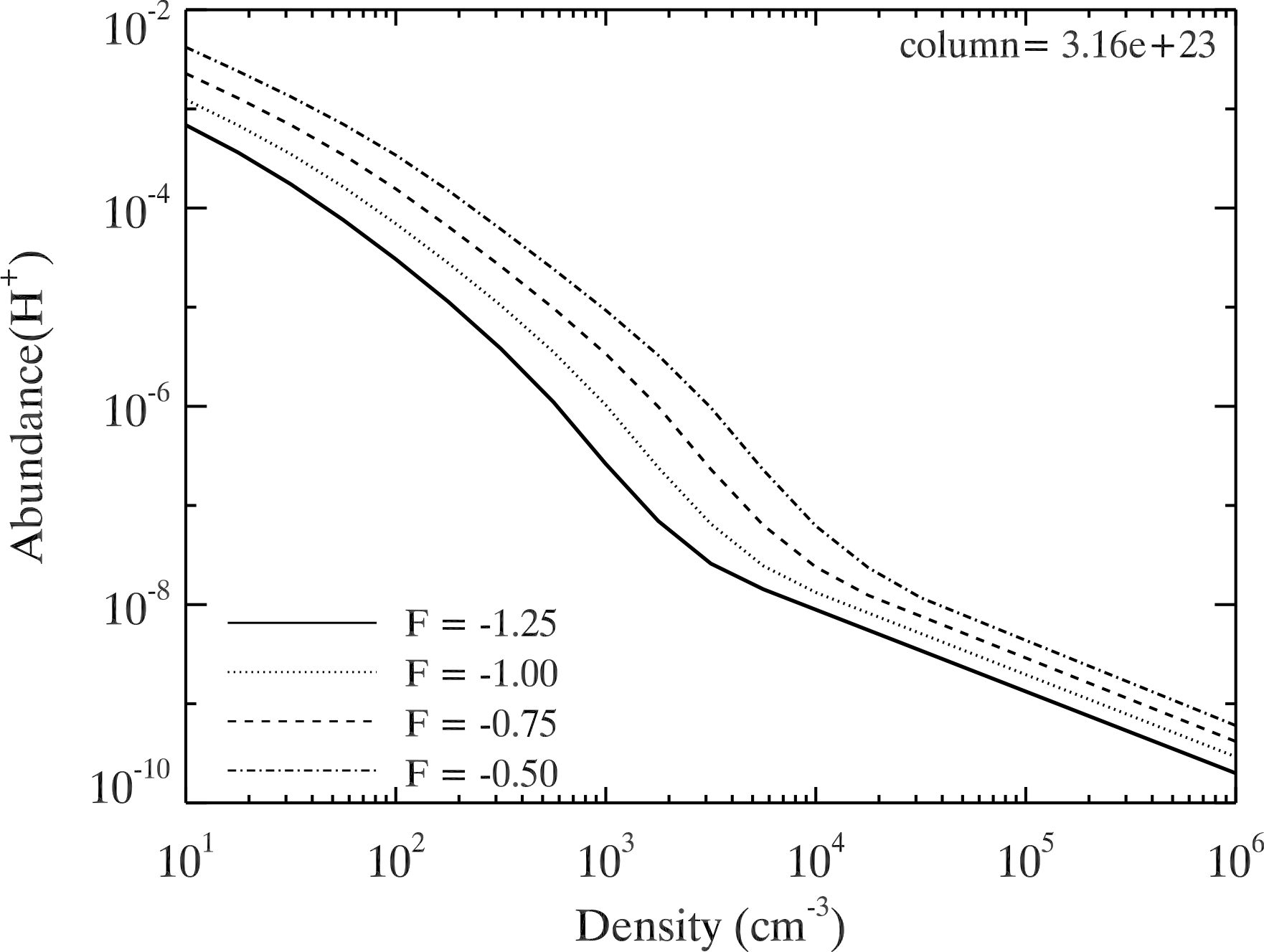}\\
\includegraphics[angle=0,width=7cm]{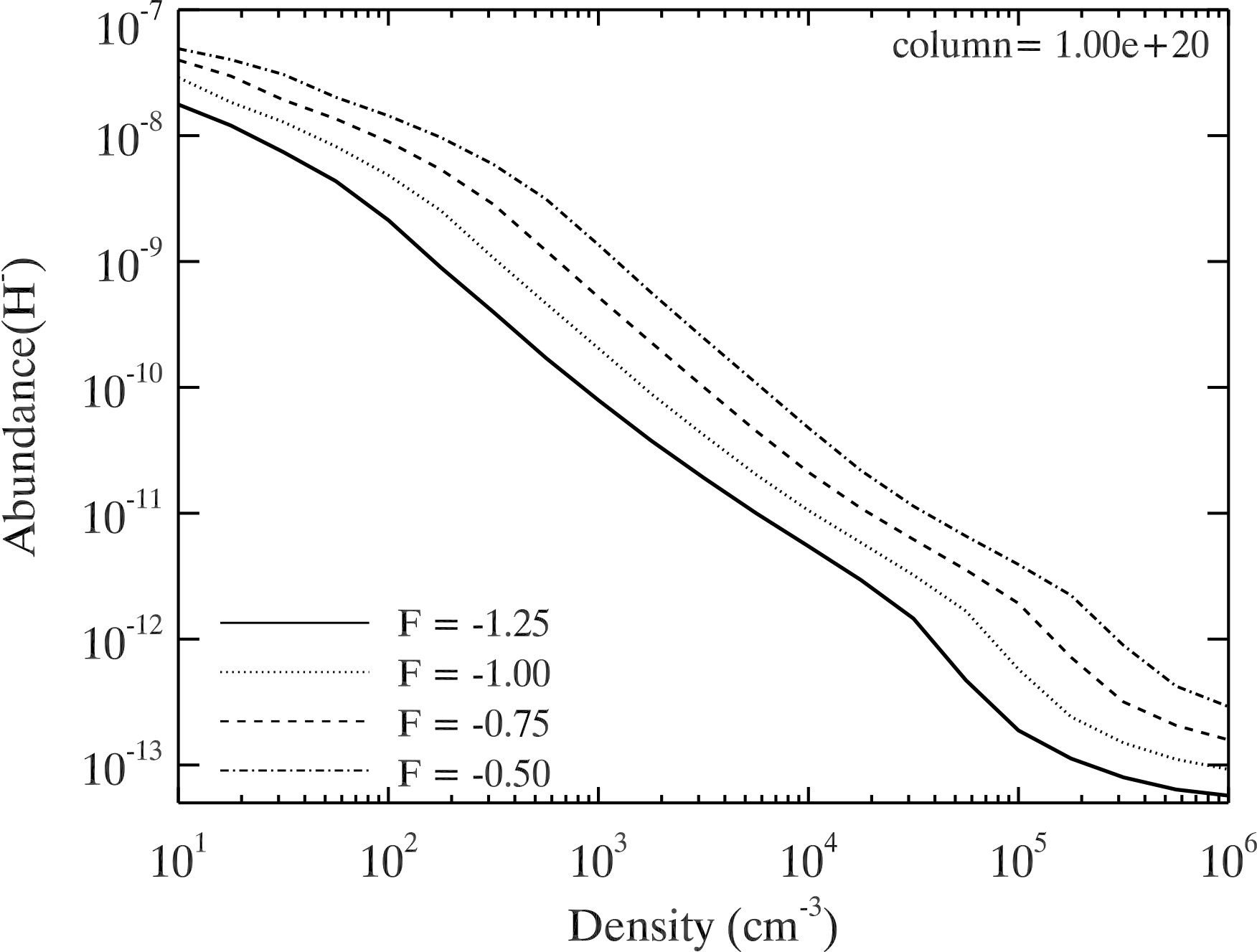}
\vspace{0.05cm}
\includegraphics[angle=0,width=7cm]{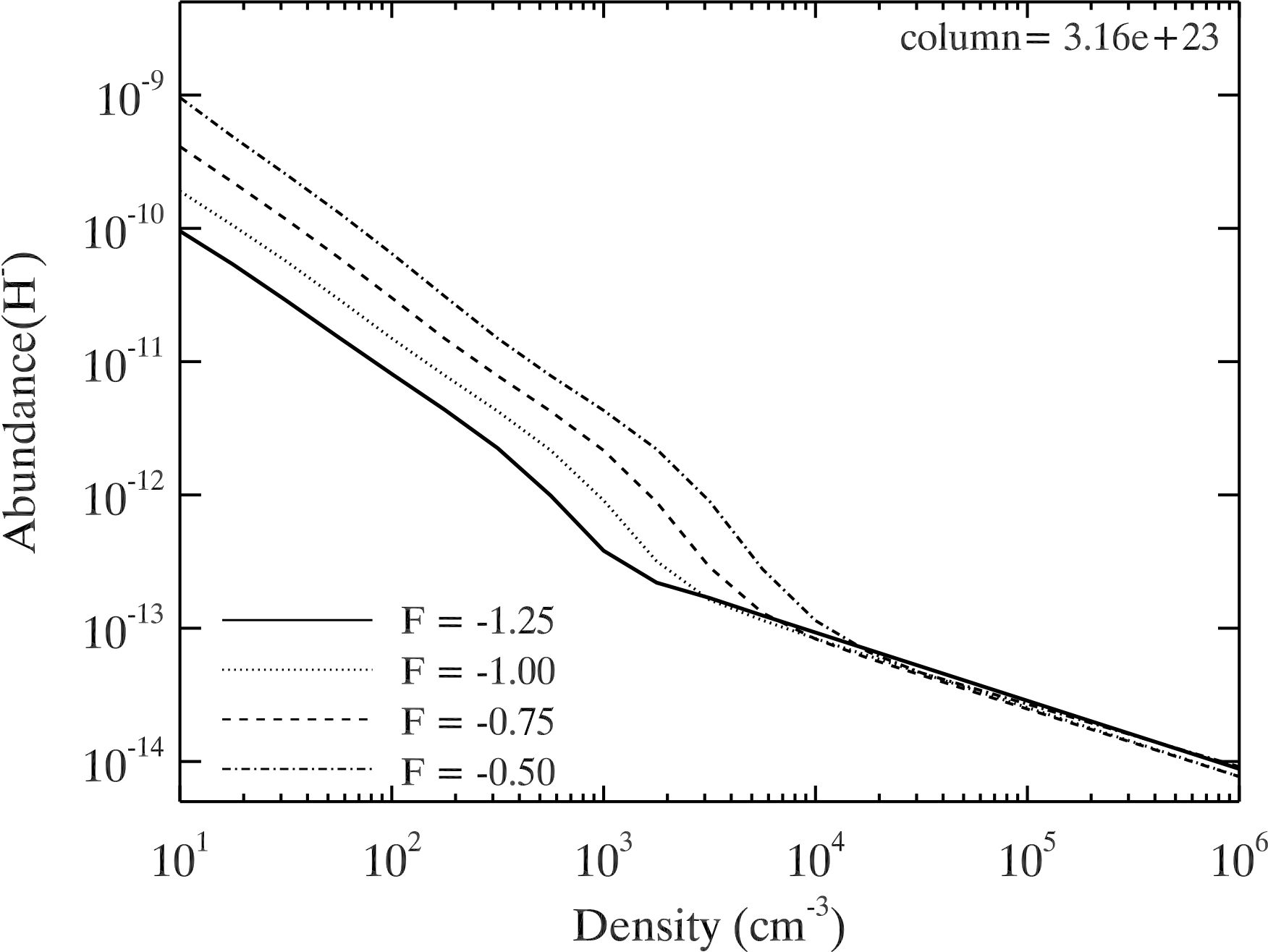}\\
\end{figure*}
\begin{figure*}
\includegraphics[angle=0,width=7cm]{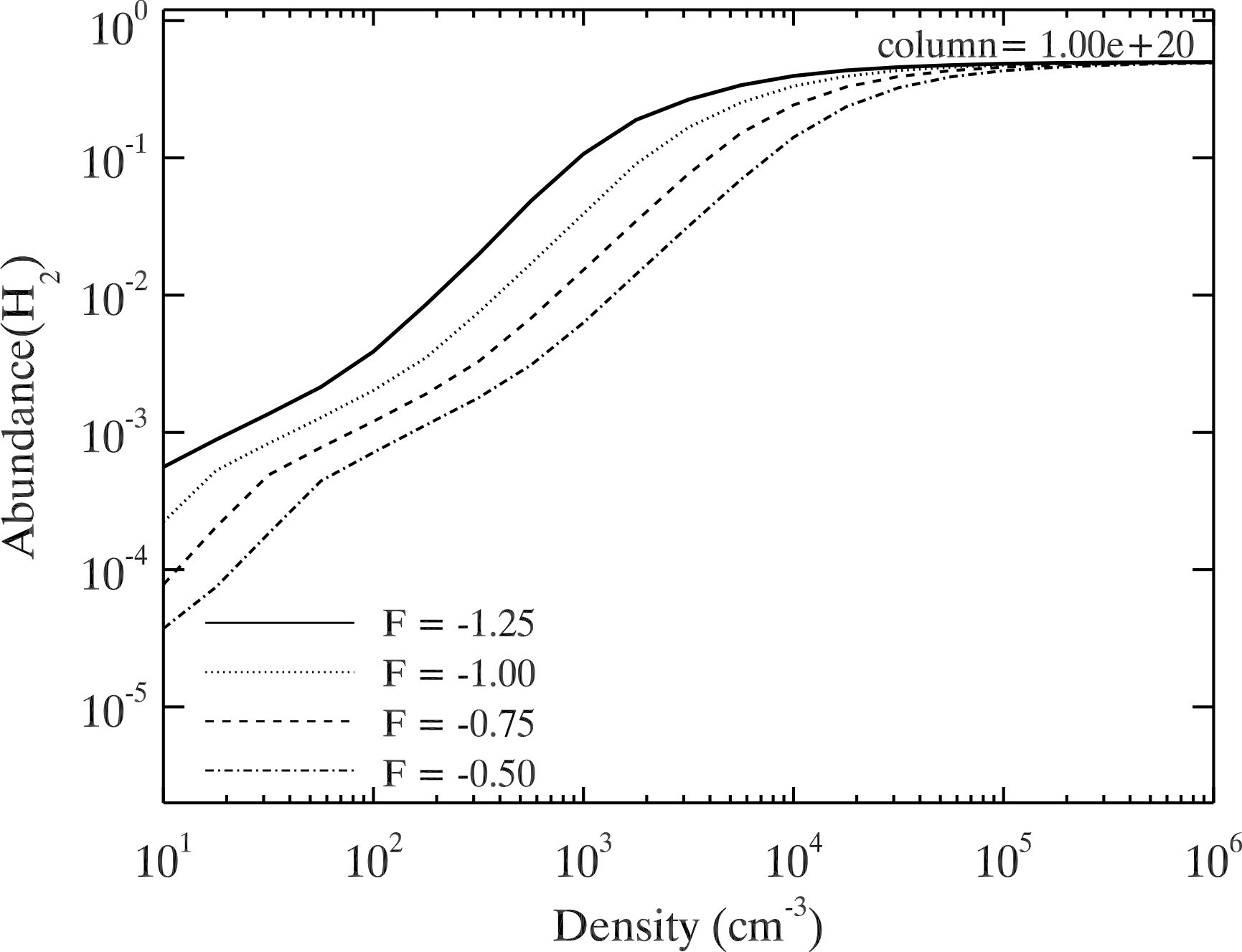}
\vspace{0.05cm}
\includegraphics[angle=0,width=7cm]{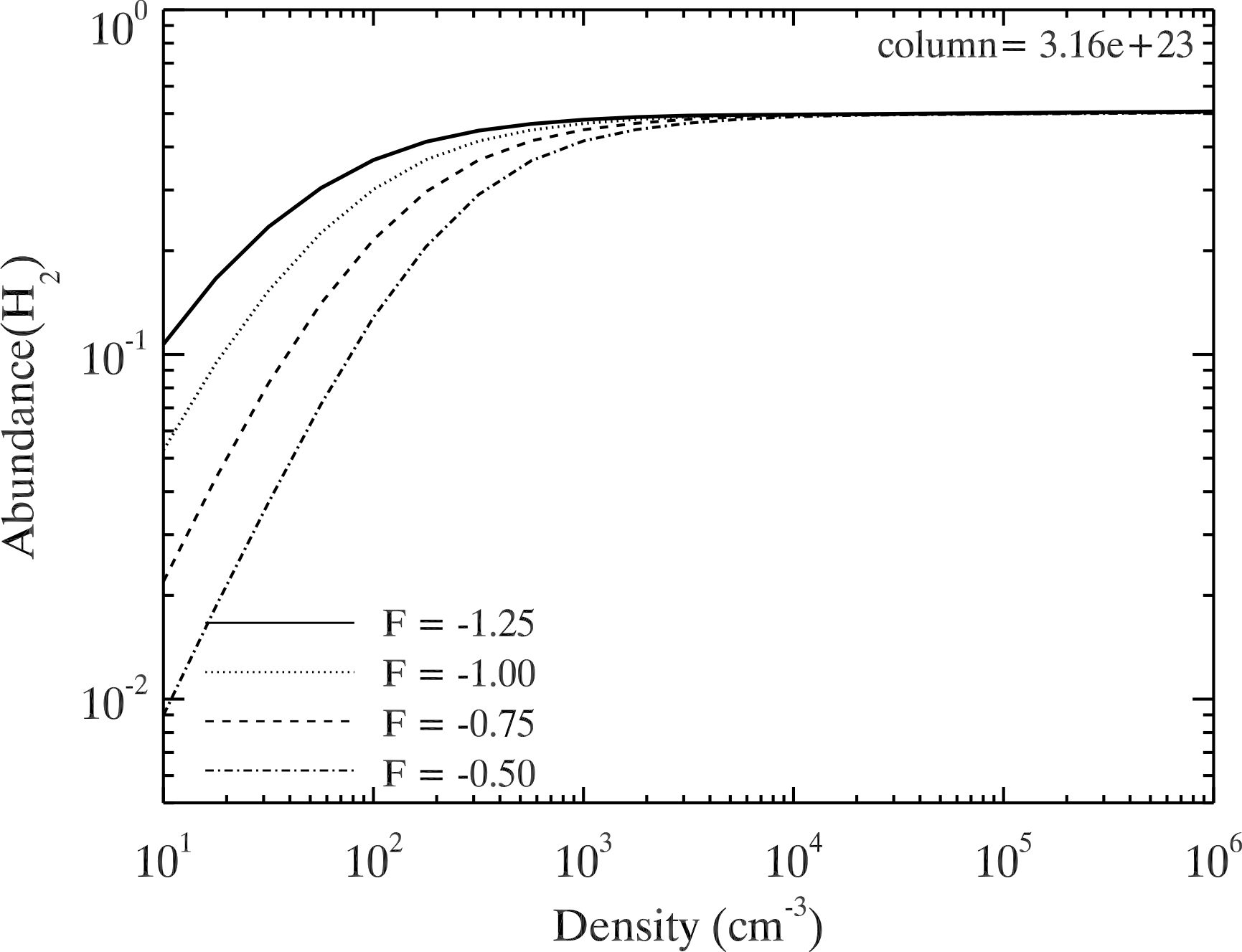}\\
\includegraphics[angle=0,width=7cm]{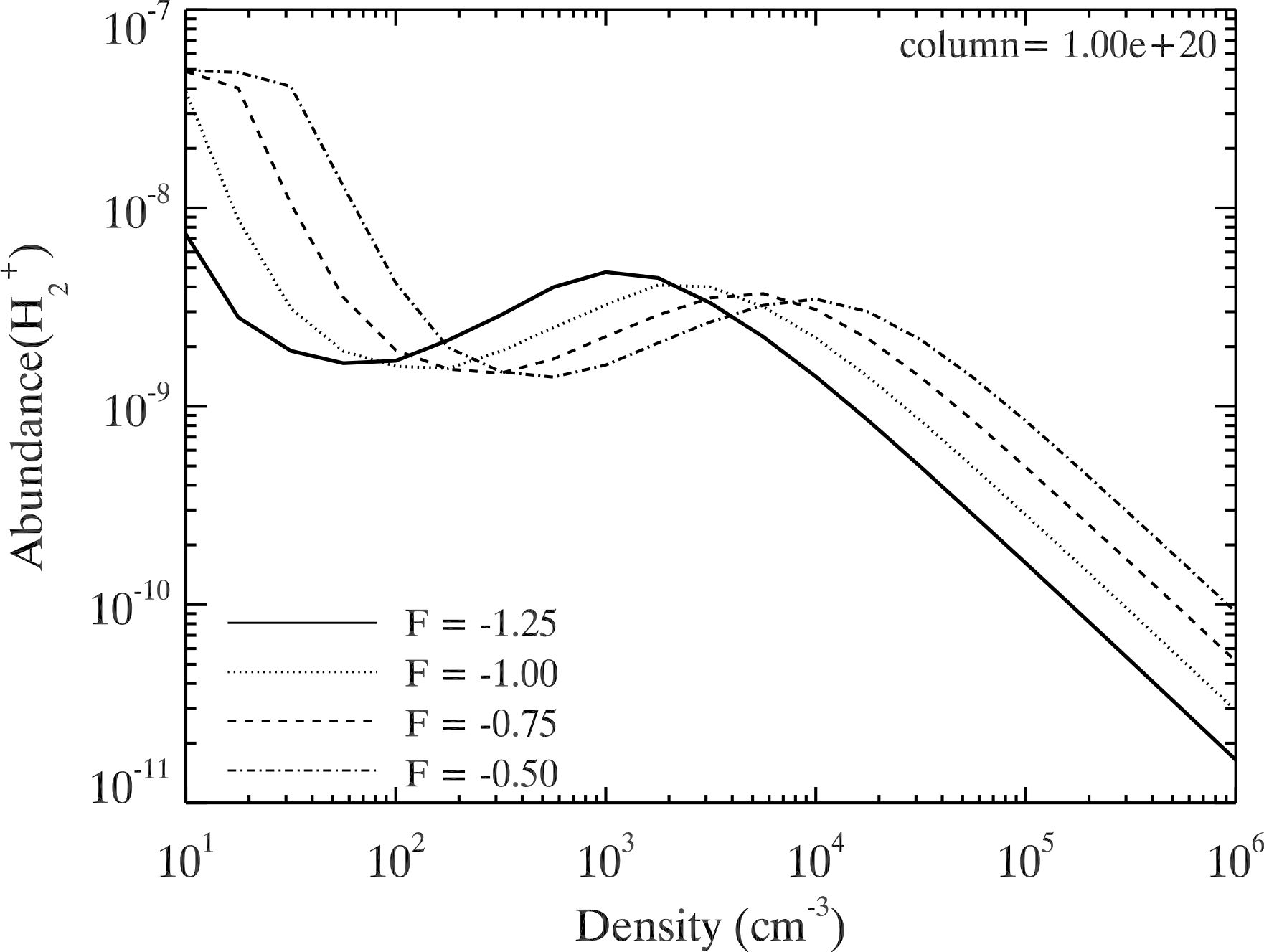}
\vspace{0.05cm}
\includegraphics[angle=0,width=7cm]{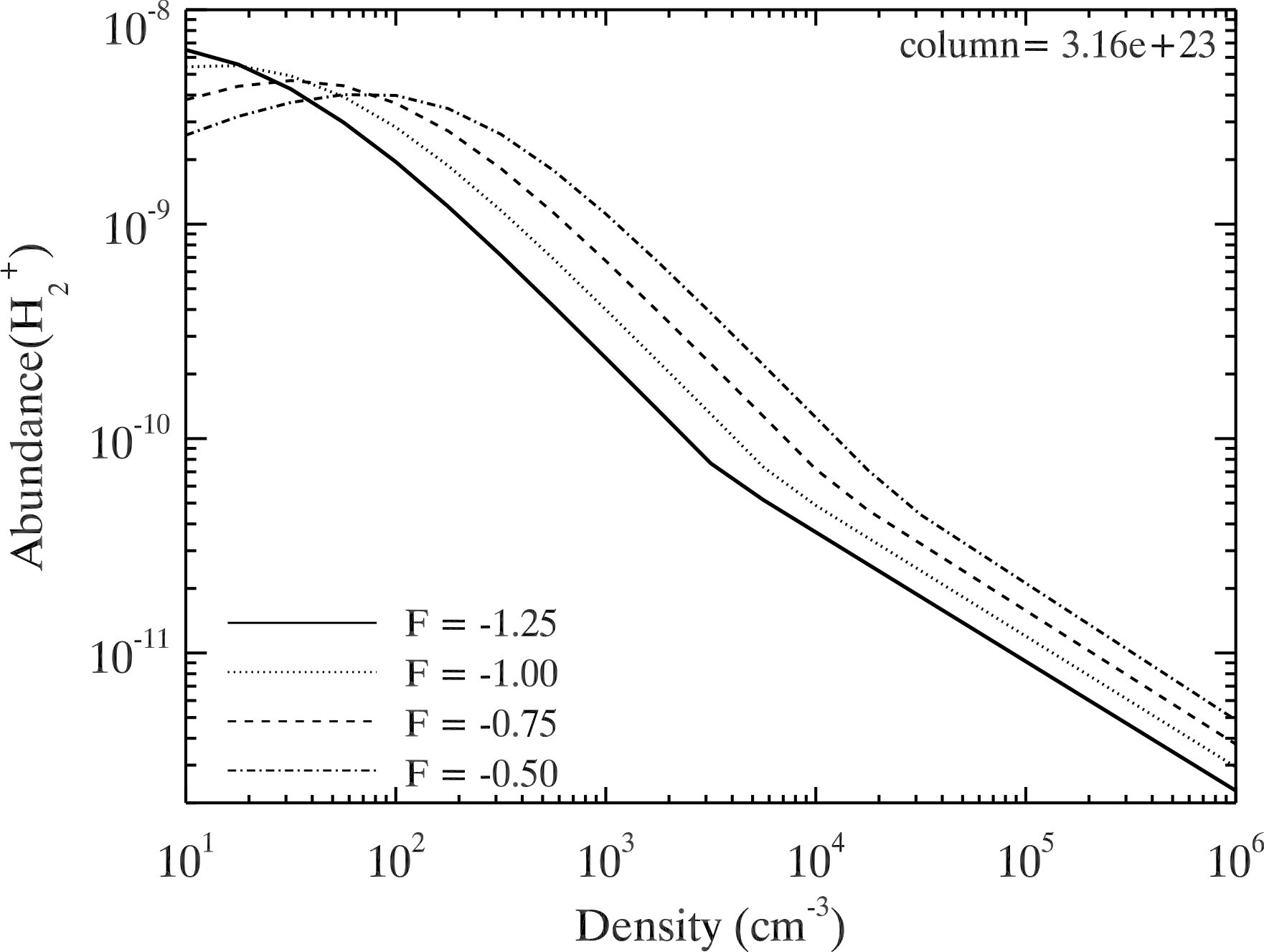}\\
\includegraphics[angle=0,width=7cm]{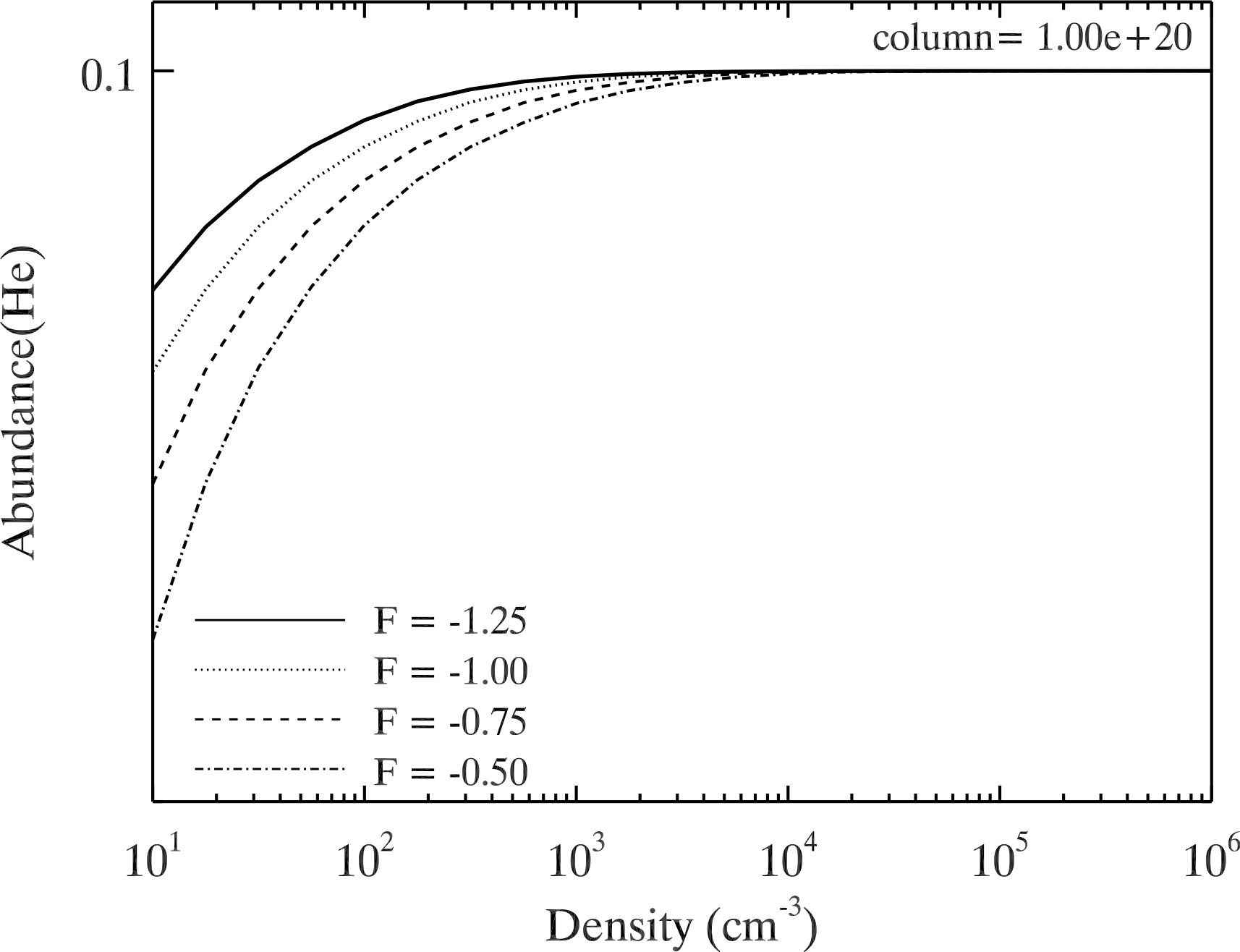}
\vspace{0.05cm}
\includegraphics[angle=0,width=7cm]{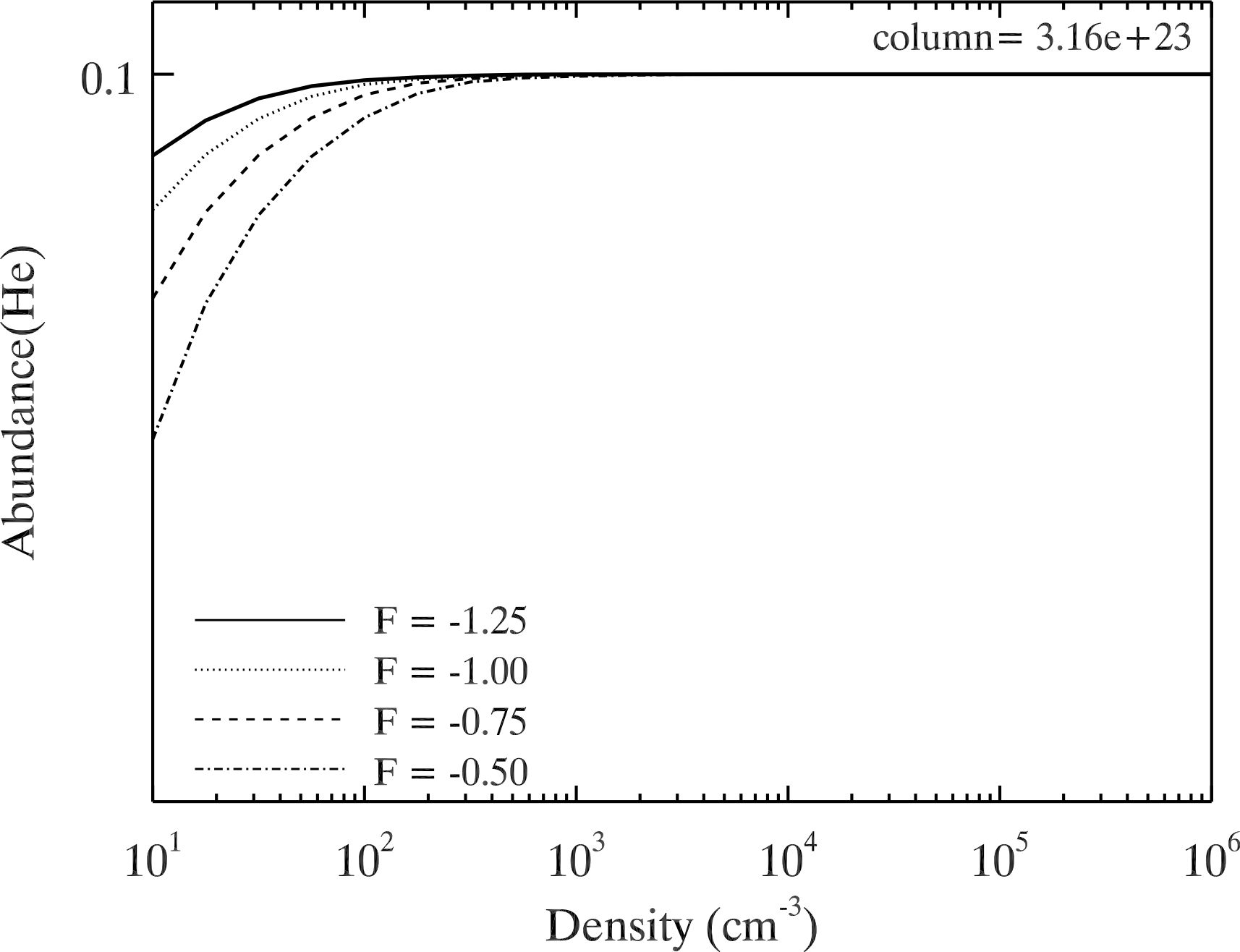}\\
\includegraphics[angle=0,width=7cm]{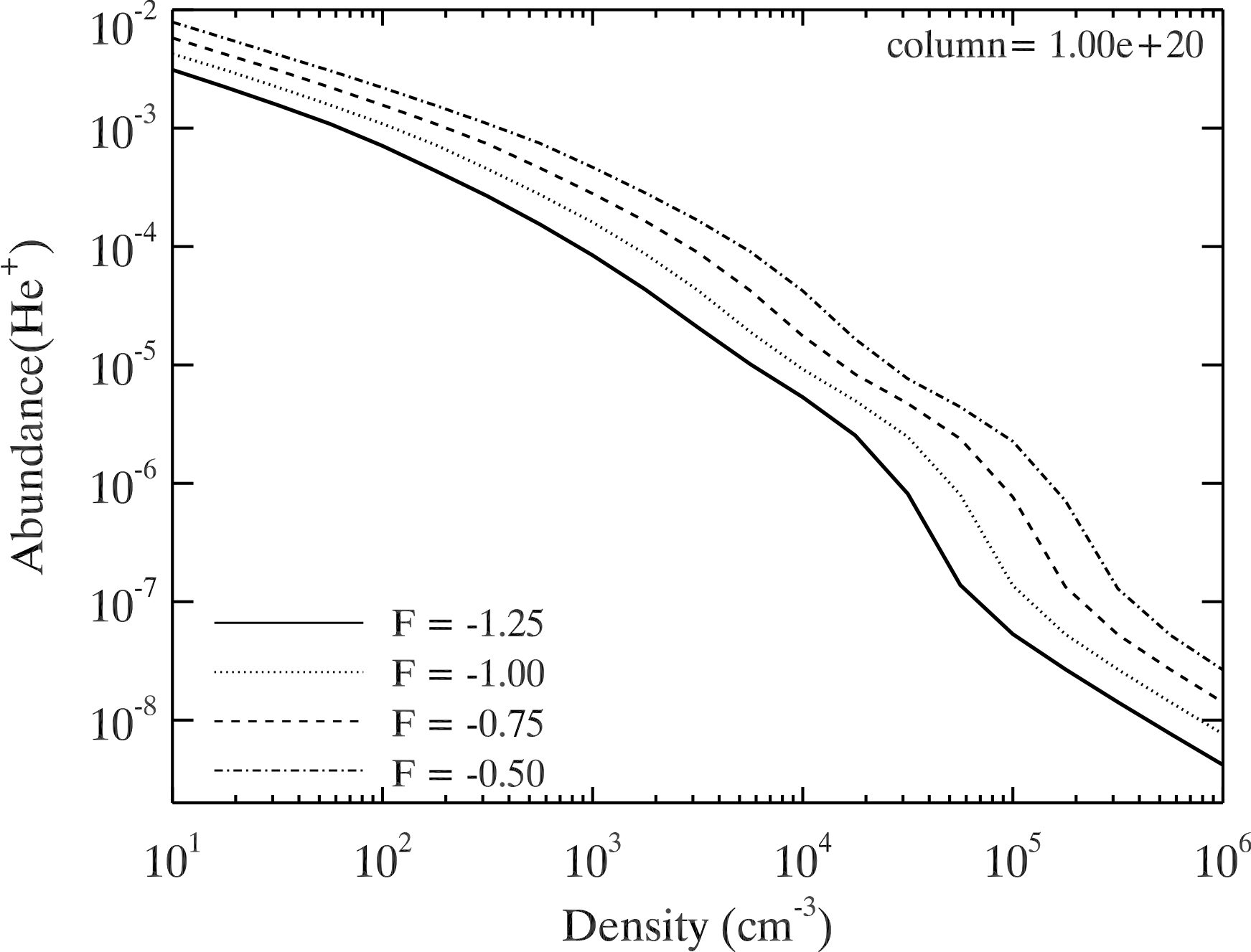}
\vspace{0.05cm}
\includegraphics[angle=0,width=7cm]{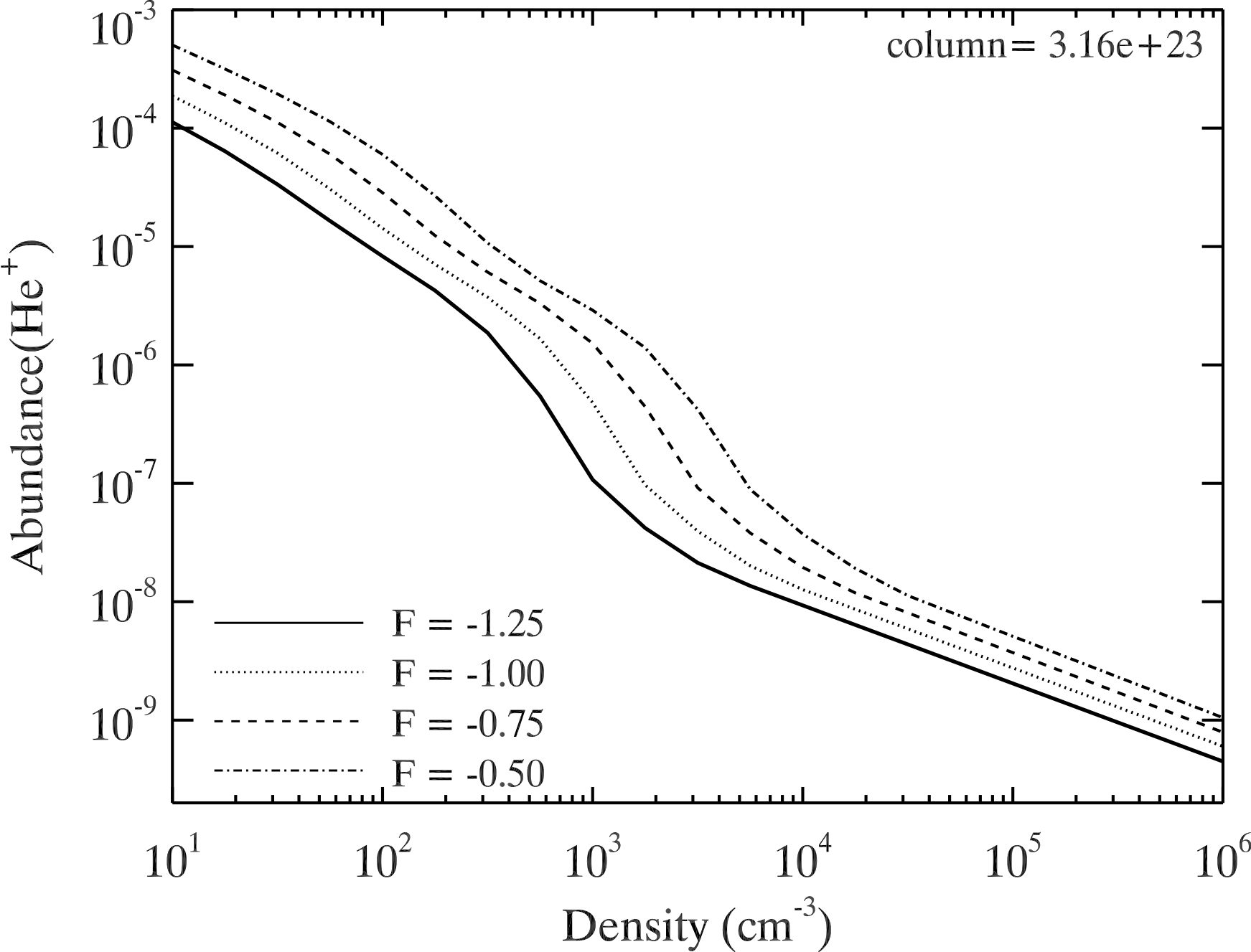}\\
\end{figure*}
\begin{figure*}
\includegraphics[angle=0,width=7cm]{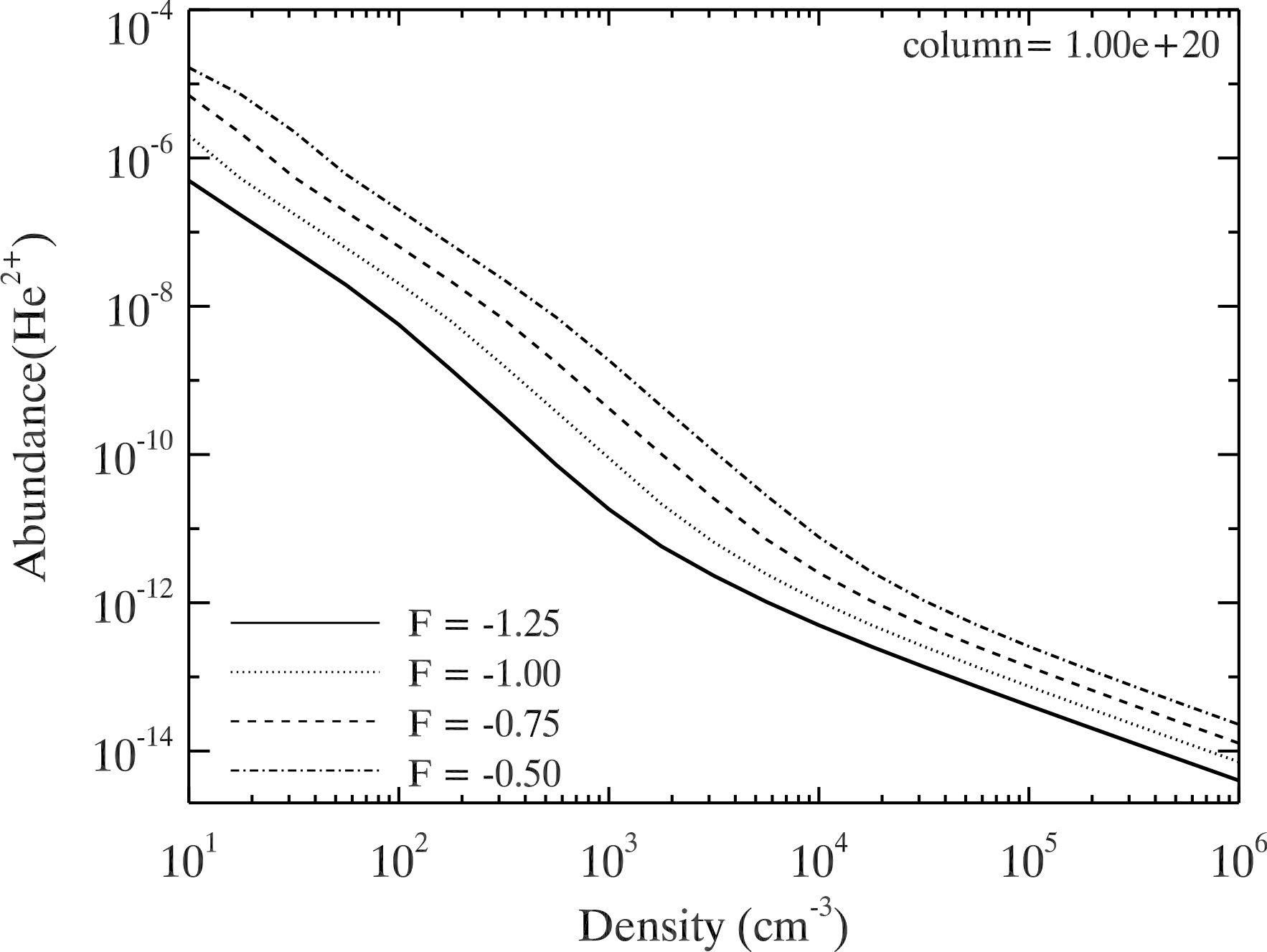}
\vspace{0.05cm}
\includegraphics[angle=0,width=7cm]{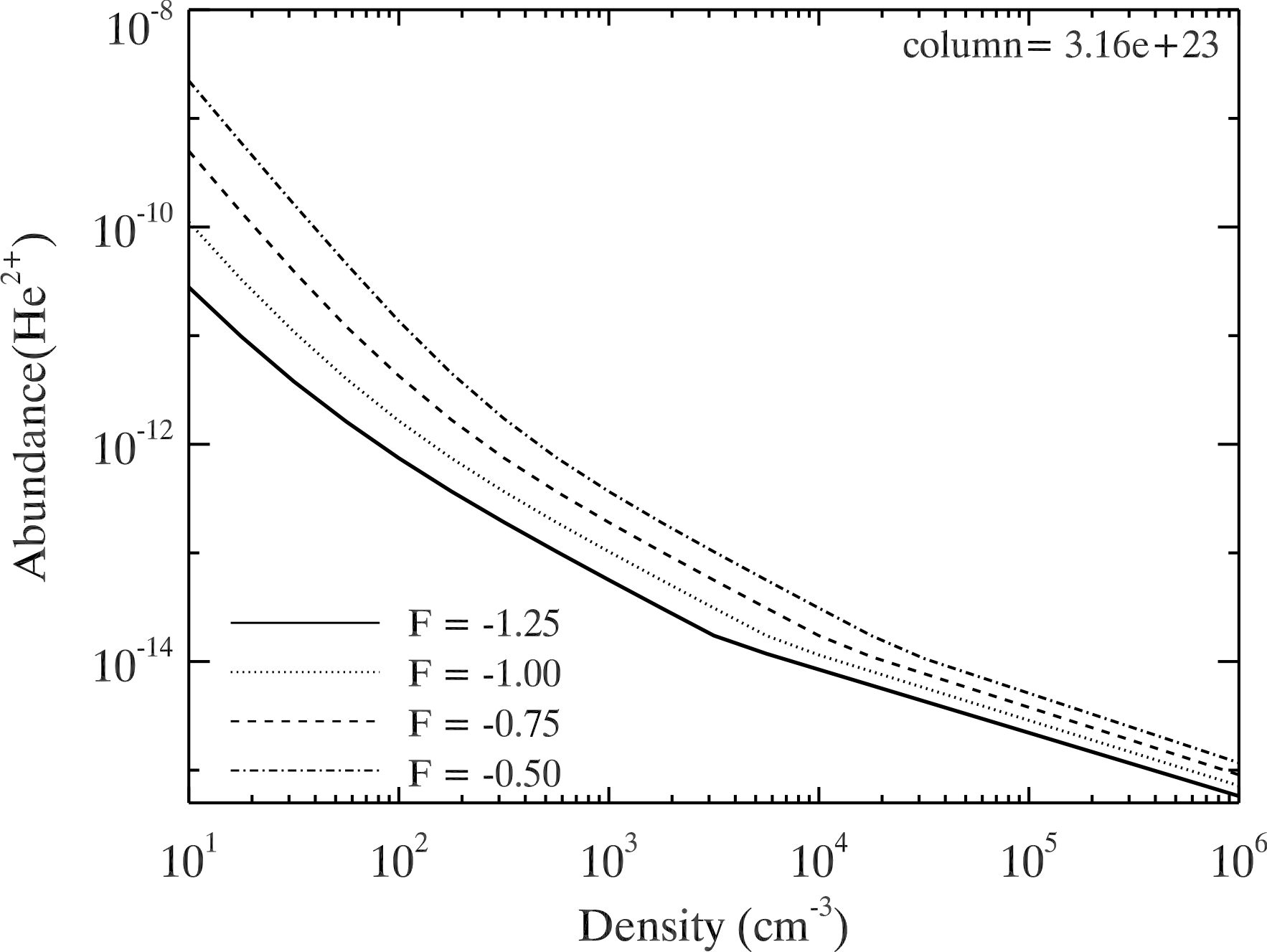}\\
\includegraphics[angle=0,width=7cm]{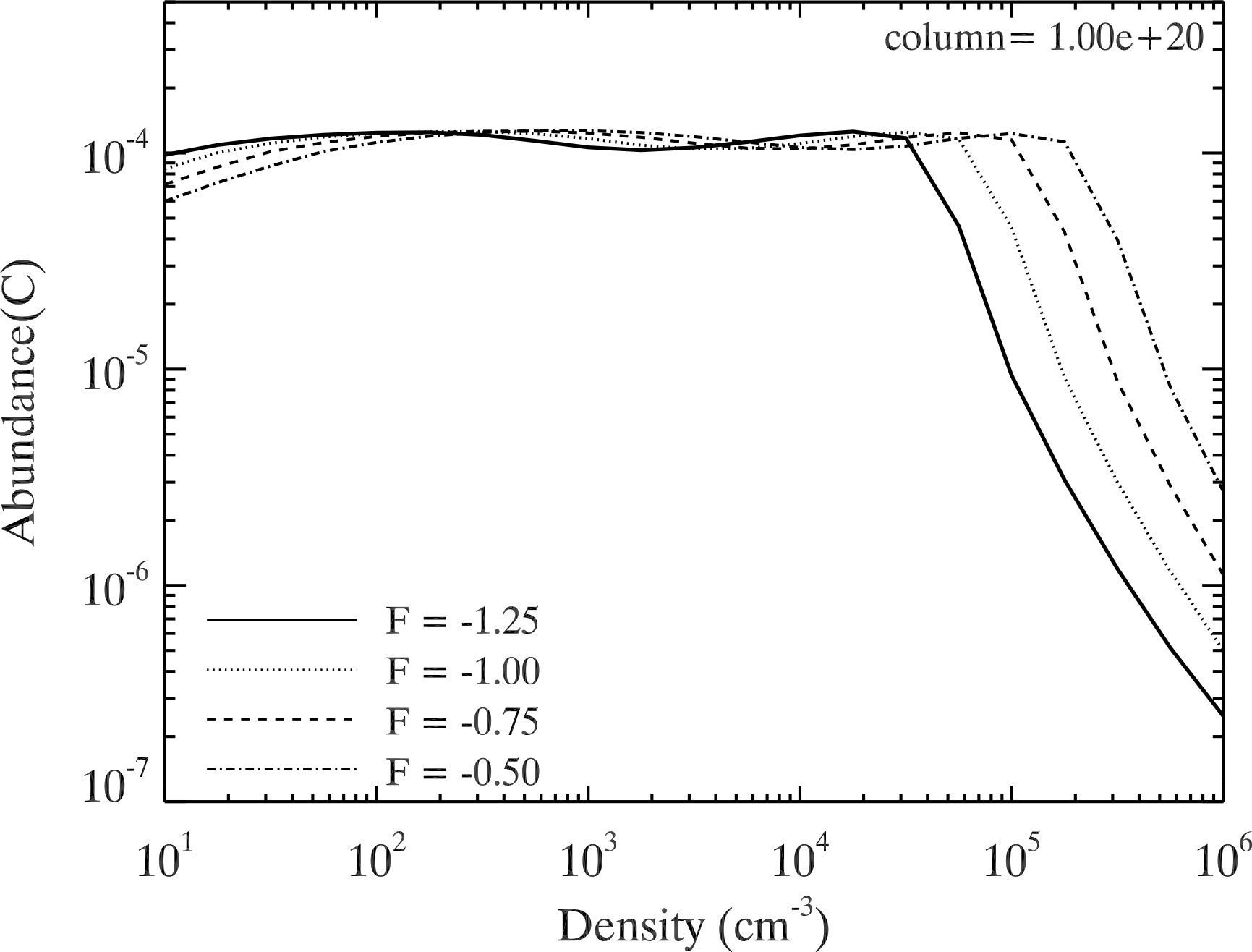}
\vspace{0.05cm}
\includegraphics[angle=0,width=7cm]{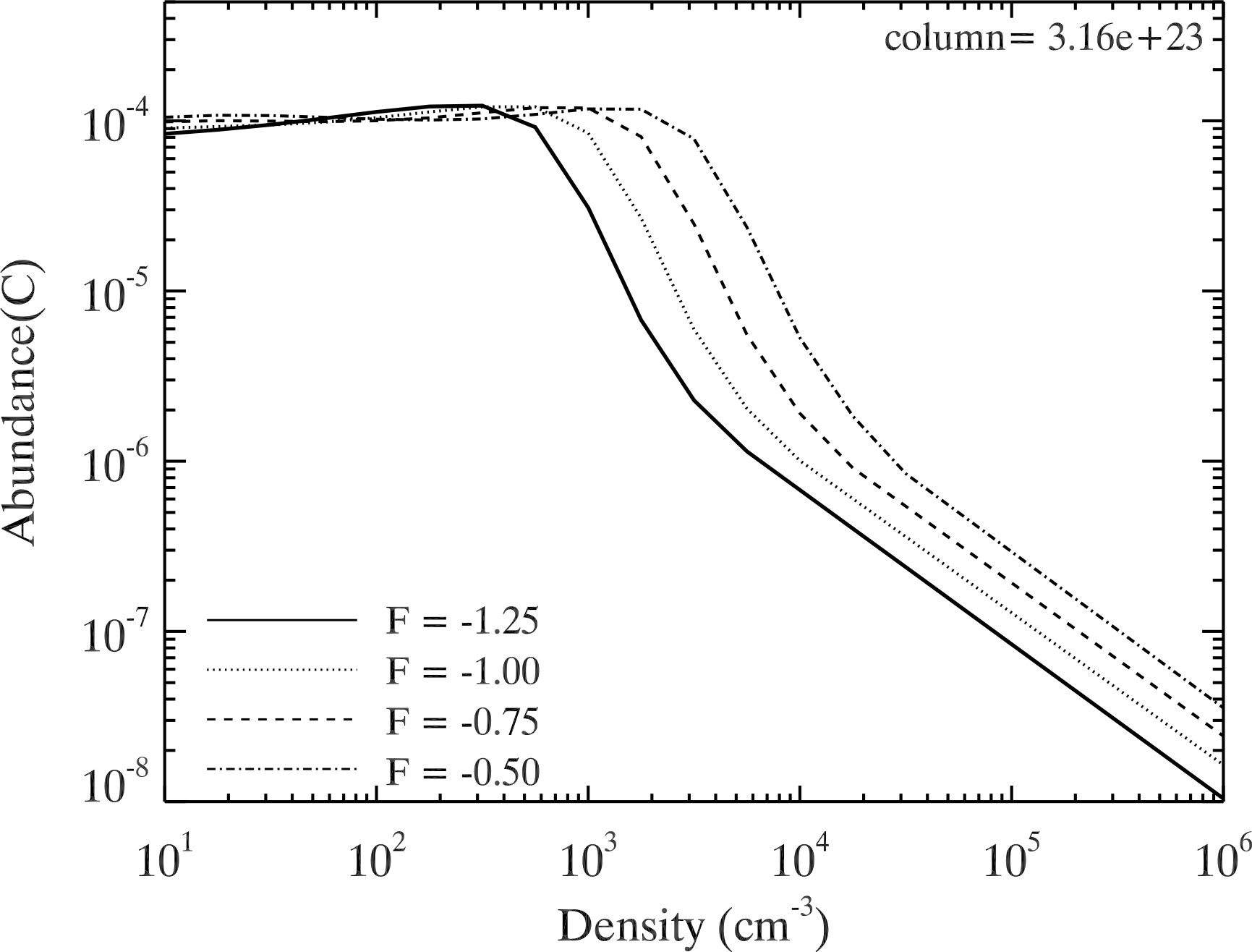}\\
\includegraphics[angle=0,width=7cm]{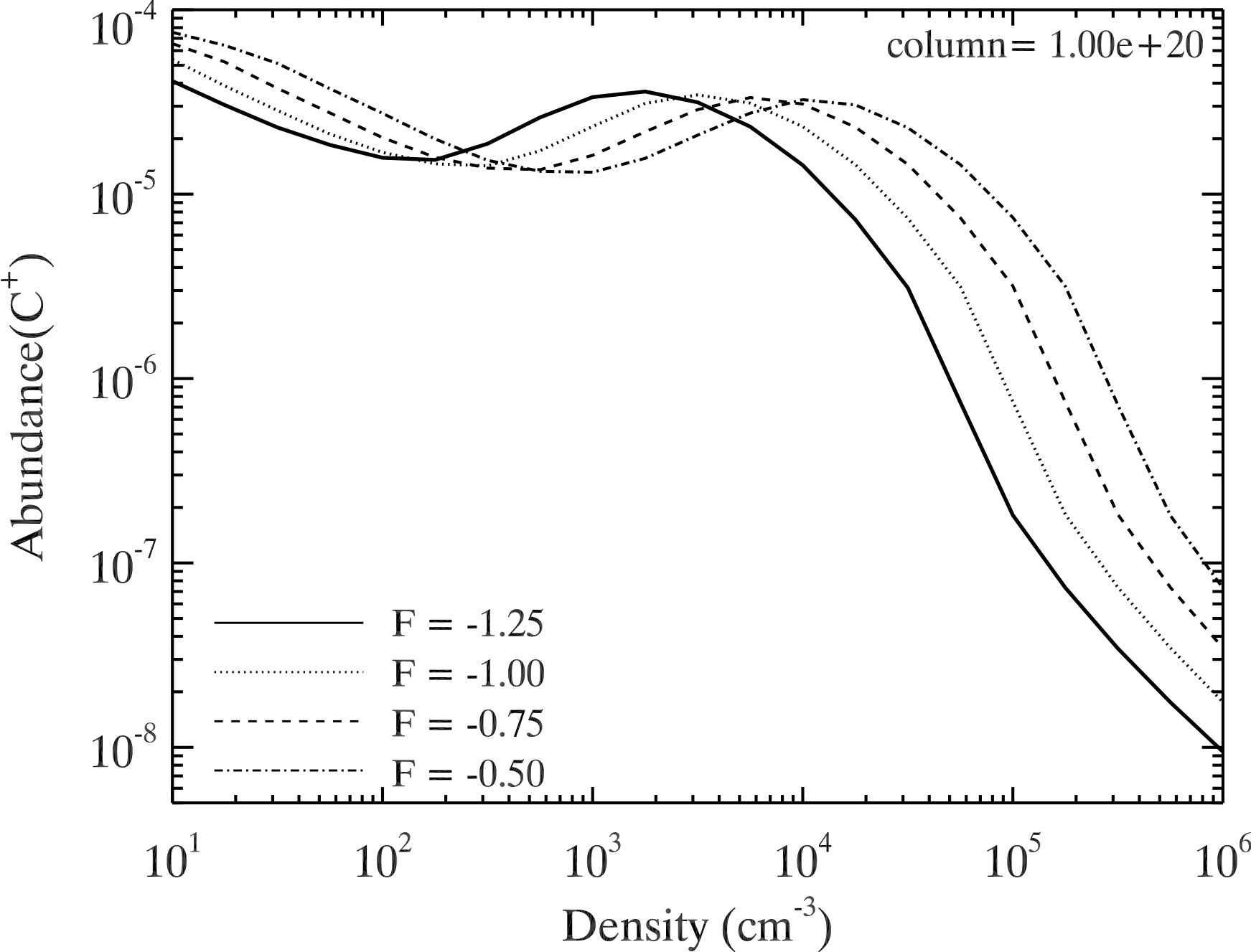}
\vspace{0.05cm}
\includegraphics[angle=0,width=7cm]{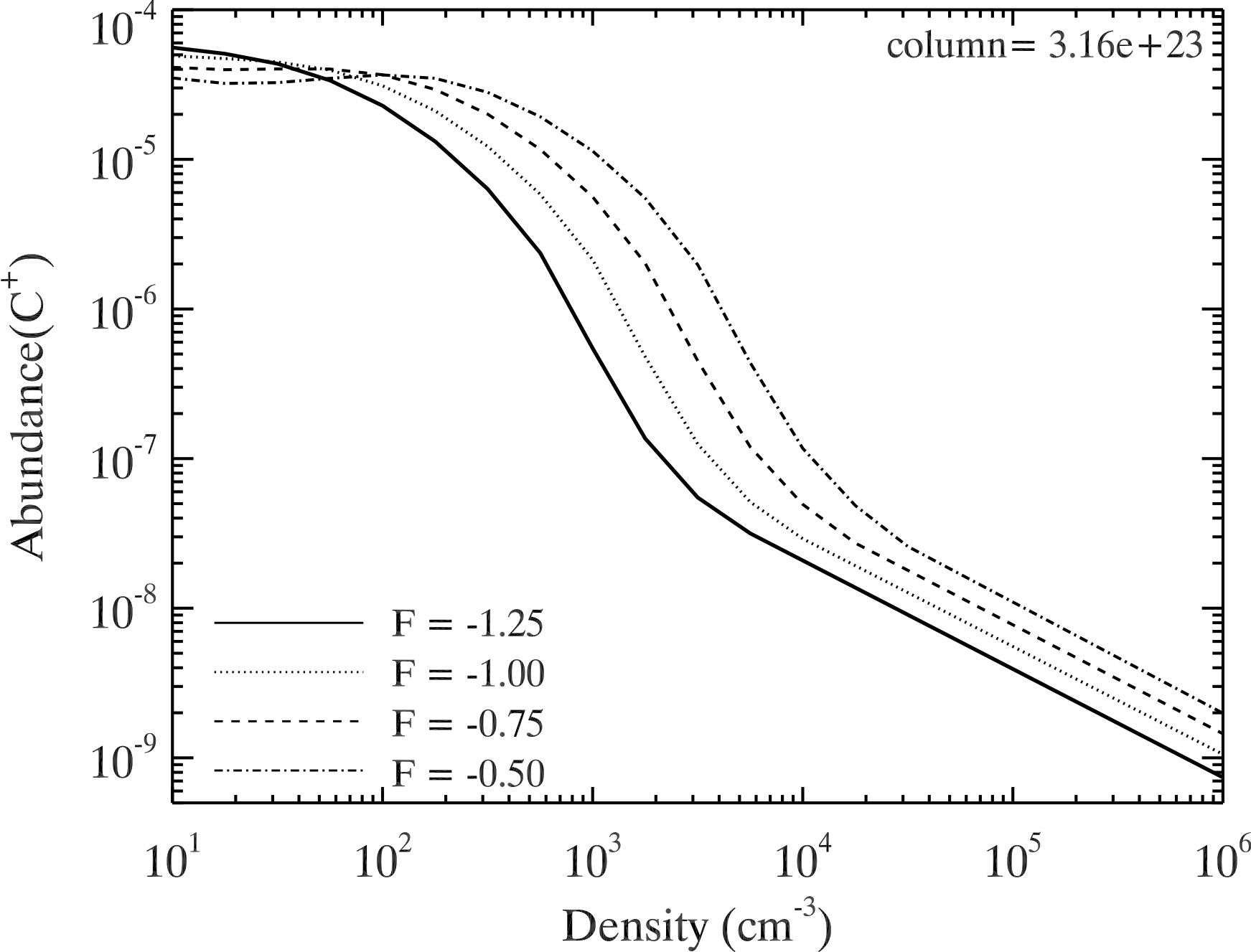}\\
\includegraphics[angle=0,width=7cm]{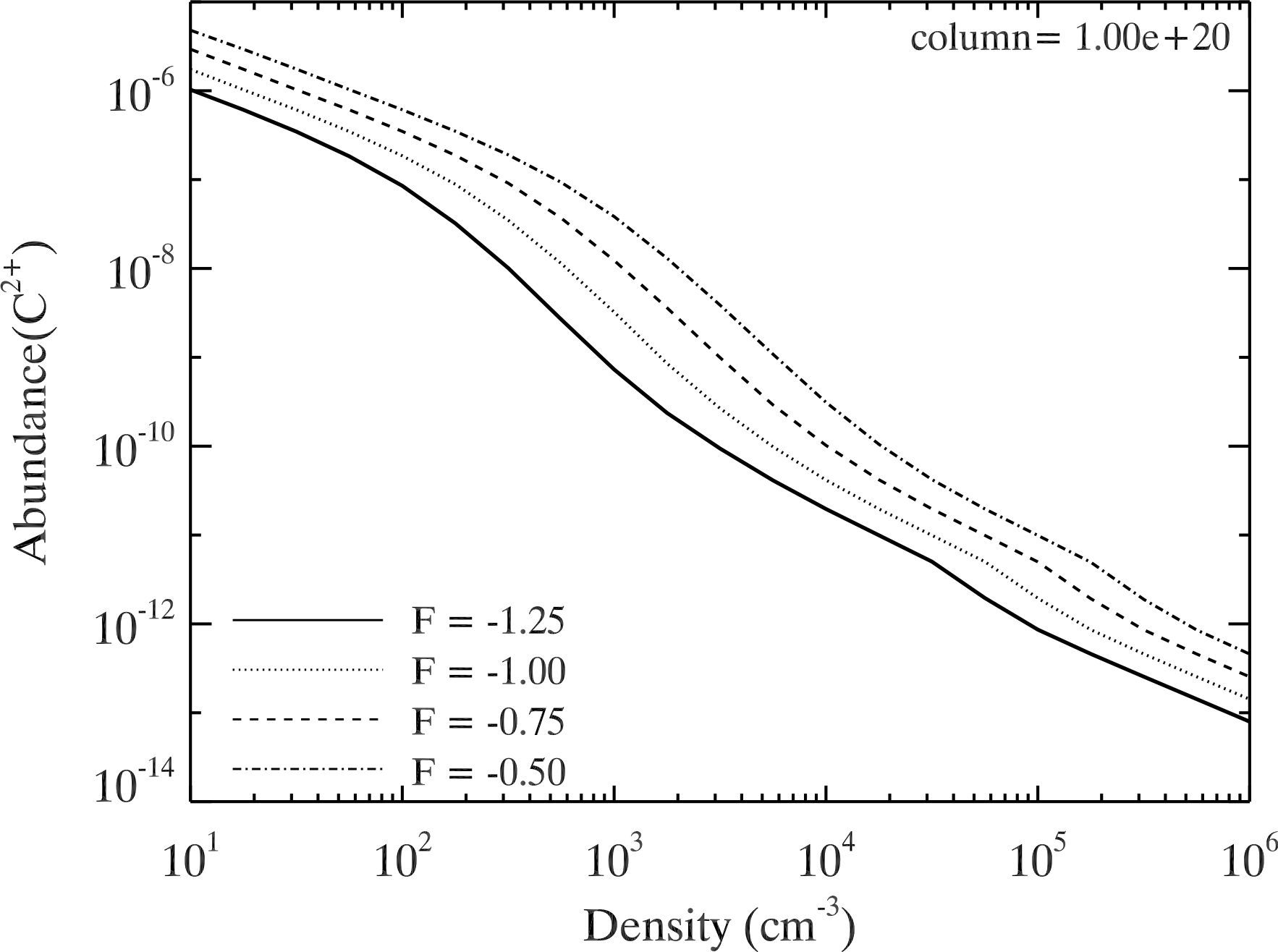}
\vspace{0.05cm}
\includegraphics[angle=0,width=7cm]{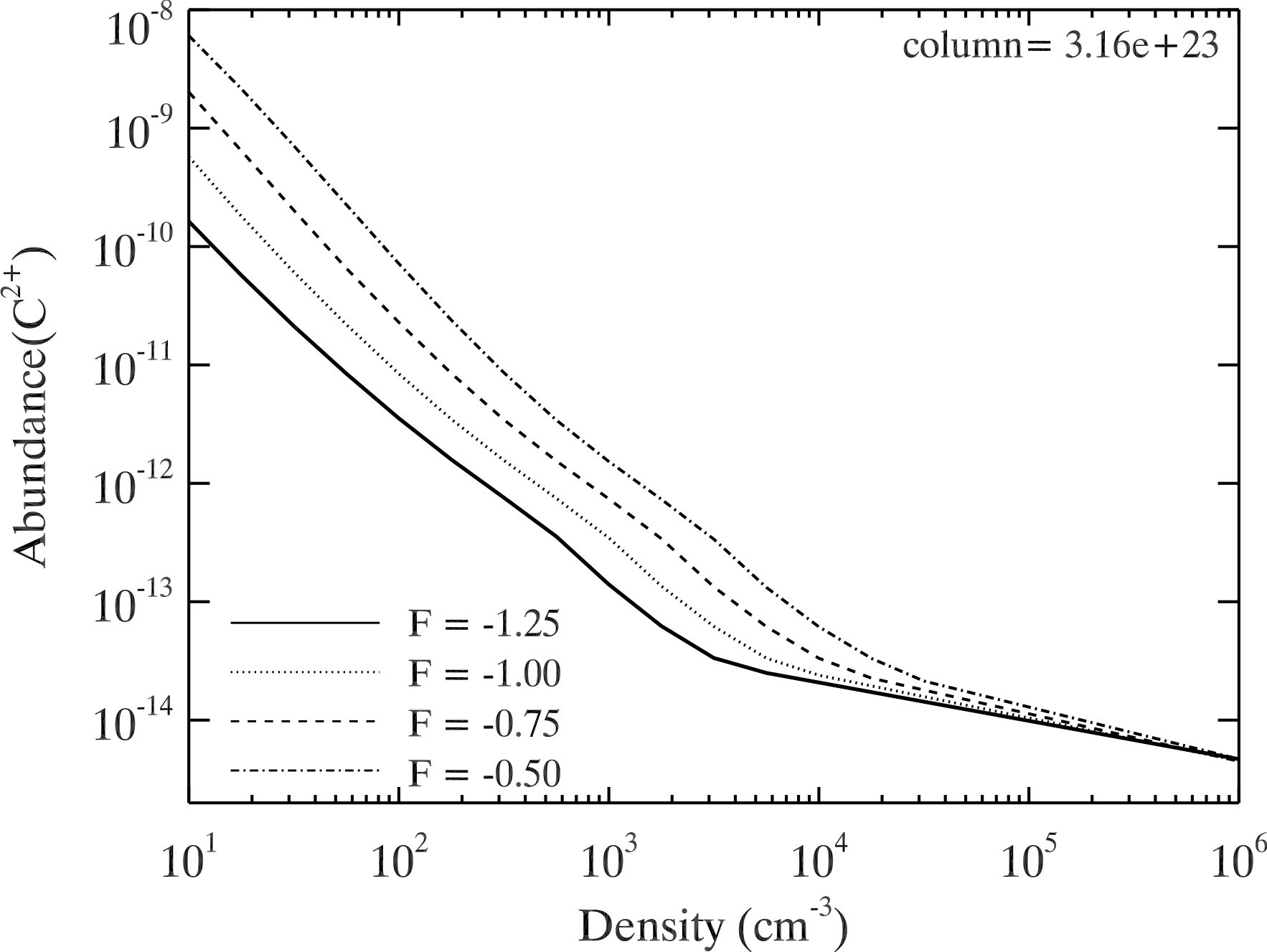}\\
\end{figure*}
\begin{figure*}
\includegraphics[angle=0,width=7cm]{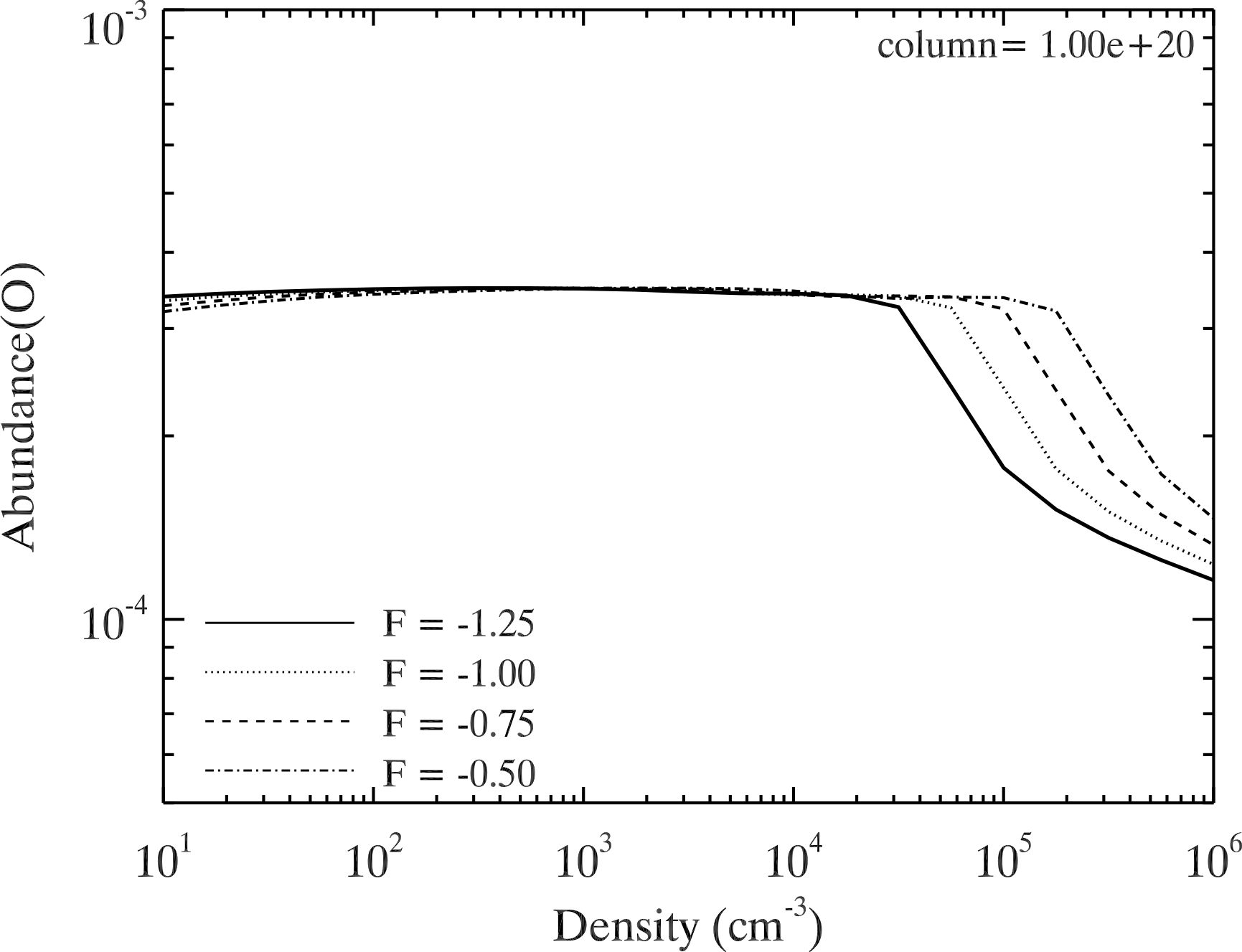}
\vspace{0.05cm}
\includegraphics[angle=0,width=7cm]{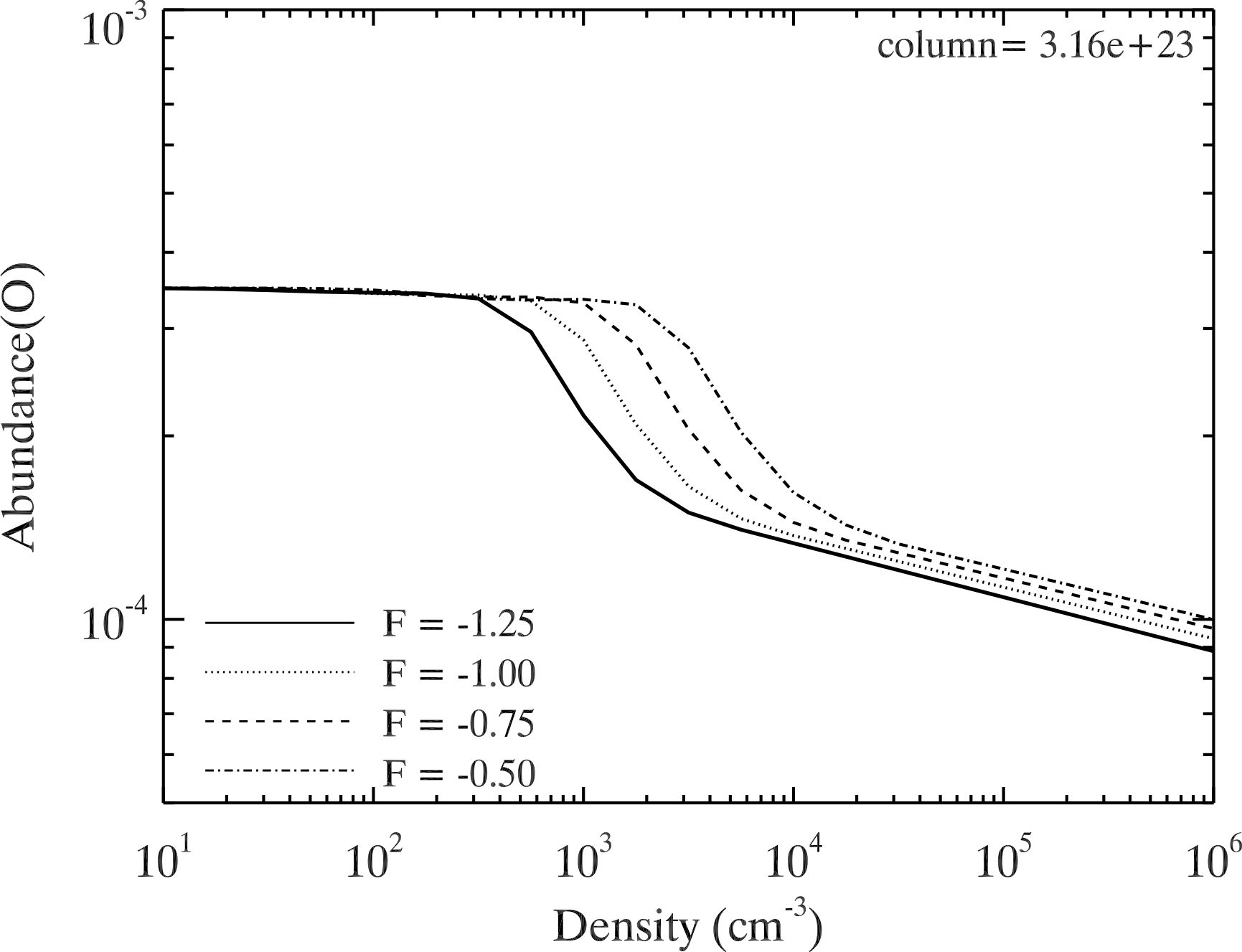}\\
\includegraphics[angle=0,width=7cm]{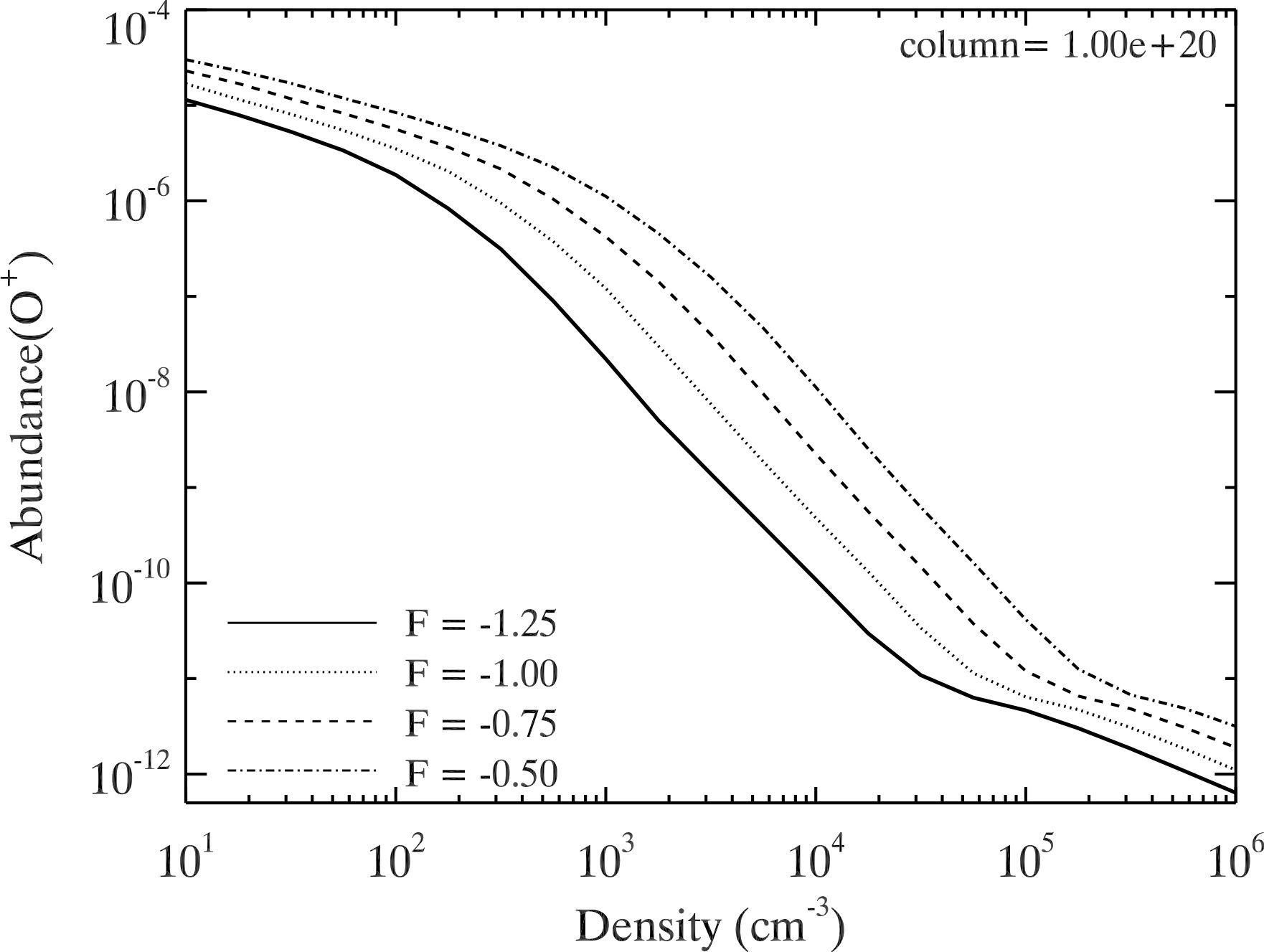}
\vspace{0.05cm}
\includegraphics[angle=0,width=7cm]{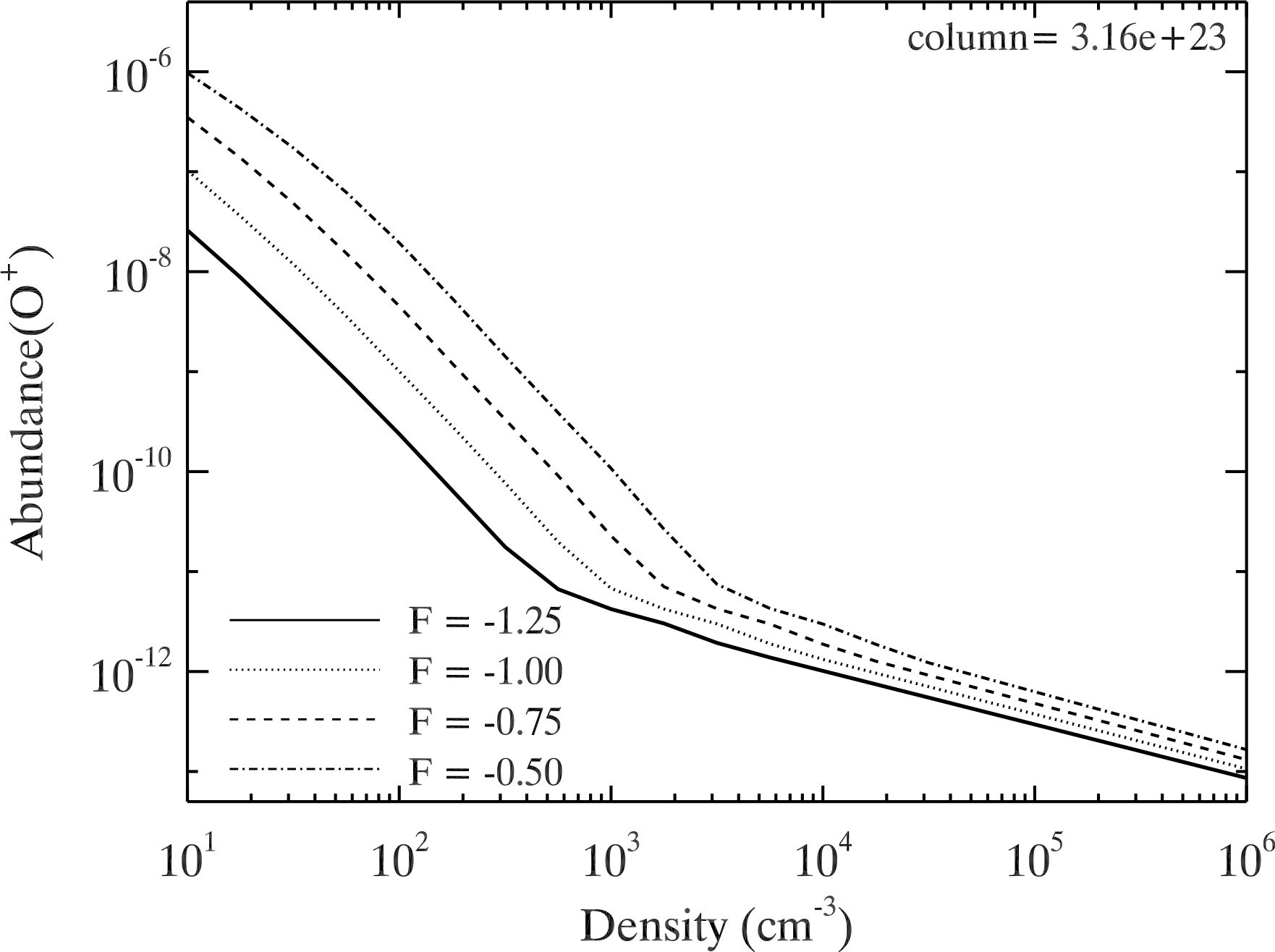}\\
\includegraphics[angle=0,width=7cm]{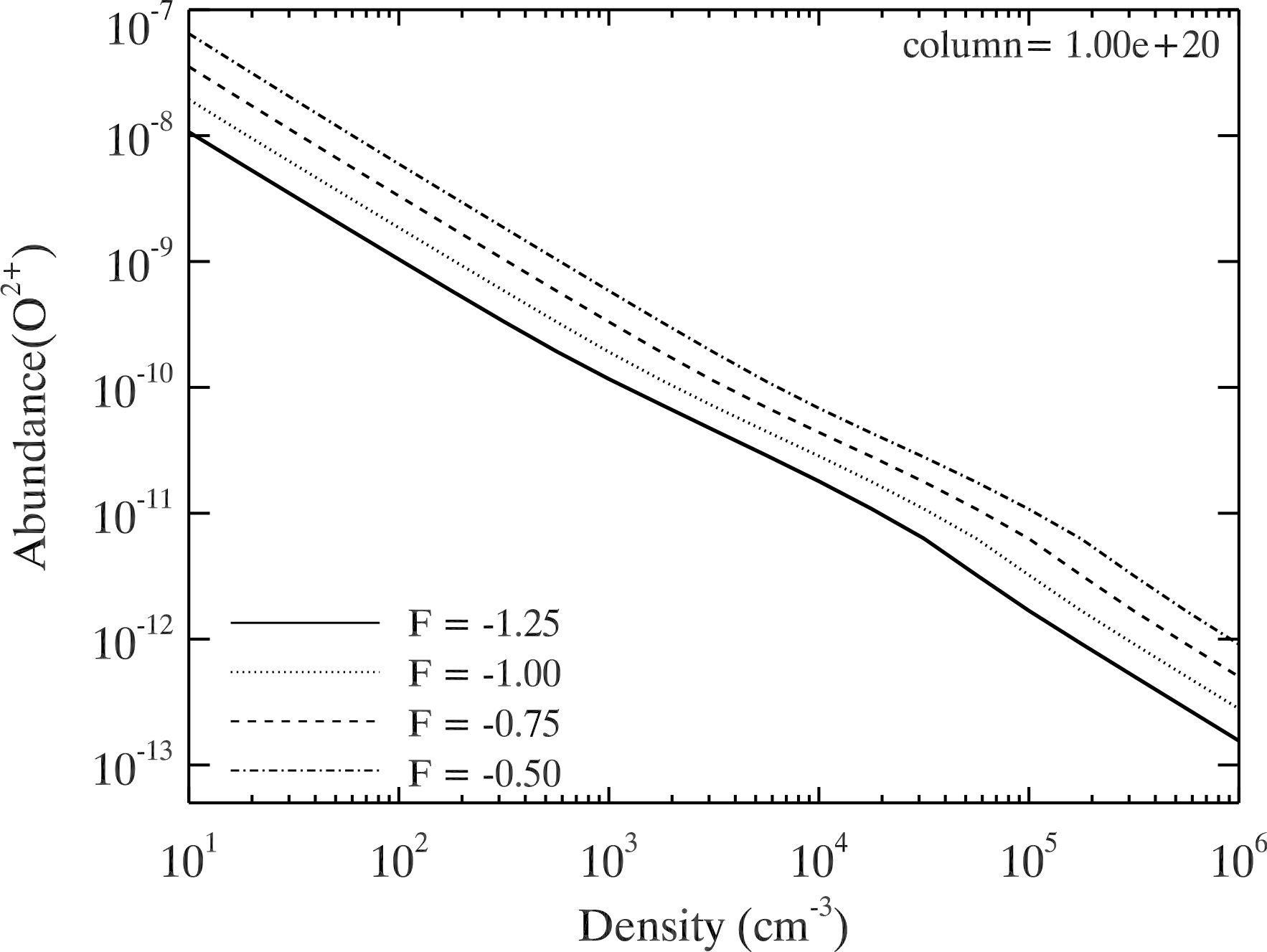}
\vspace{0.05cm}
\includegraphics[angle=0,width=7cm]{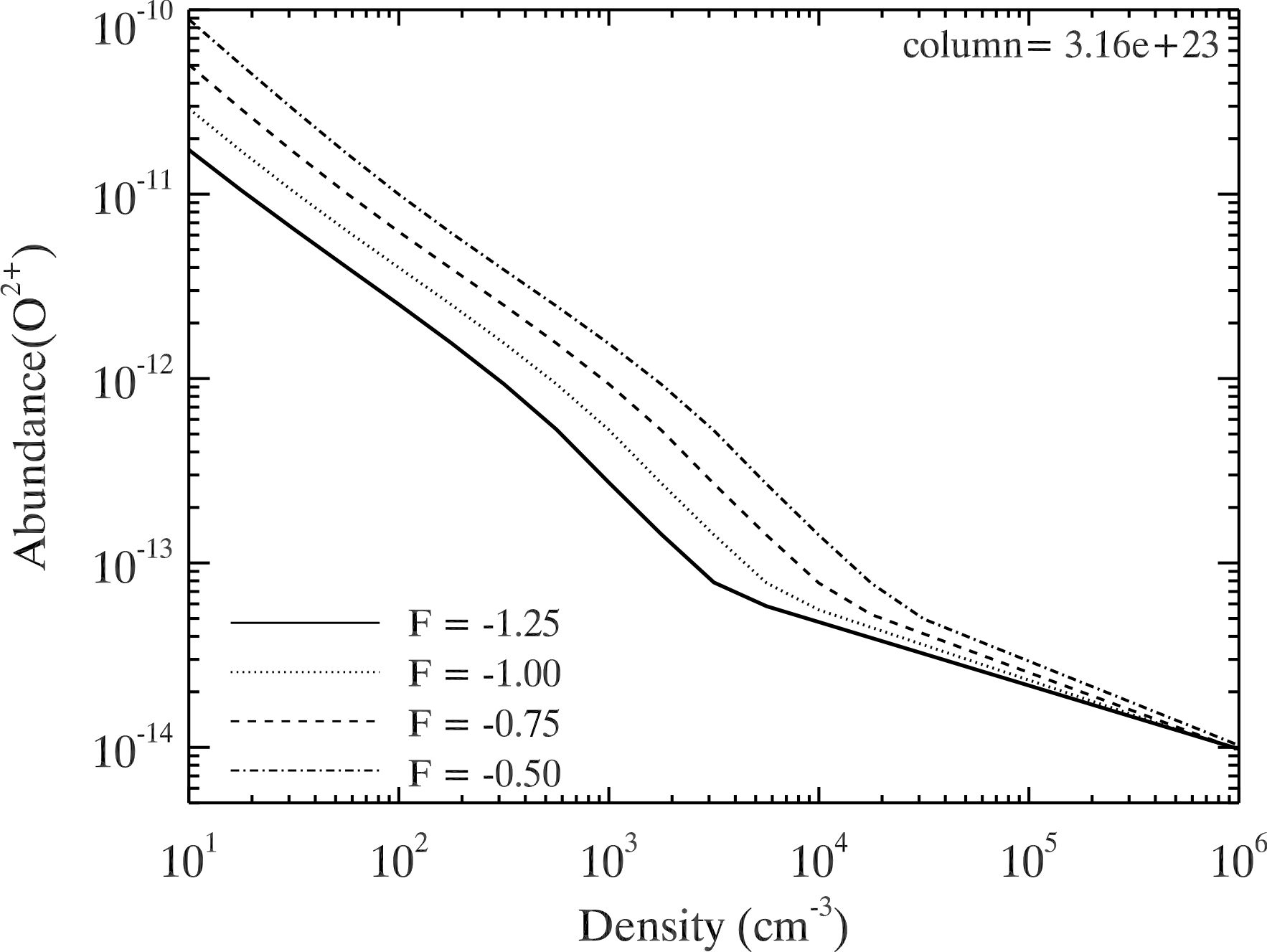}\\
\includegraphics[angle=0,width=7cm]{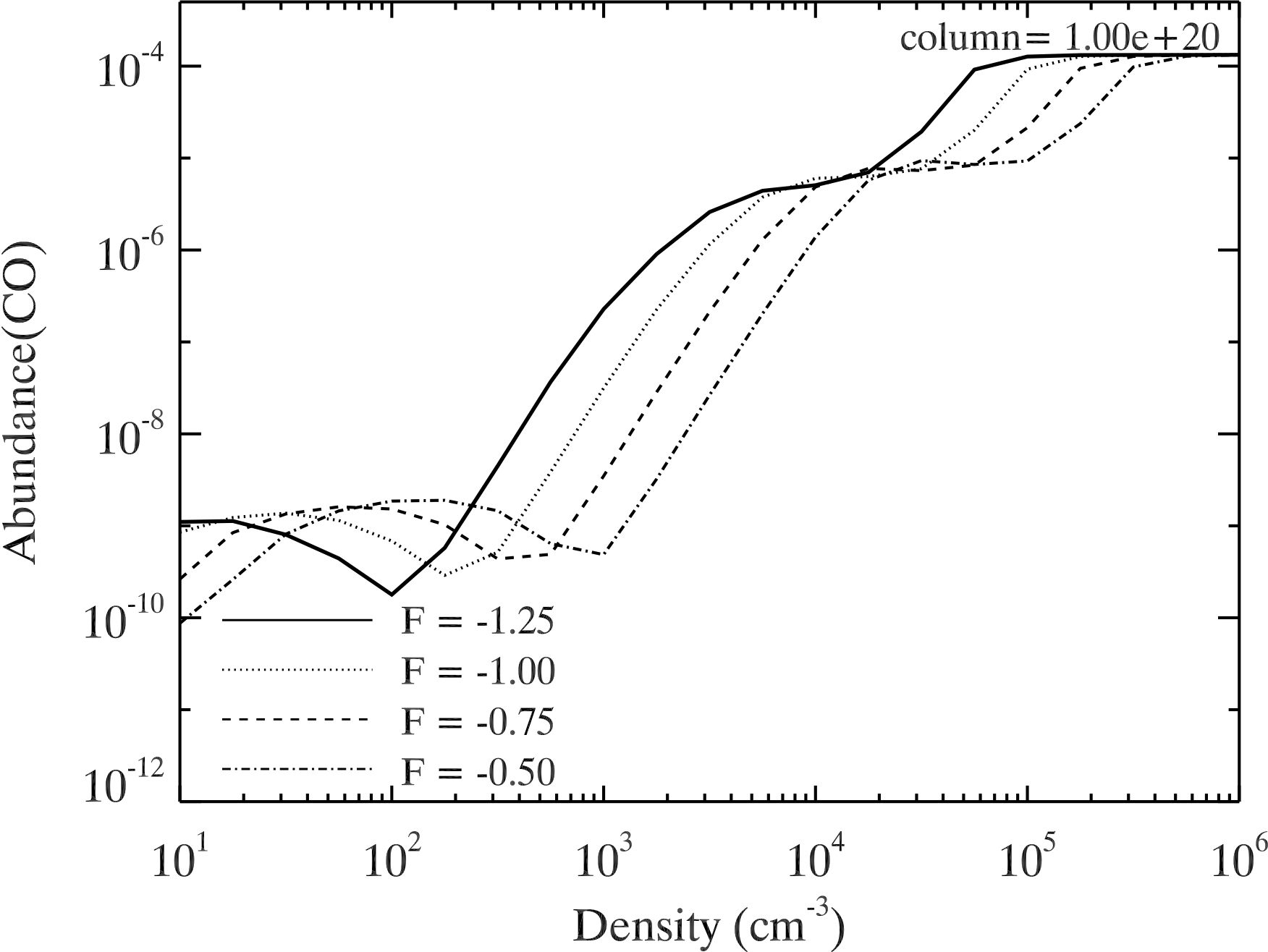}
\vspace{0.05cm}
\includegraphics[angle=0,width=7cm]{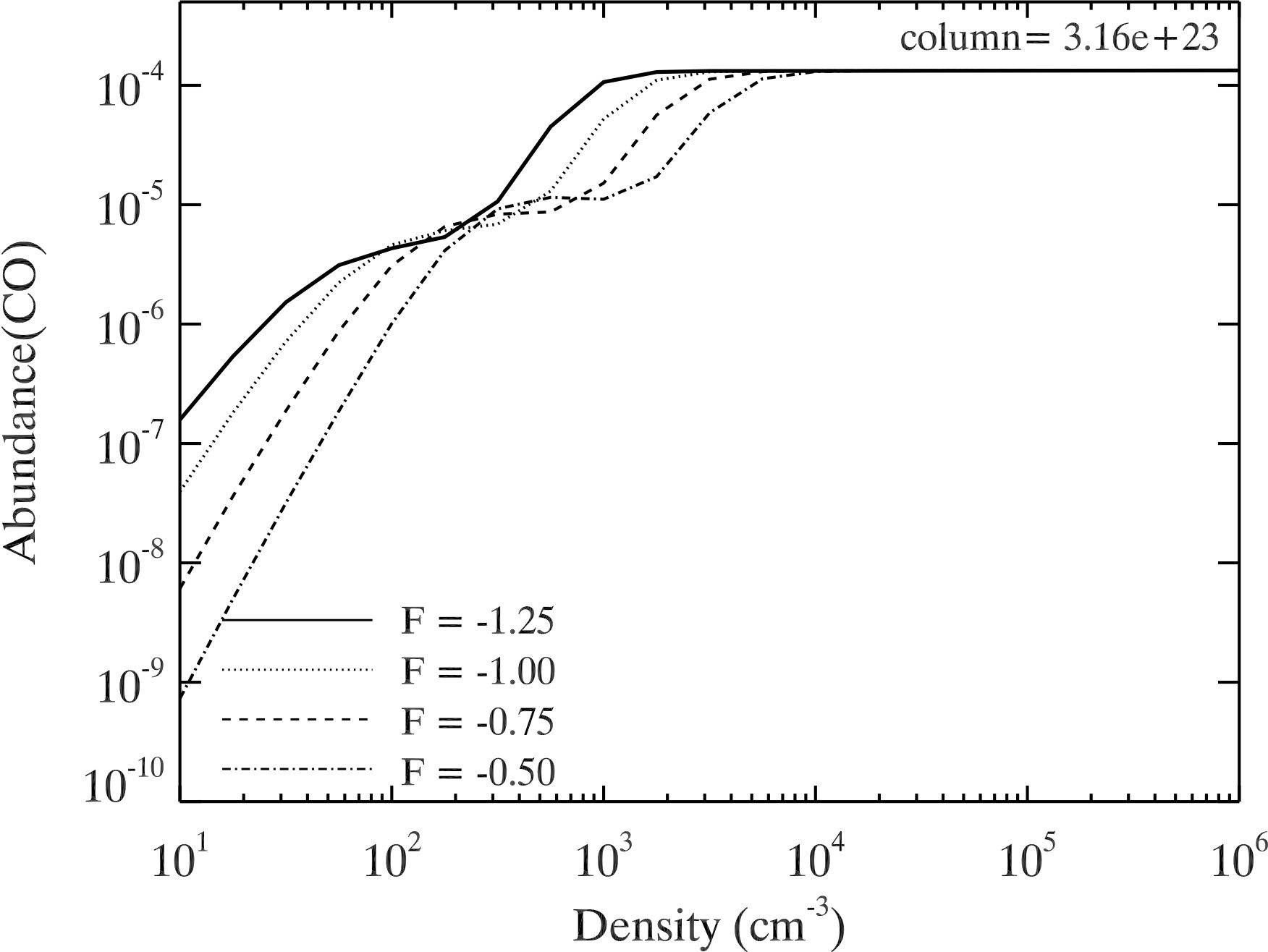}\\
\end{figure*}
\begin{figure*}
\includegraphics[angle=0,width=7cm]{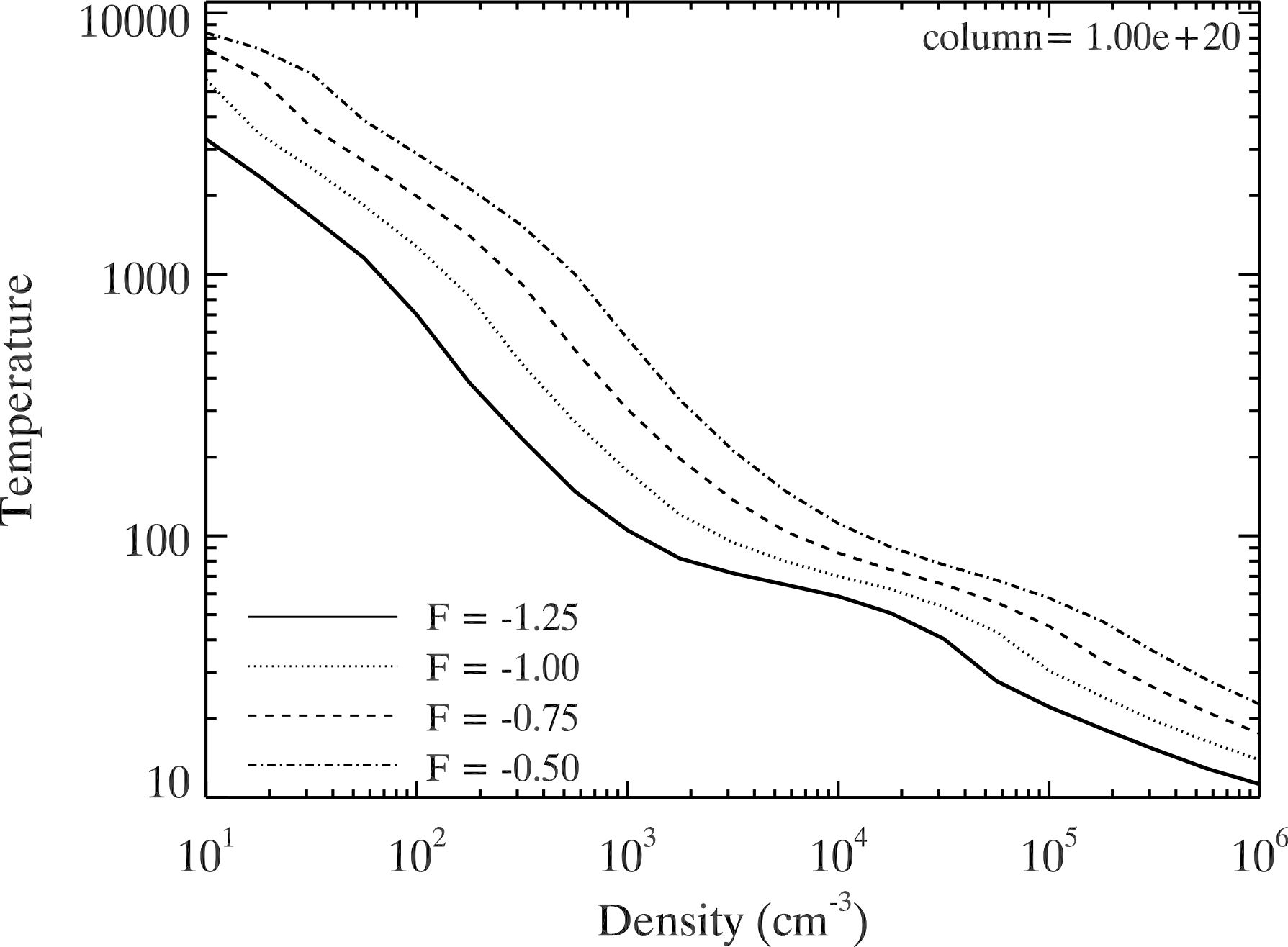}
\vspace{0.05cm}
\includegraphics[angle=0,width=7cm]{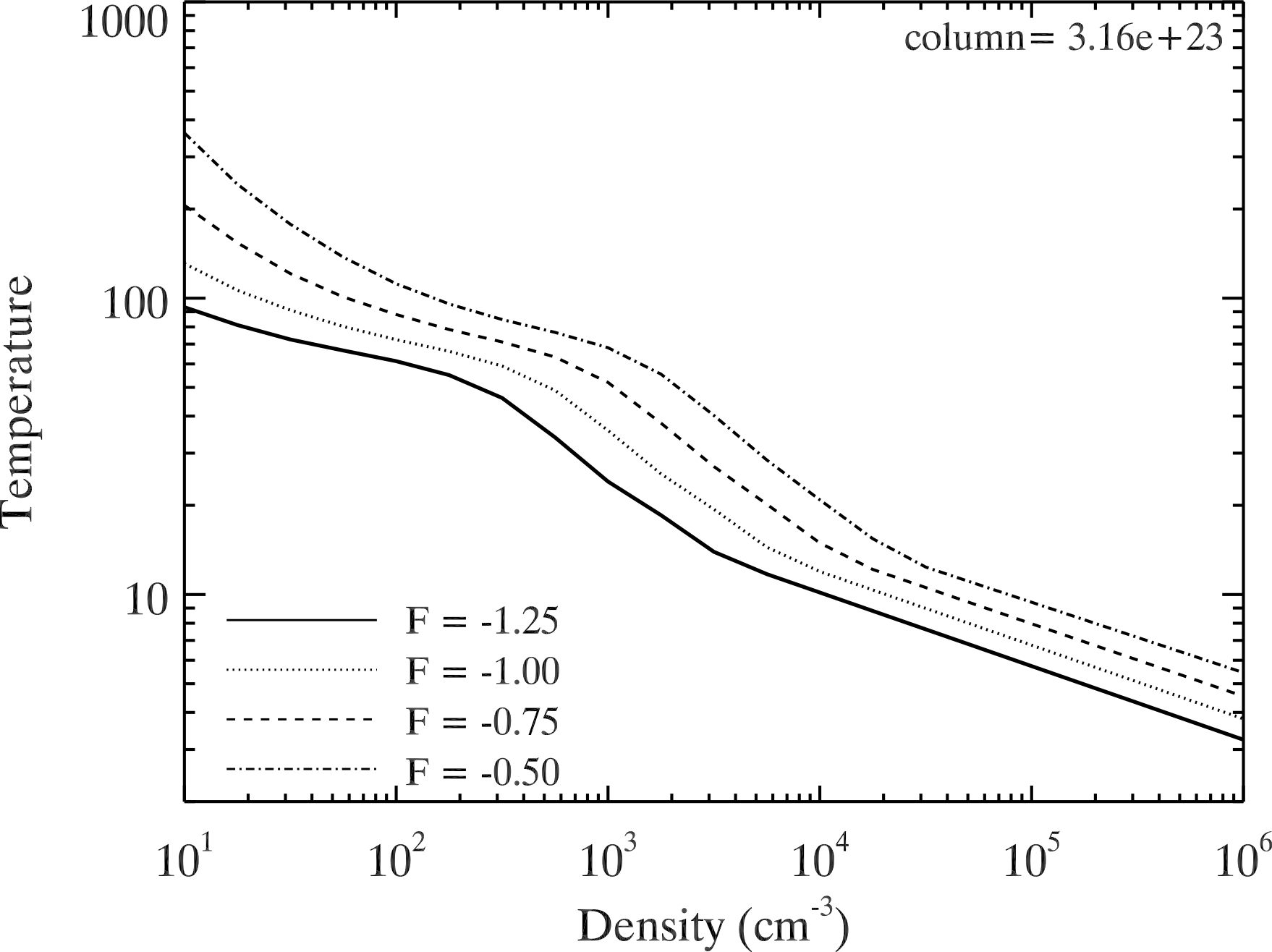}\\
\caption{Abundances of e$^-$, H, H$^+$, H$^-$, H$_2$, H$_2^+$, He, He$^+$, He$^{2+}$, C, C$^+$, C$^{2+}$, O, O$^+$, O$^{2+}$, CO and temperature for column densities of 10$^{20}$ cm$^{-2}$ (left column) and $3.16\times10^{23}$ cm$^{-2}$ (right column) for a X-ray flux of log F = -1.25 - -0.5 erg s$^{-1}$ cm$^{-2}$.} \label{fig16}
\end{figure*}

\begin{figure*}
\includegraphics[angle=0,width=7cm]{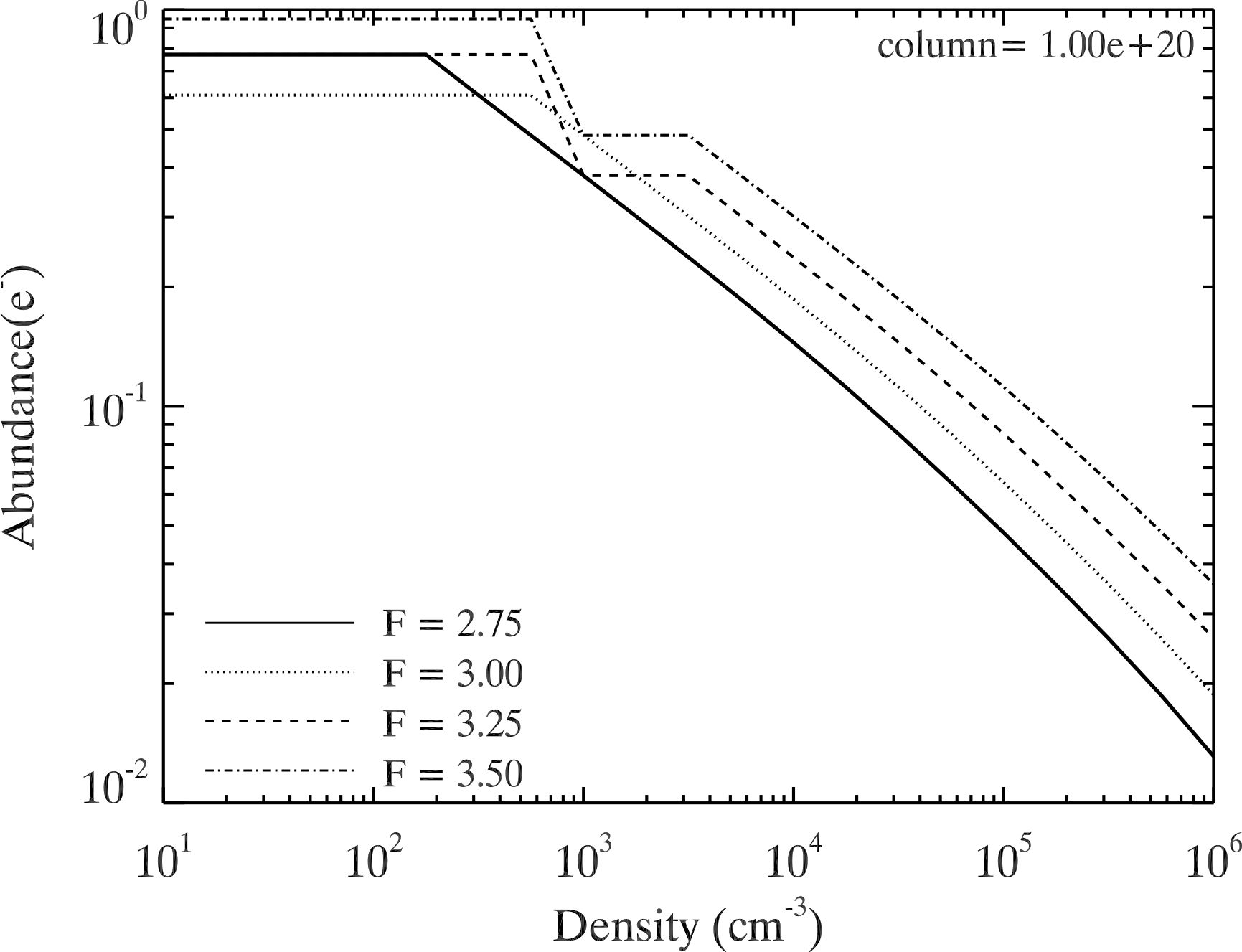}
\vspace{0.05cm}
\includegraphics[angle=0,width=7cm]{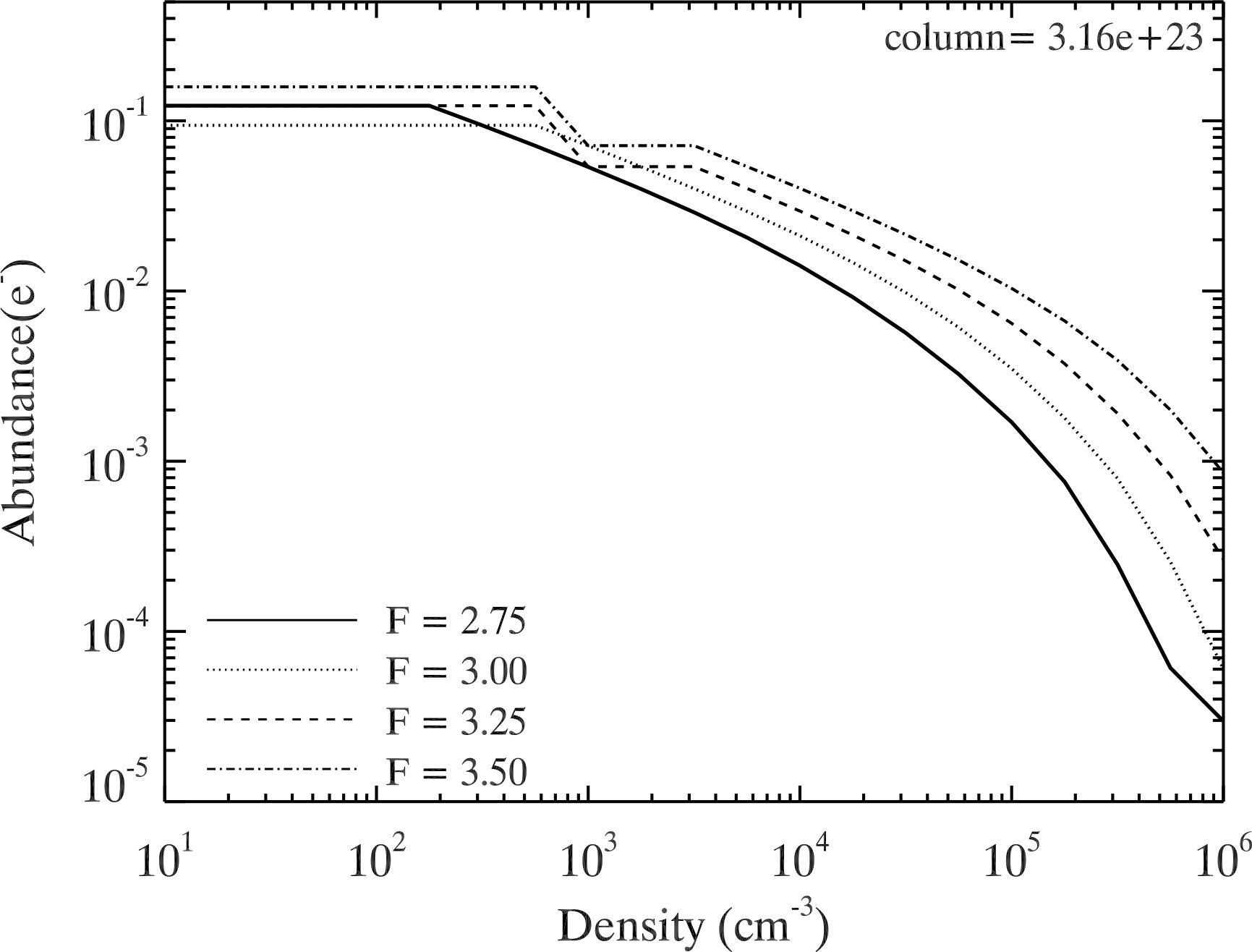}\\
\includegraphics[angle=0,width=7cm]{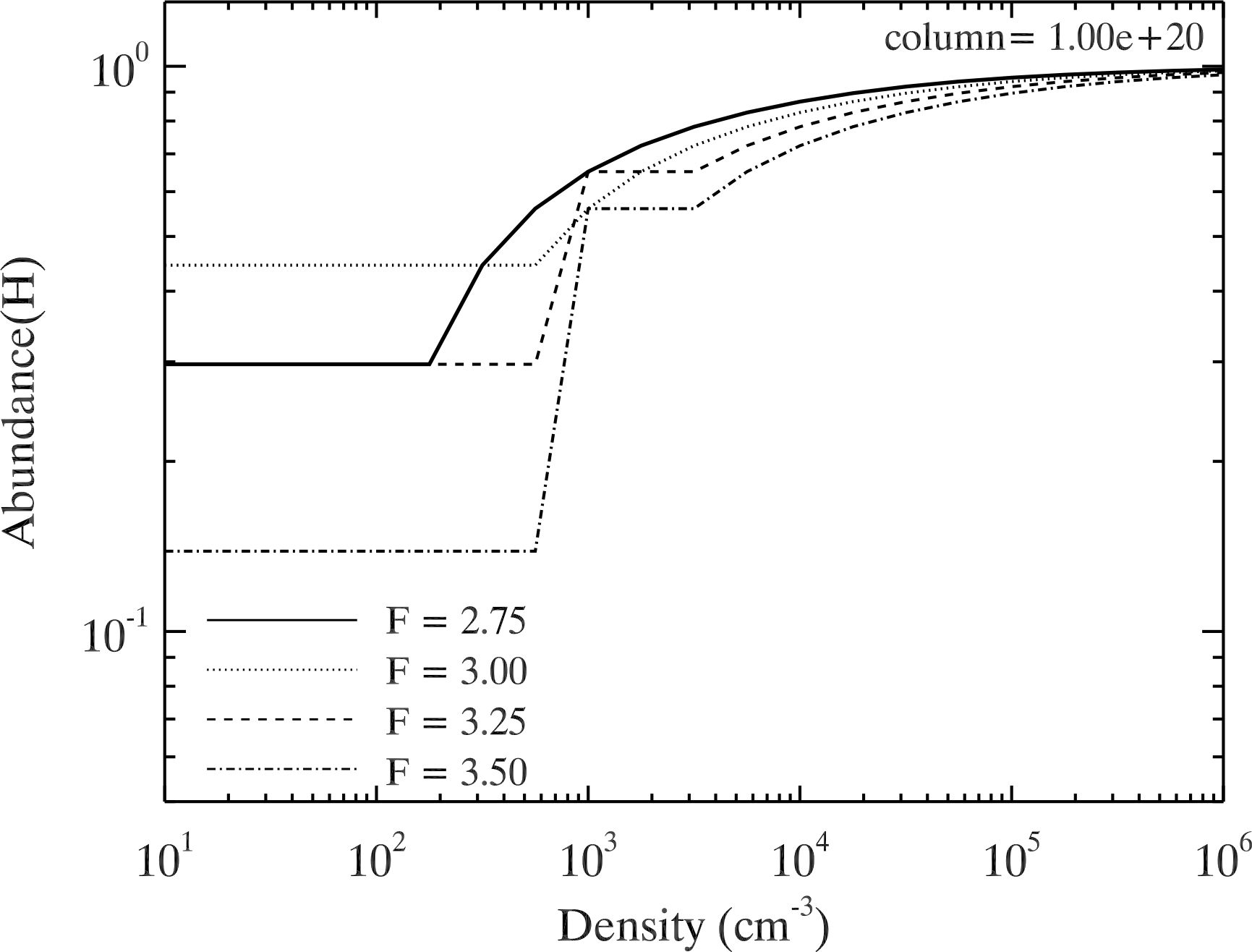}
\vspace{0.05cm}
\includegraphics[angle=0,width=7cm]{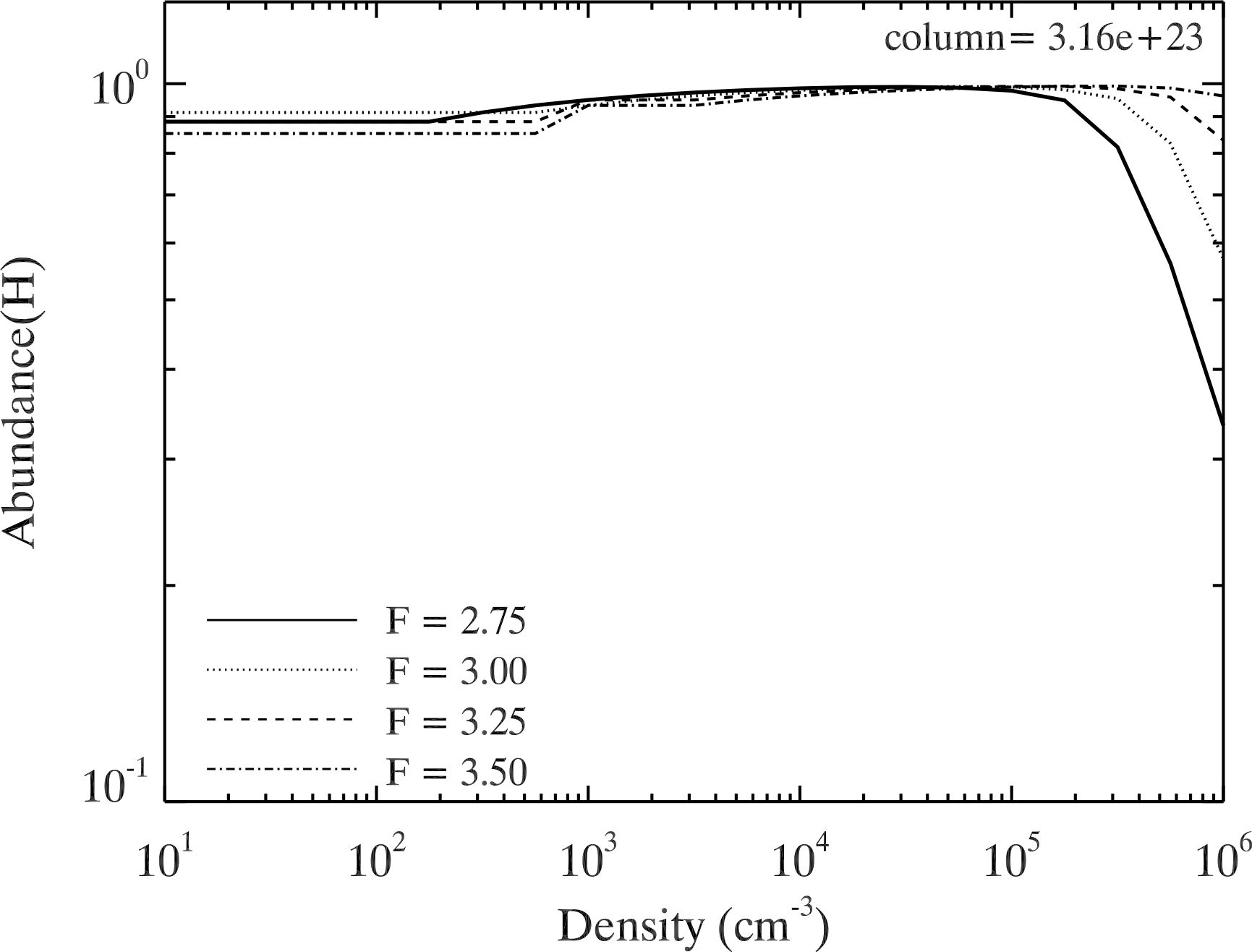}\\
\includegraphics[angle=0,width=7cm]{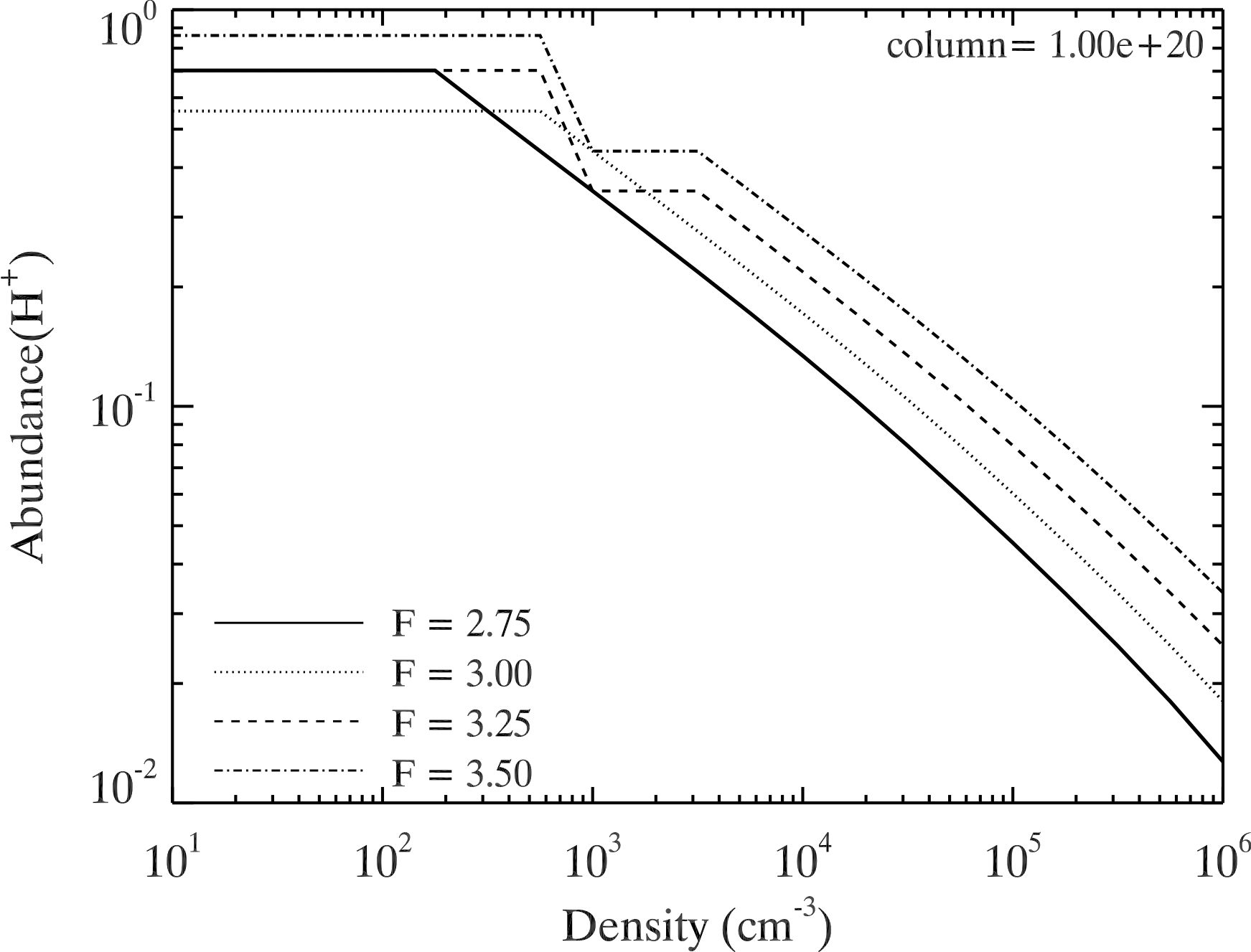}
\vspace{0.05cm}
\includegraphics[angle=0,width=7cm]{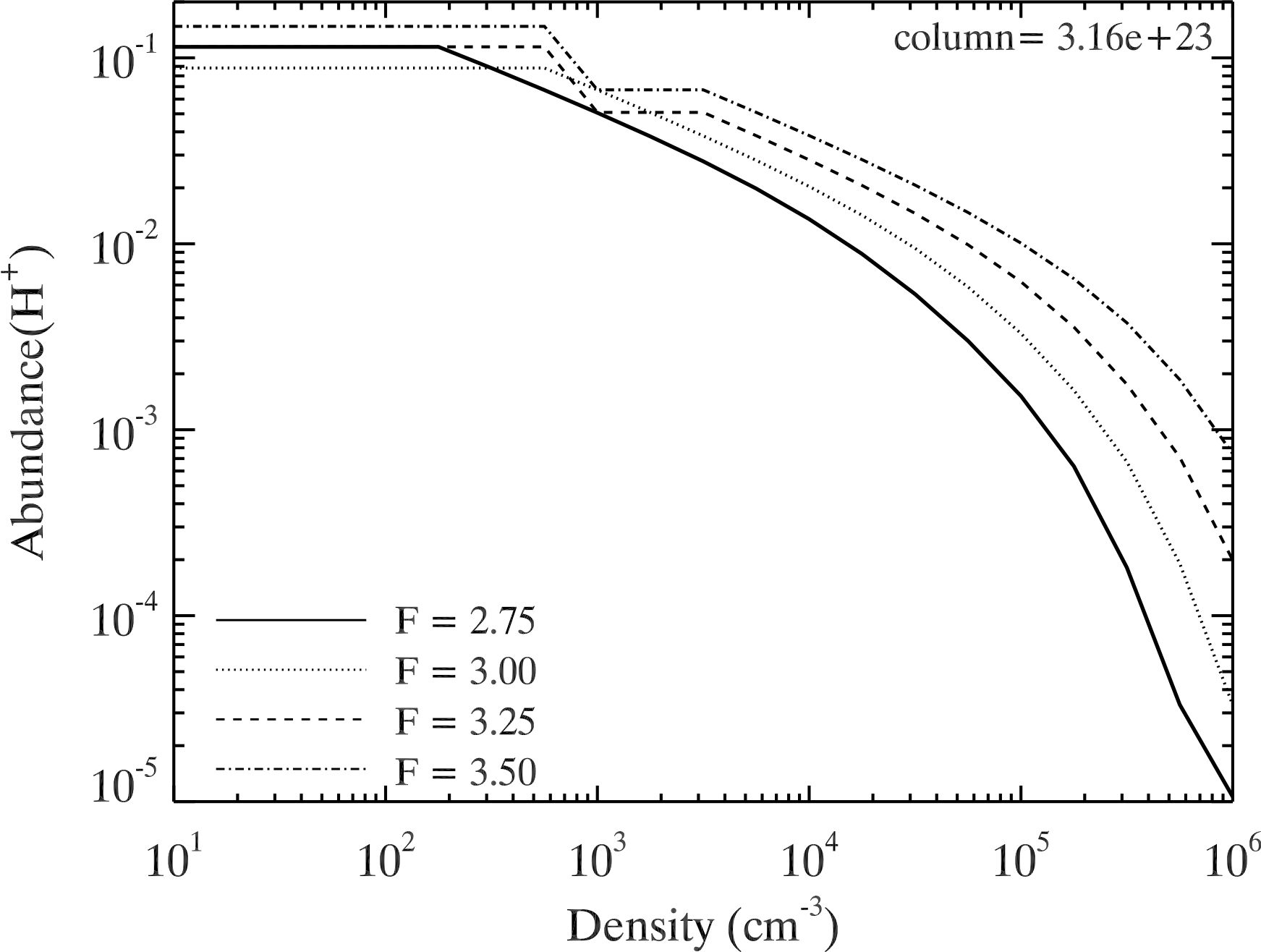}\\
\includegraphics[angle=0,width=7cm]{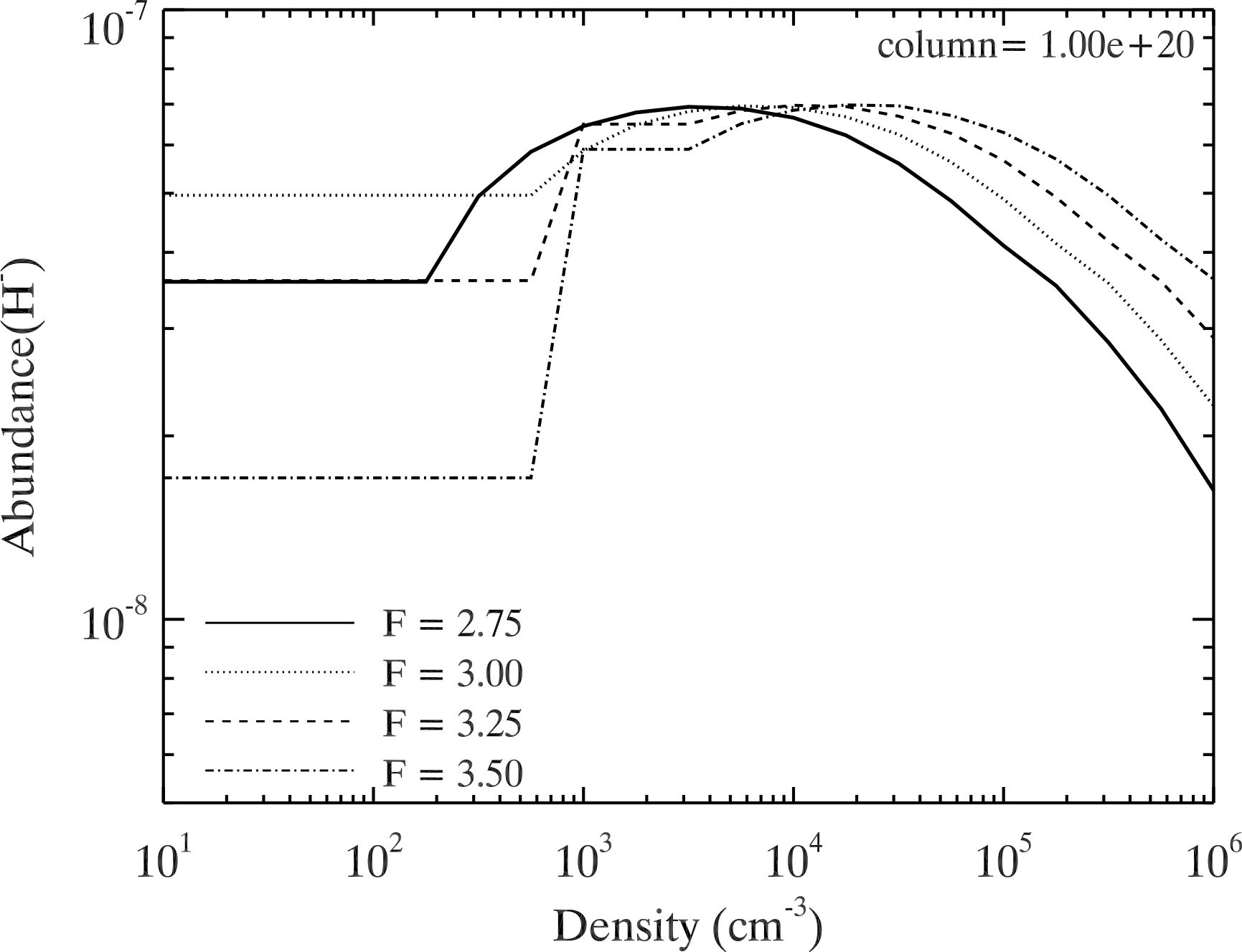}
\vspace{0.05cm}
\includegraphics[angle=0,width=7cm]{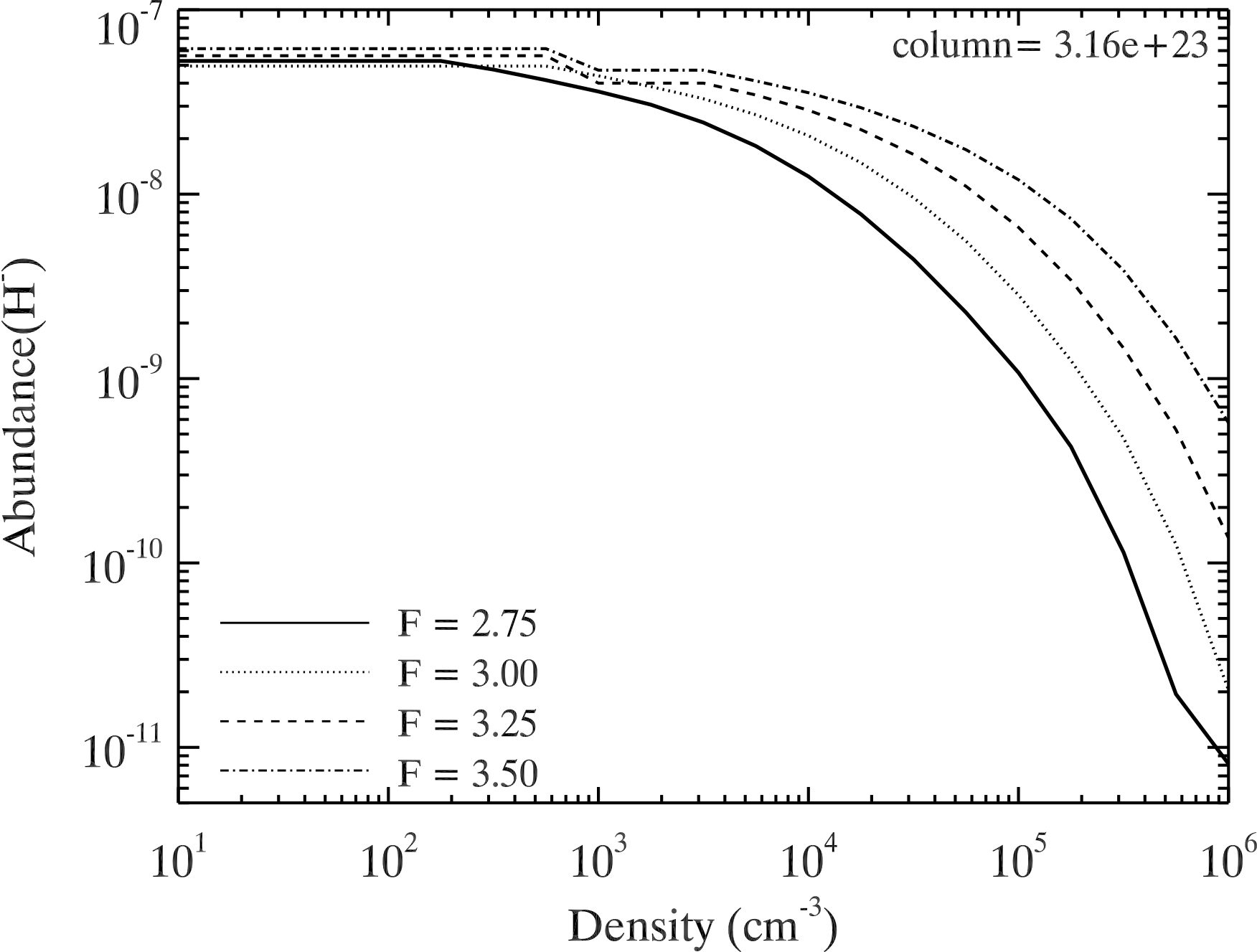}\\
\end{figure*}
\begin{figure*}
\includegraphics[angle=0,width=7cm]{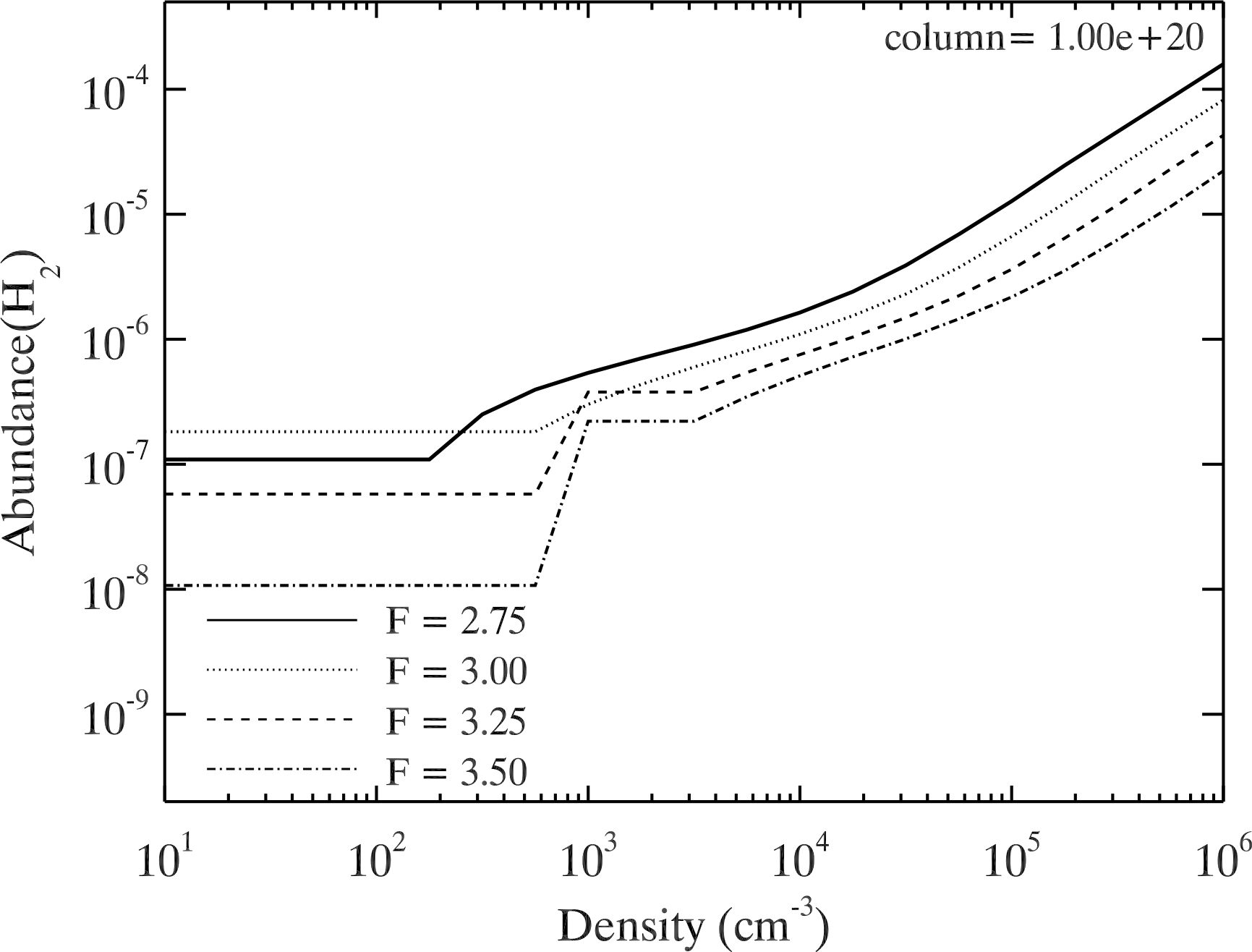}
\vspace{0.05cm}
\includegraphics[angle=0,width=7cm]{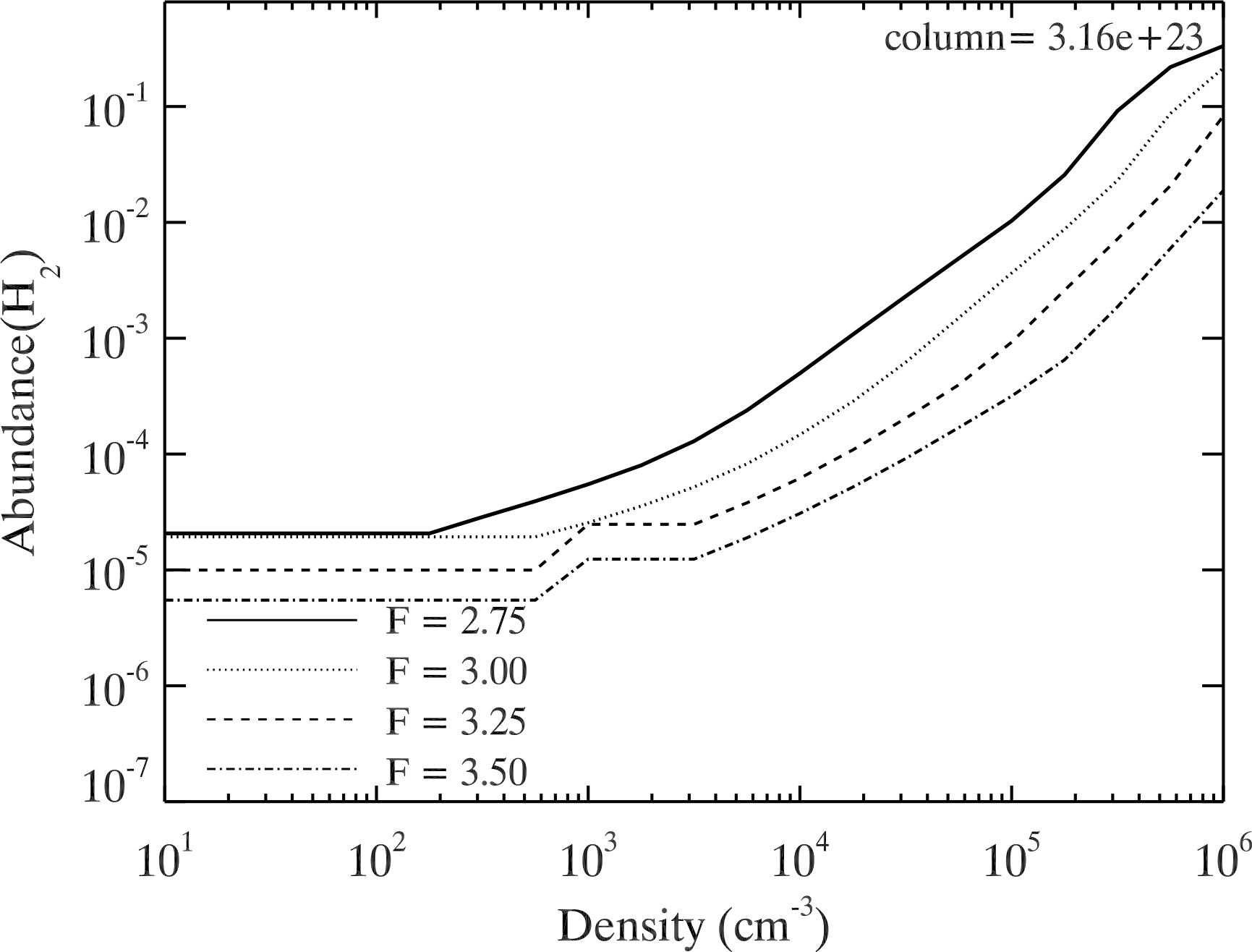}\\
\includegraphics[angle=0,width=7cm]{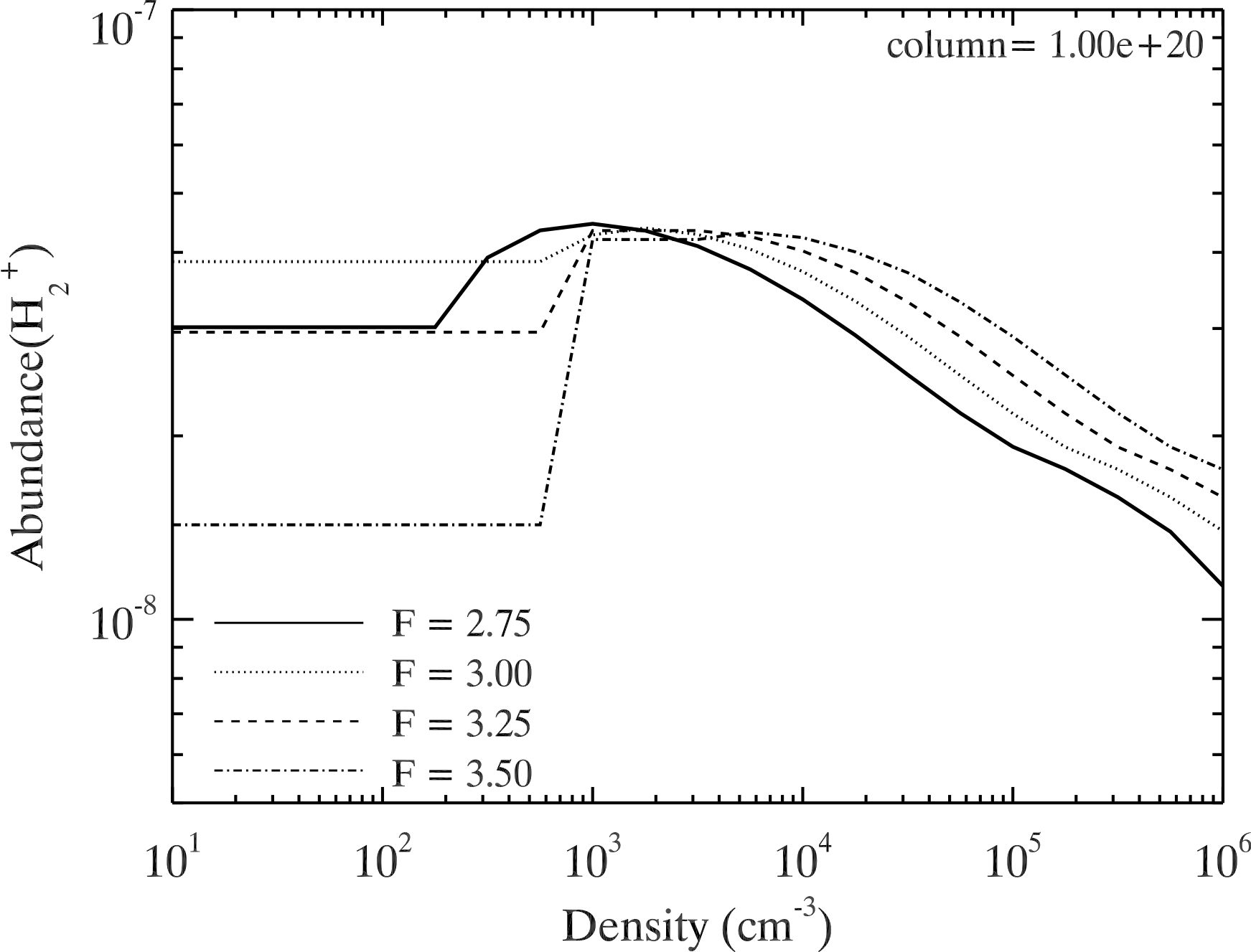}
\vspace{0.05cm}
\includegraphics[angle=0,width=7cm]{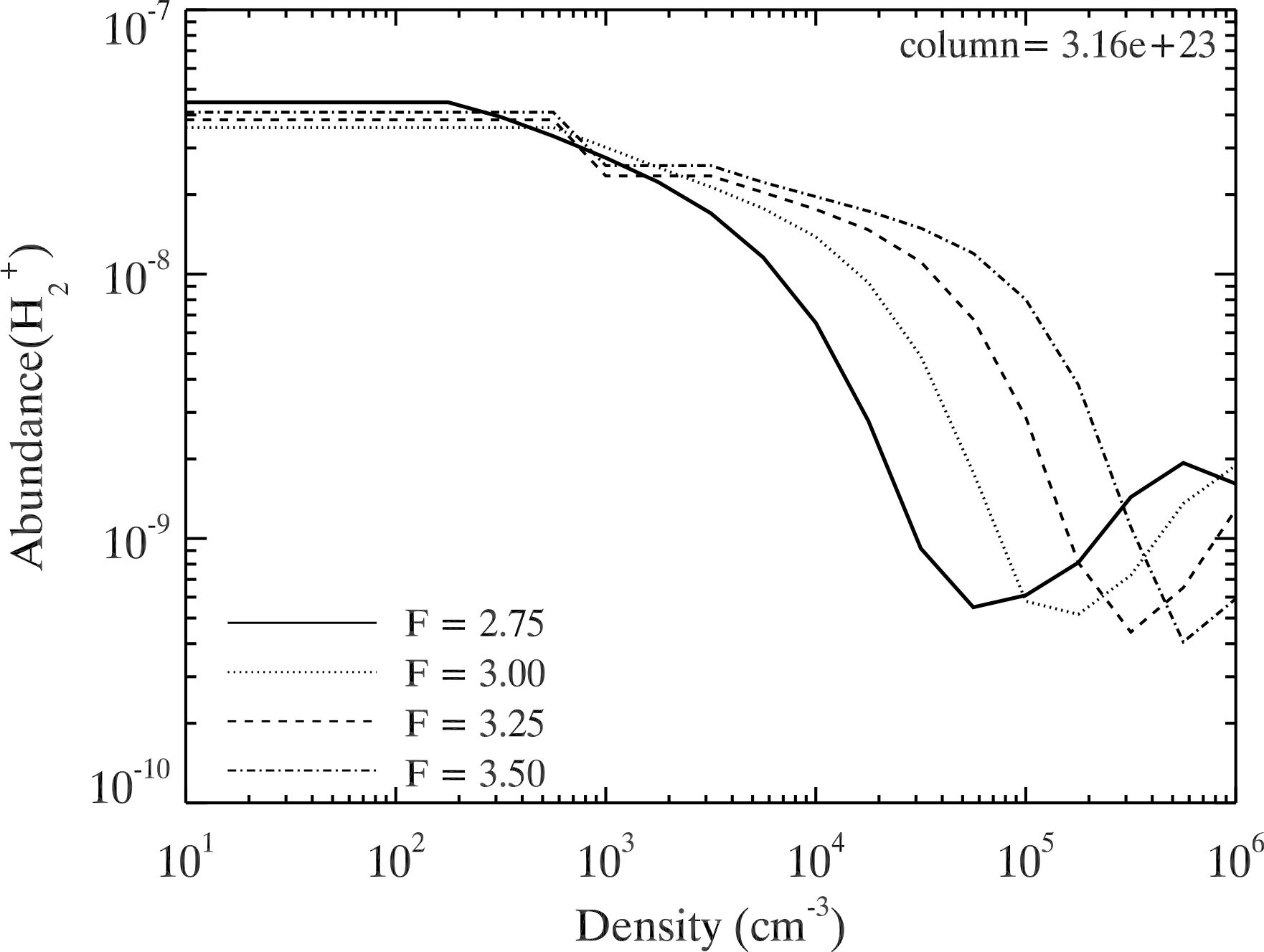}\\
\includegraphics[angle=0,width=7cm]{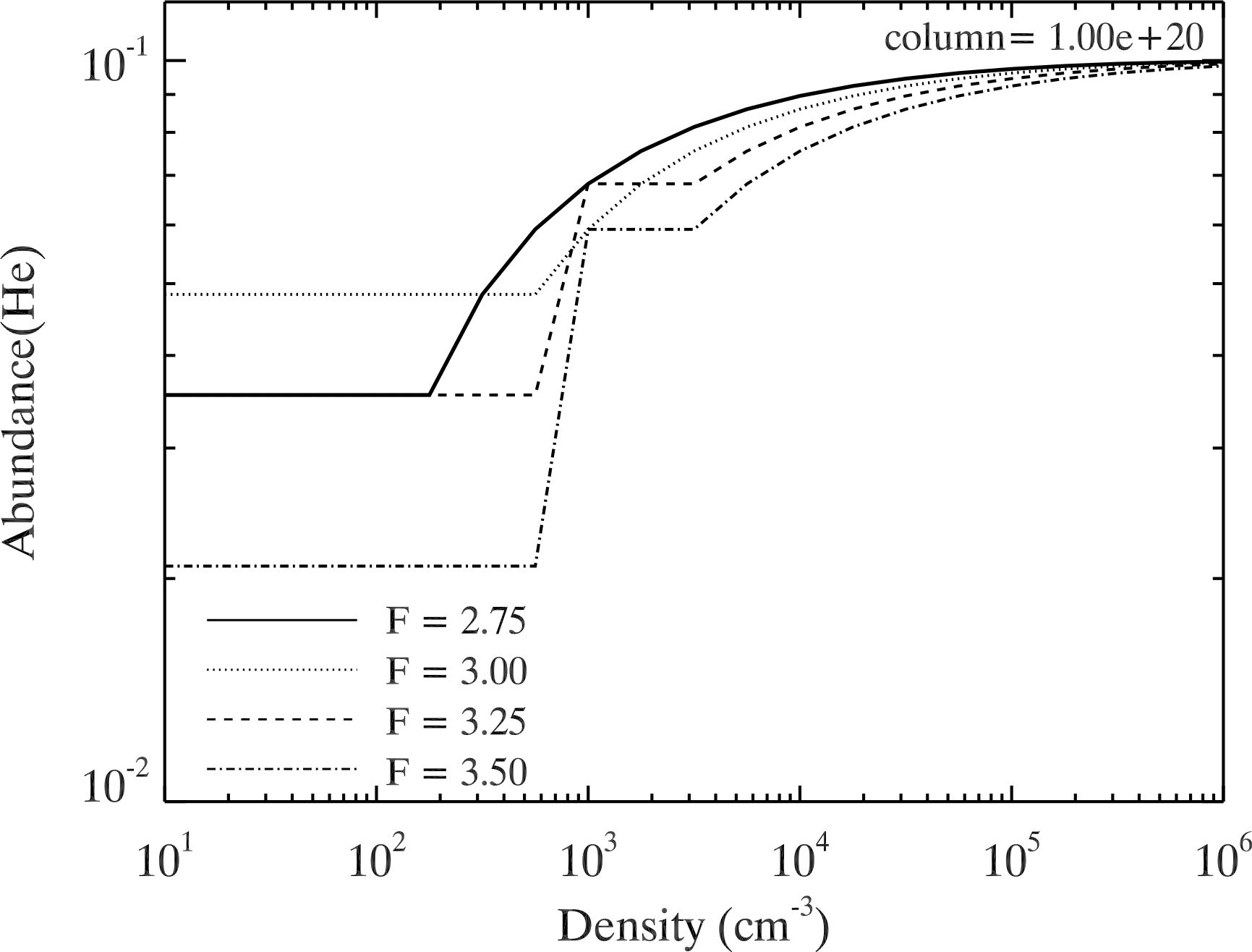}
\vspace{0.05cm}
\includegraphics[angle=0,width=7cm]{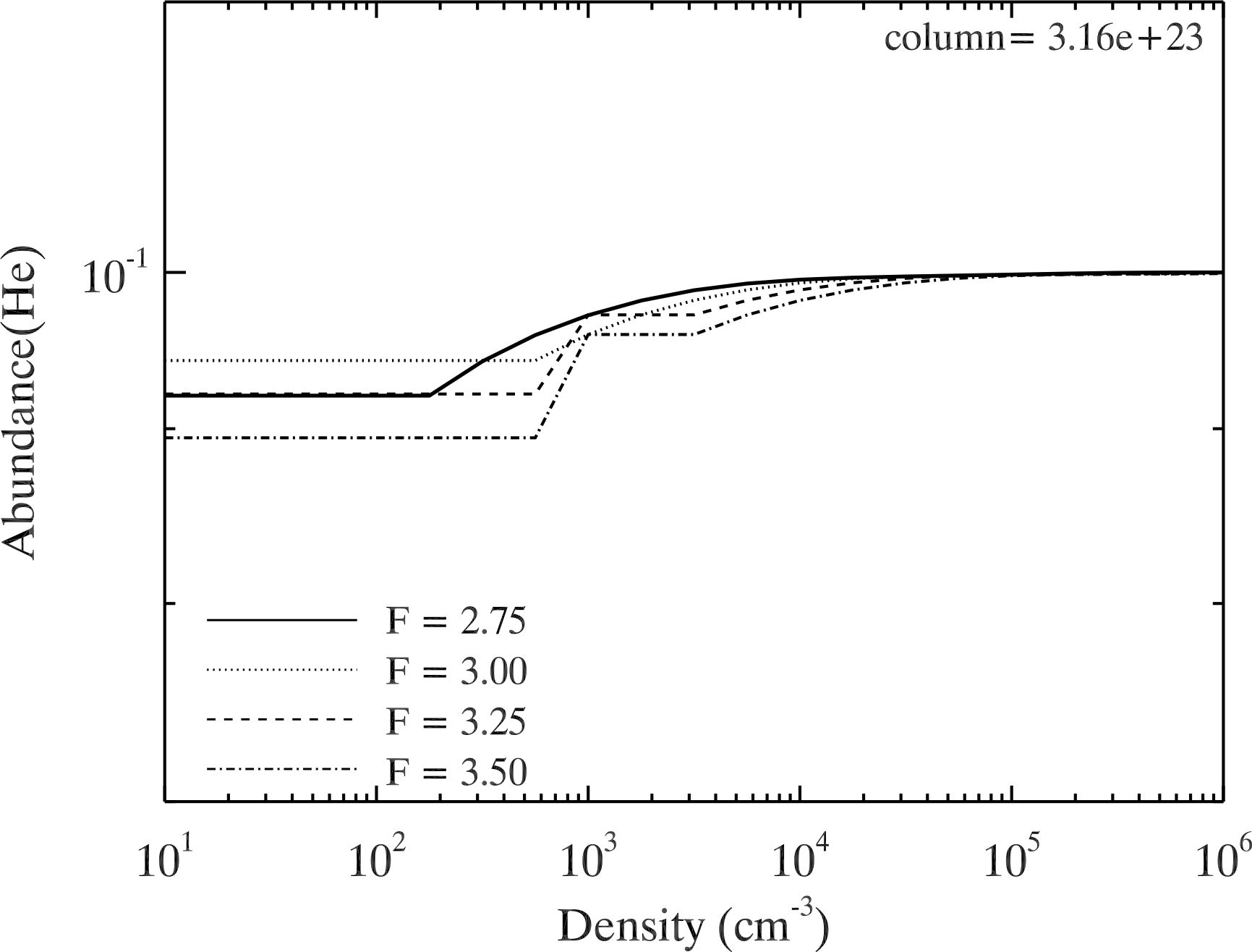}\\
\includegraphics[angle=0,width=7cm]{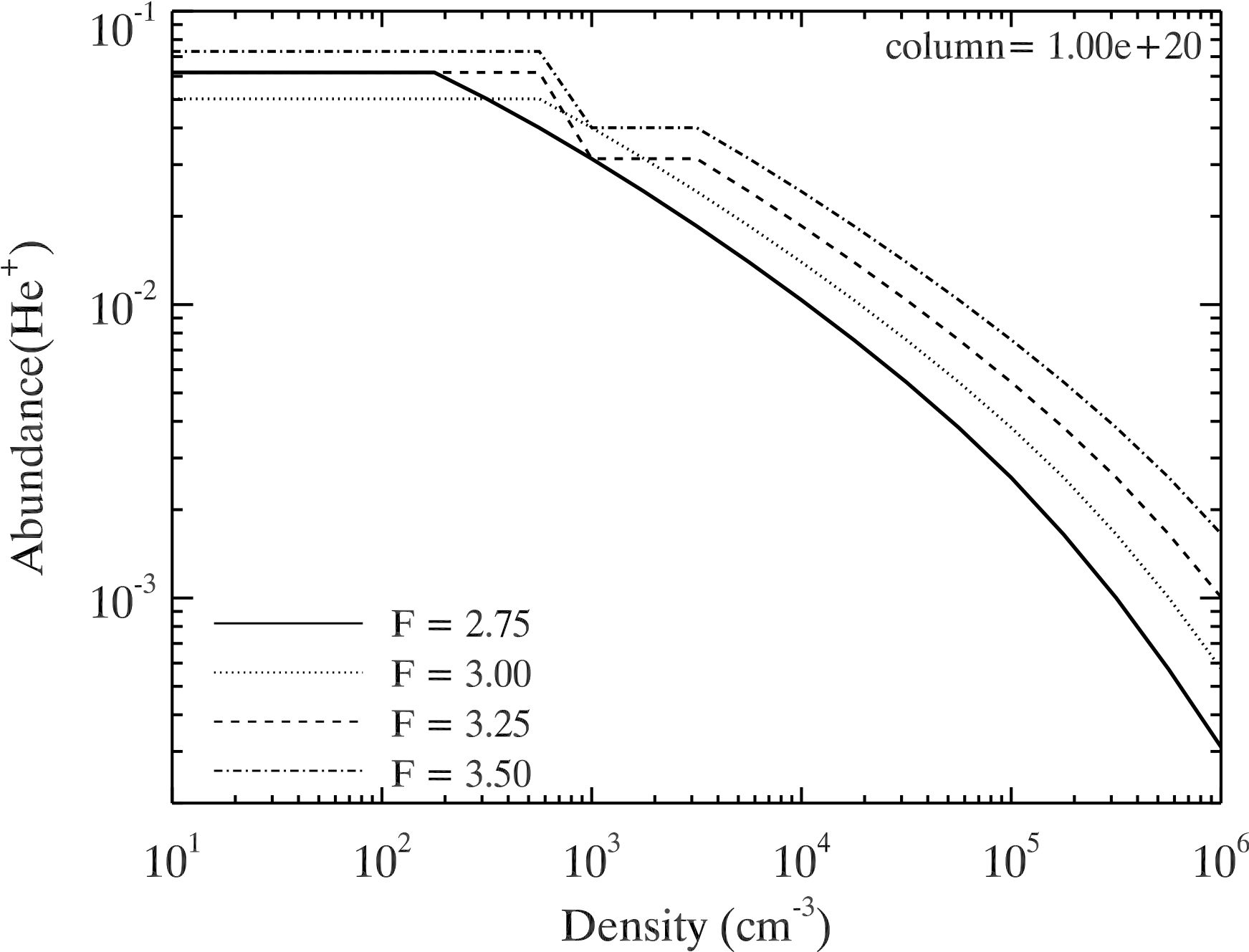}
\vspace{0.05cm}
\includegraphics[angle=0,width=7cm]{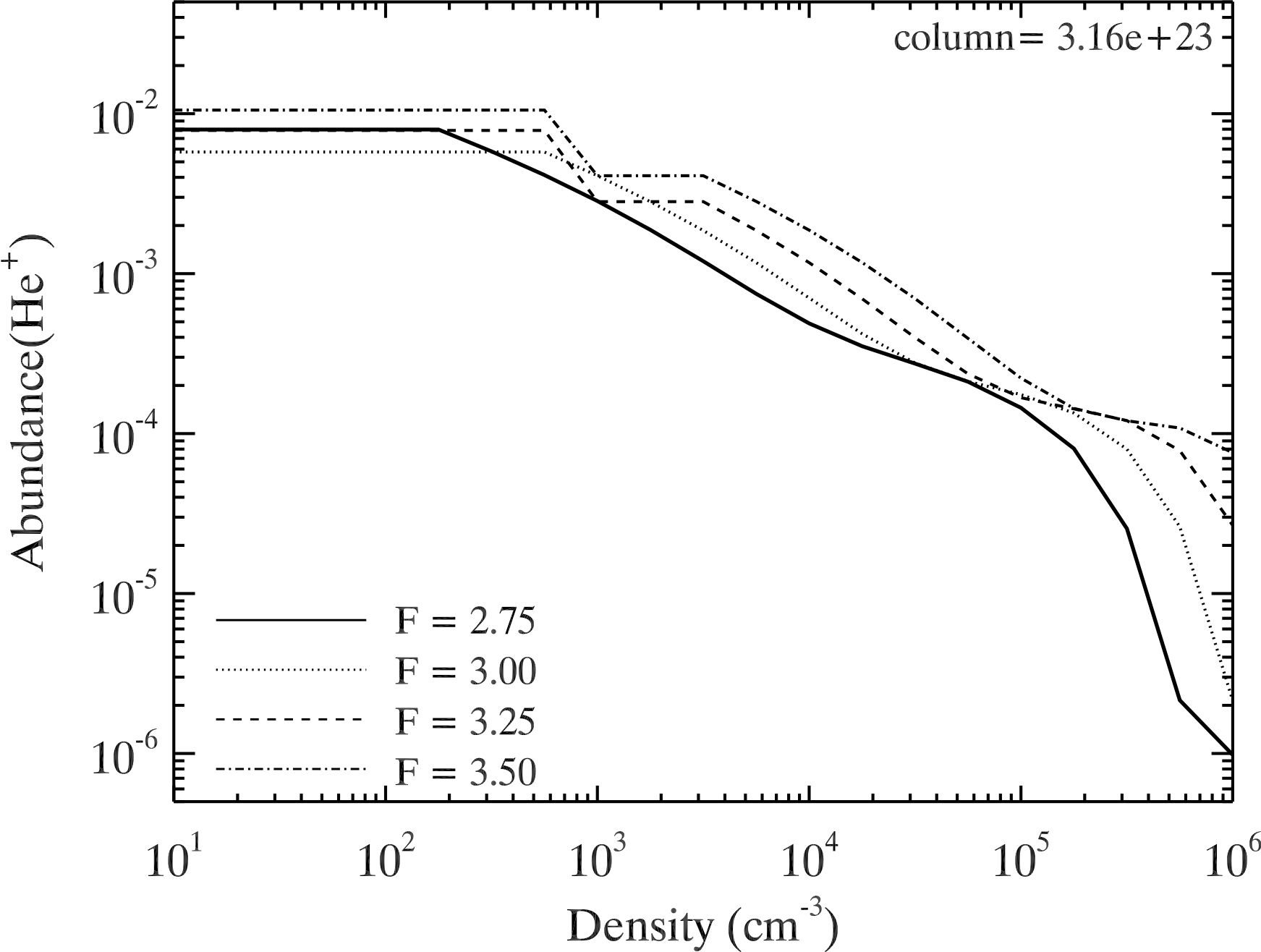}\\
\end{figure*}
\begin{figure*}
\includegraphics[angle=0,width=7cm]{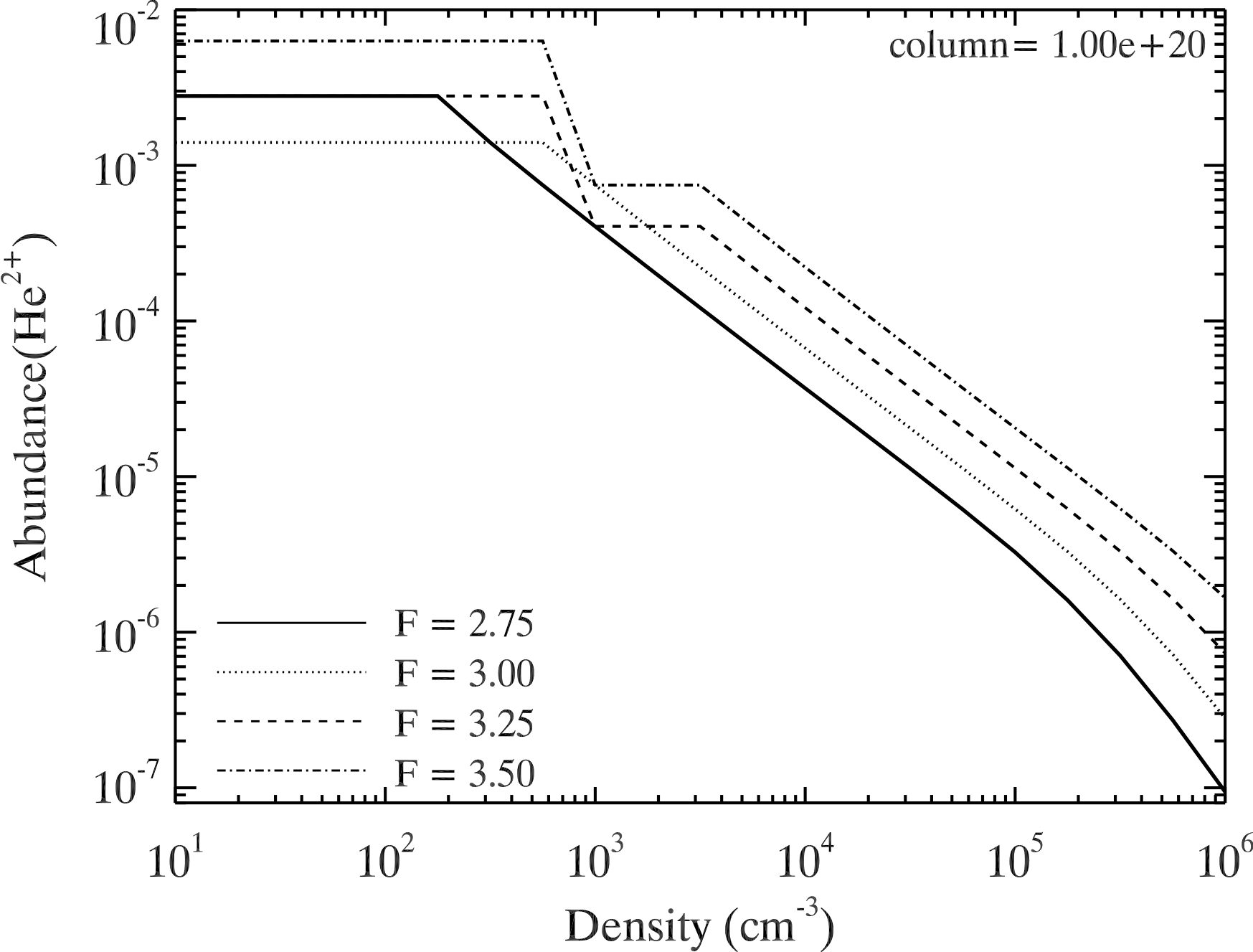}
\vspace{0.05cm}
\includegraphics[angle=0,width=7cm]{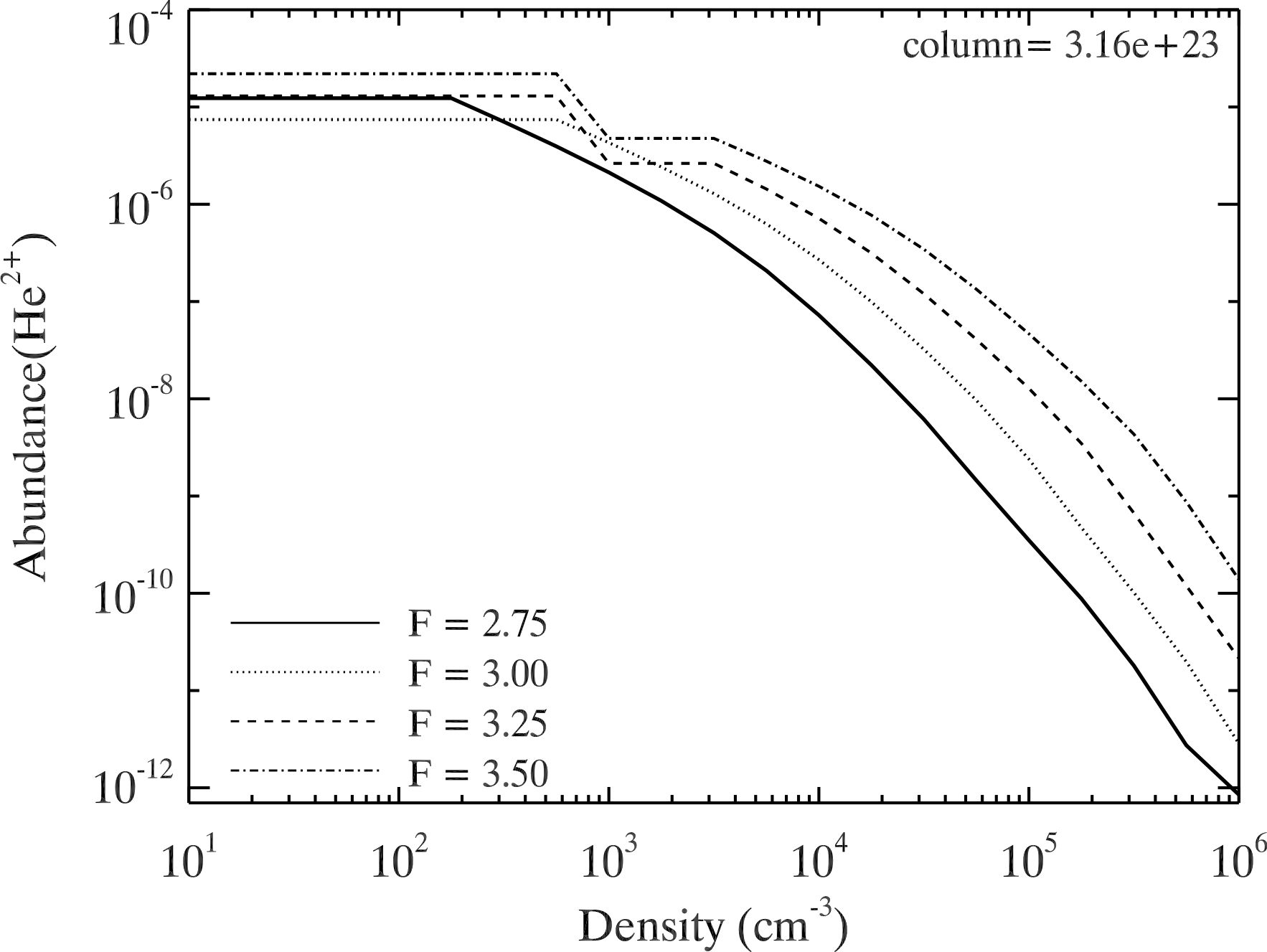}\\
\includegraphics[angle=0,width=7cm]{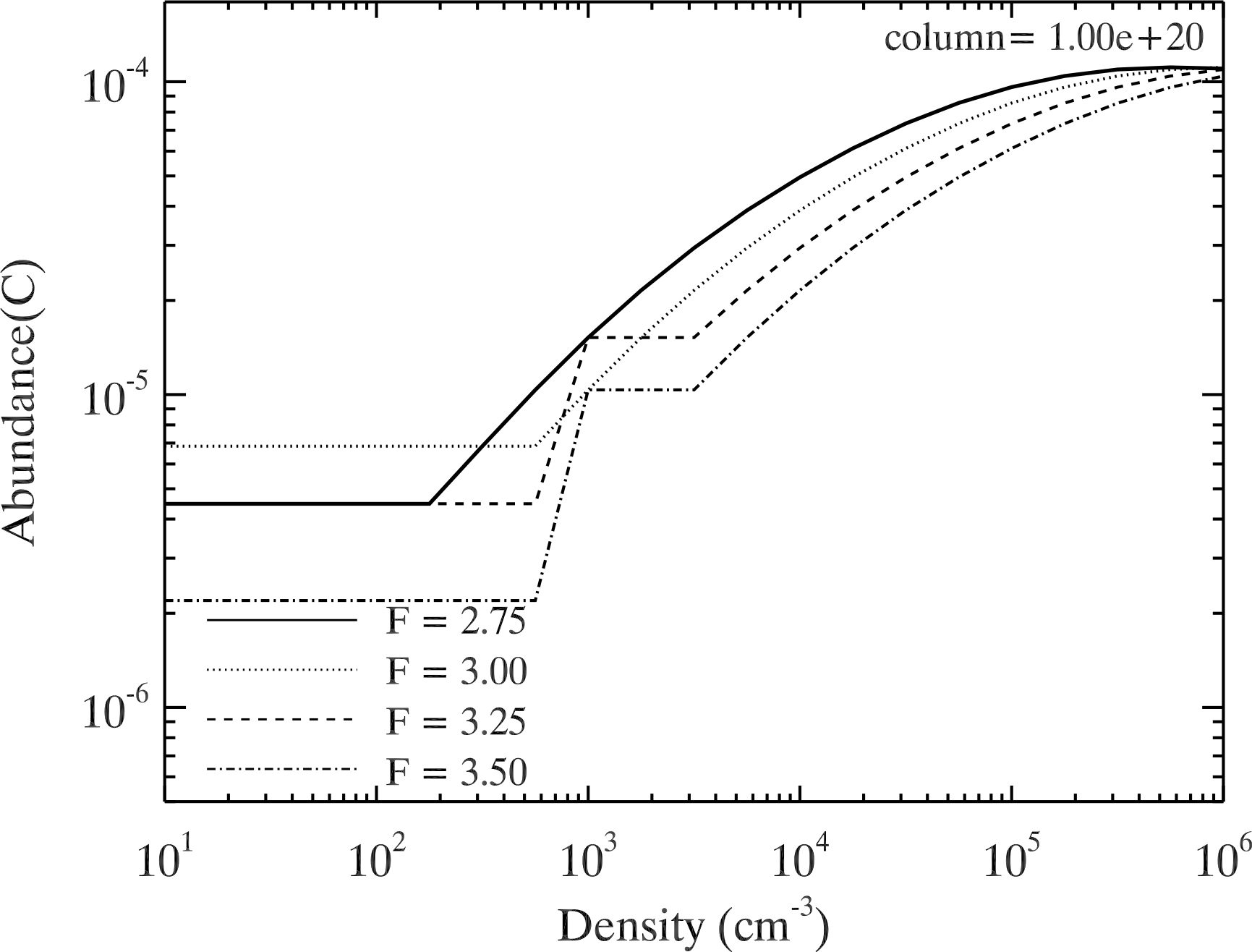}
\vspace{0.05cm}
\includegraphics[angle=0,width=7cm]{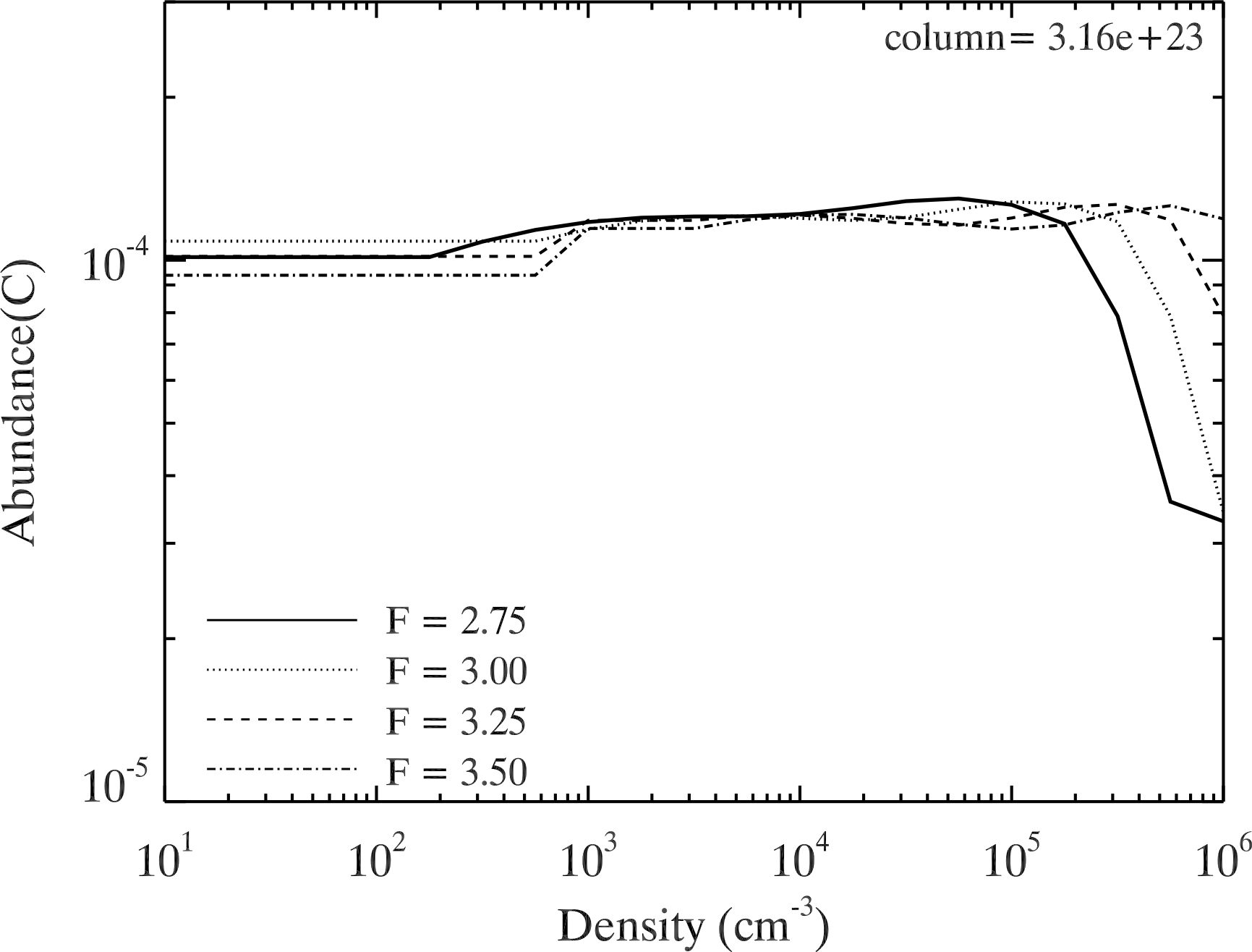}\\
\includegraphics[angle=0,width=7cm]{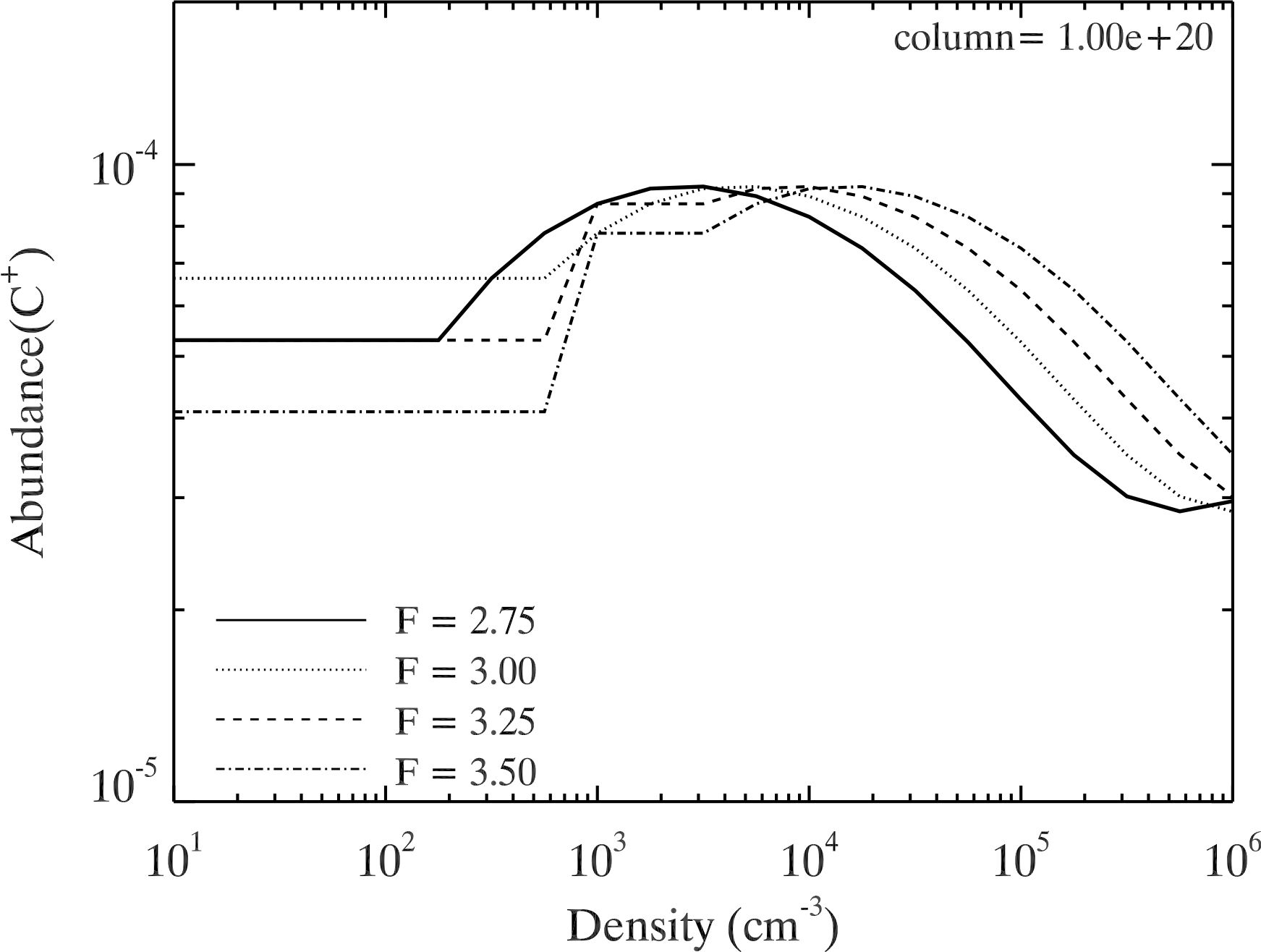}
\vspace{0.05cm}
\includegraphics[angle=0,width=7cm]{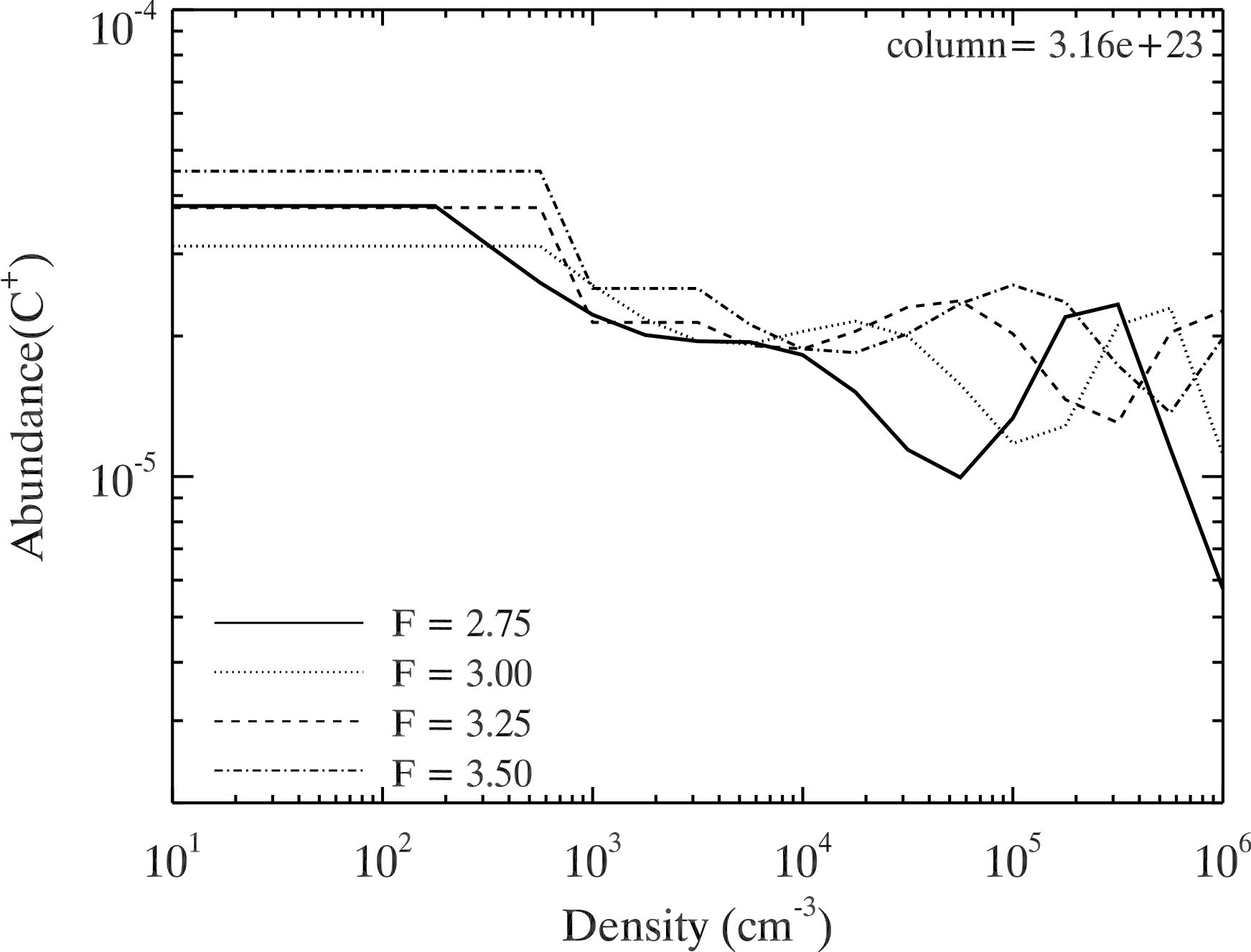}\\
\includegraphics[angle=0,width=7cm]{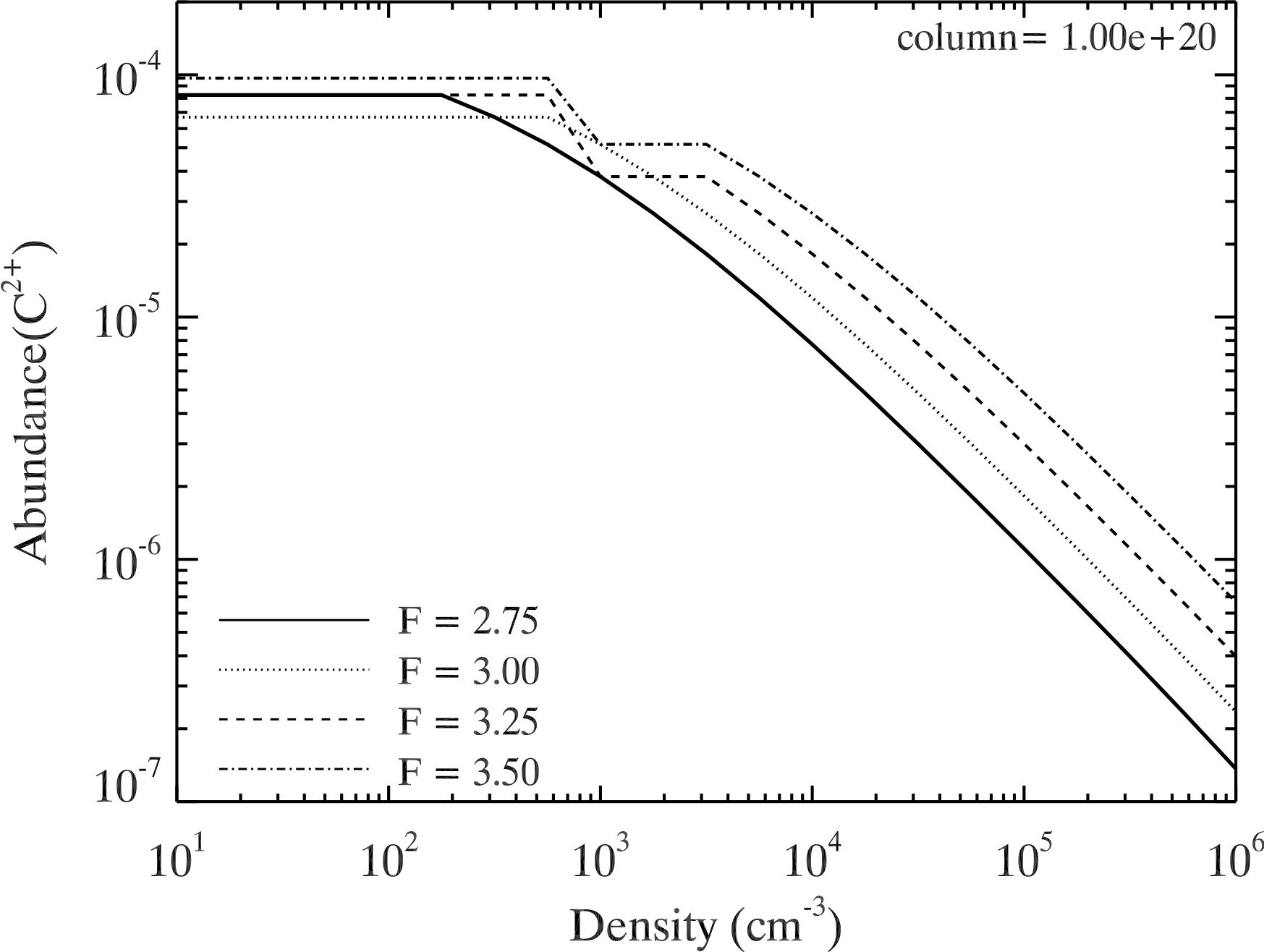}
\vspace{0.05cm}
\includegraphics[angle=0,width=7cm]{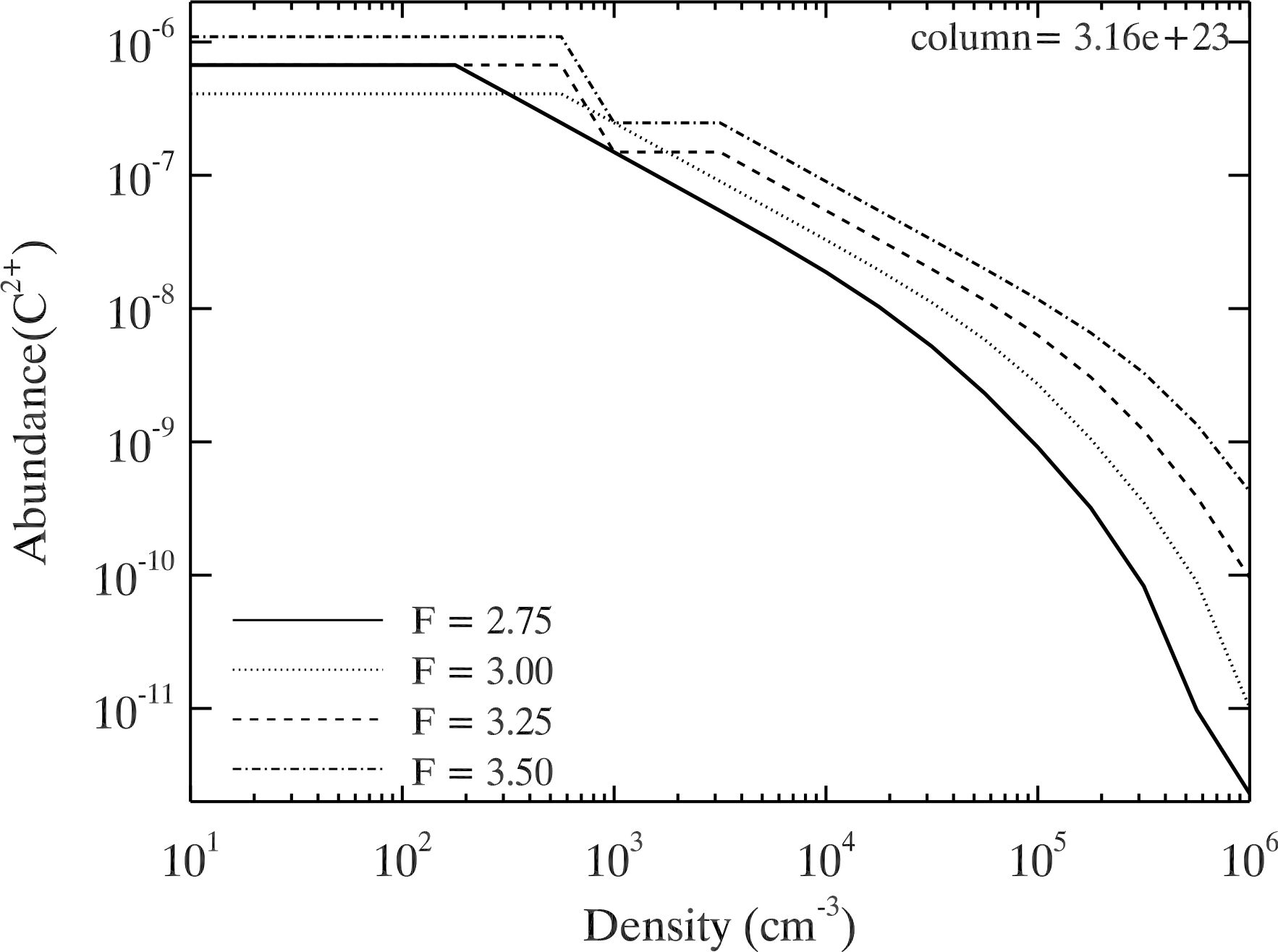}\\
\end{figure*}
\begin{figure*}
\includegraphics[angle=0,width=7cm]{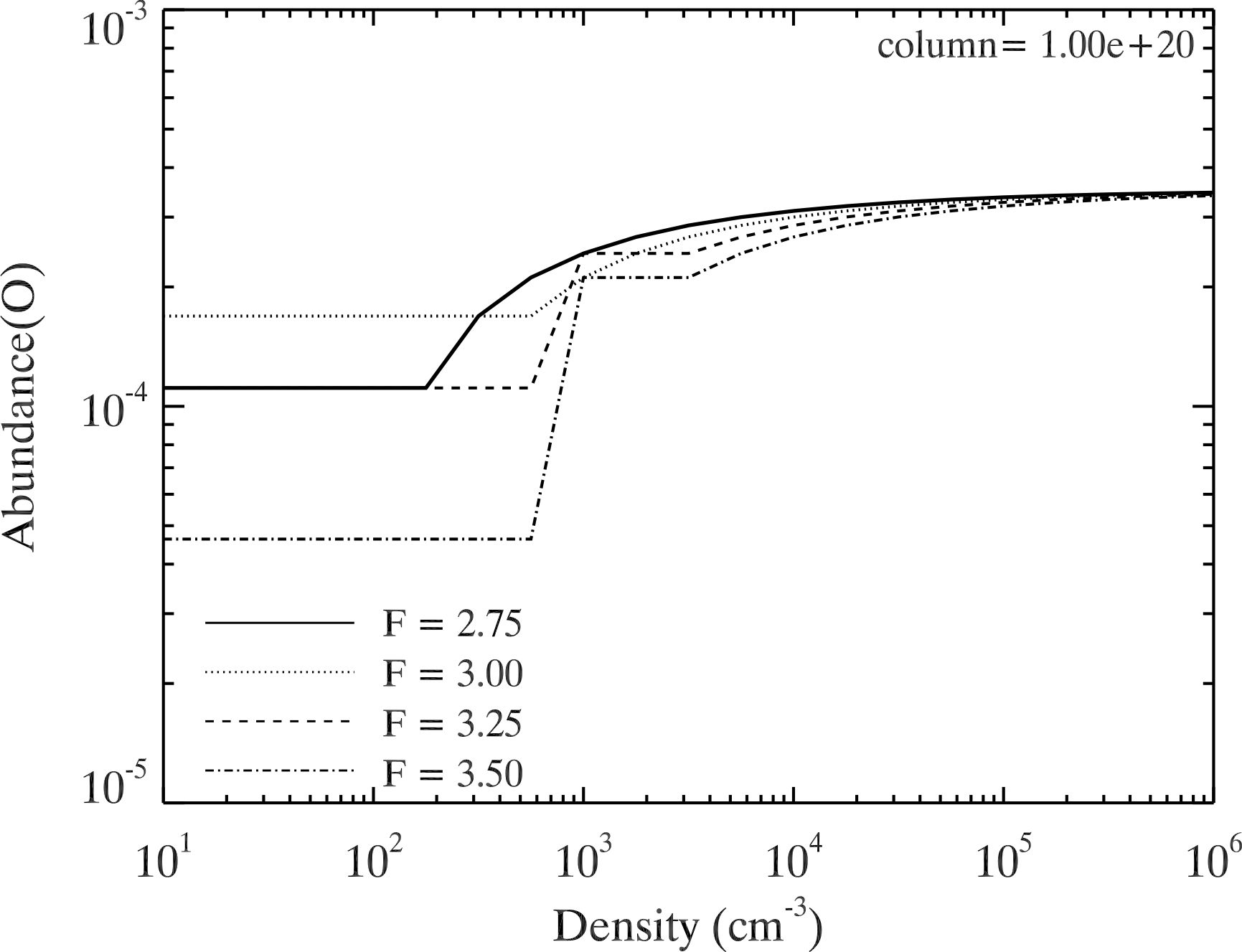}
\vspace{0.05cm}
\includegraphics[angle=0,width=7cm]{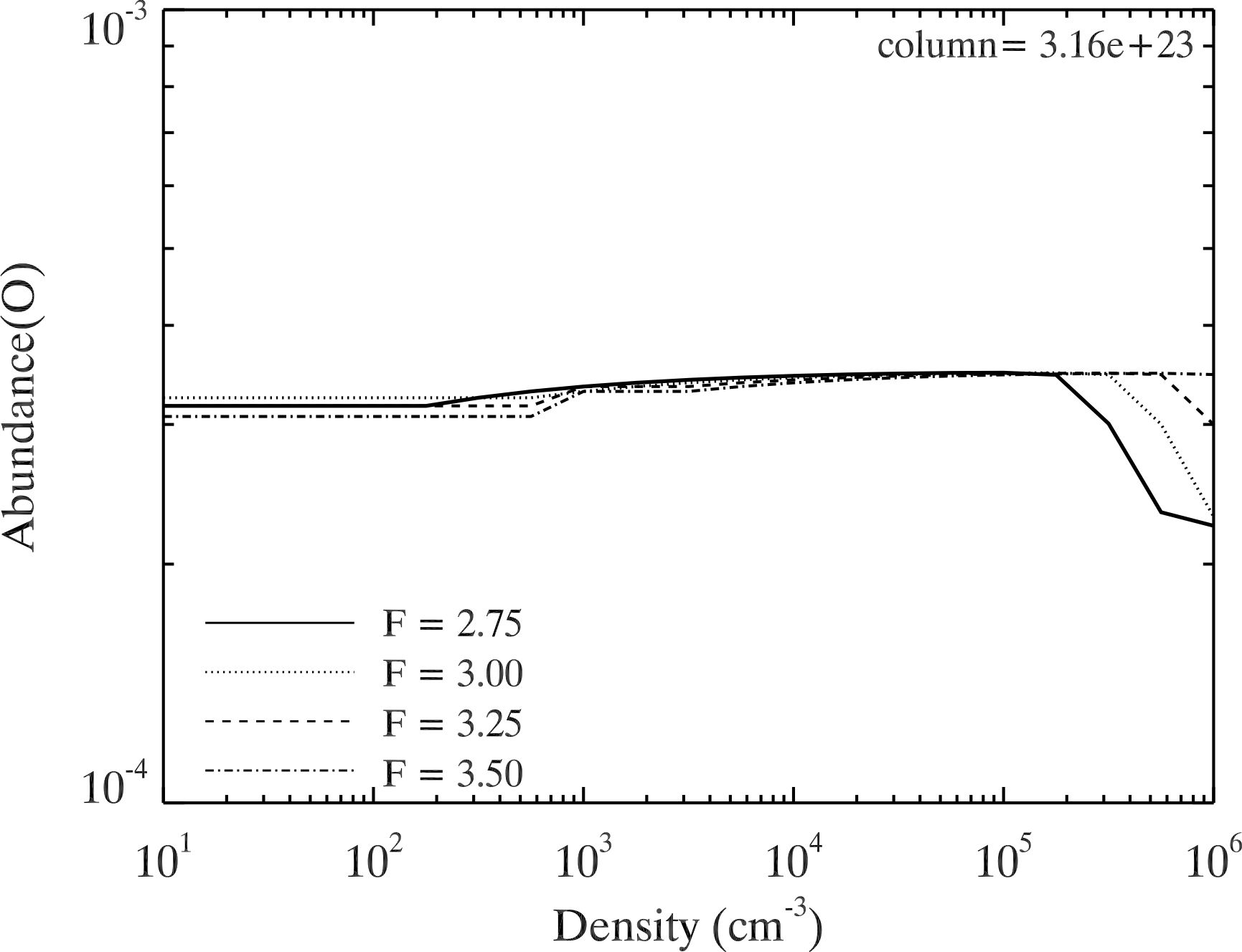}\\
\includegraphics[angle=0,width=7cm]{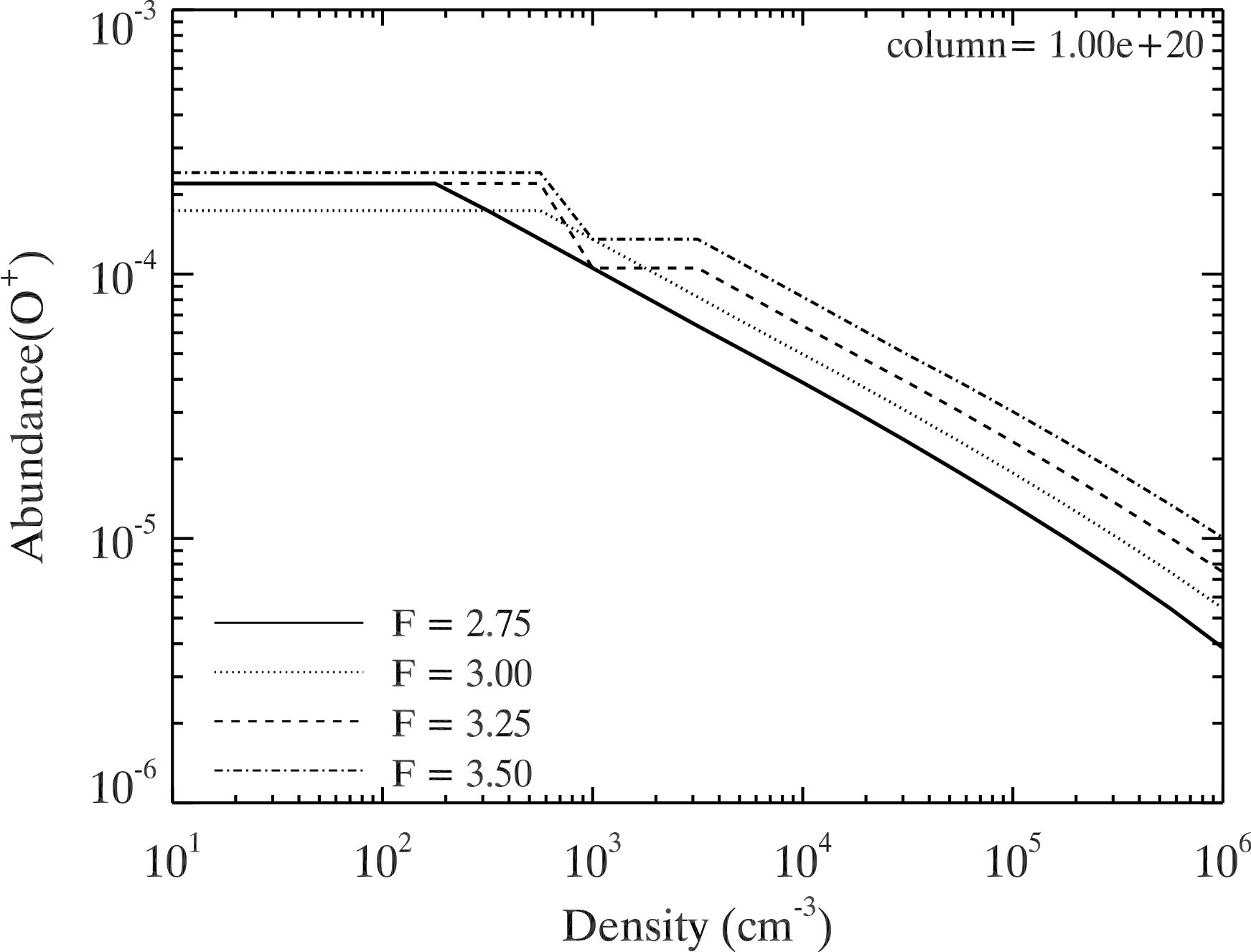}
\vspace{0.05cm}
\includegraphics[angle=0,width=7cm]{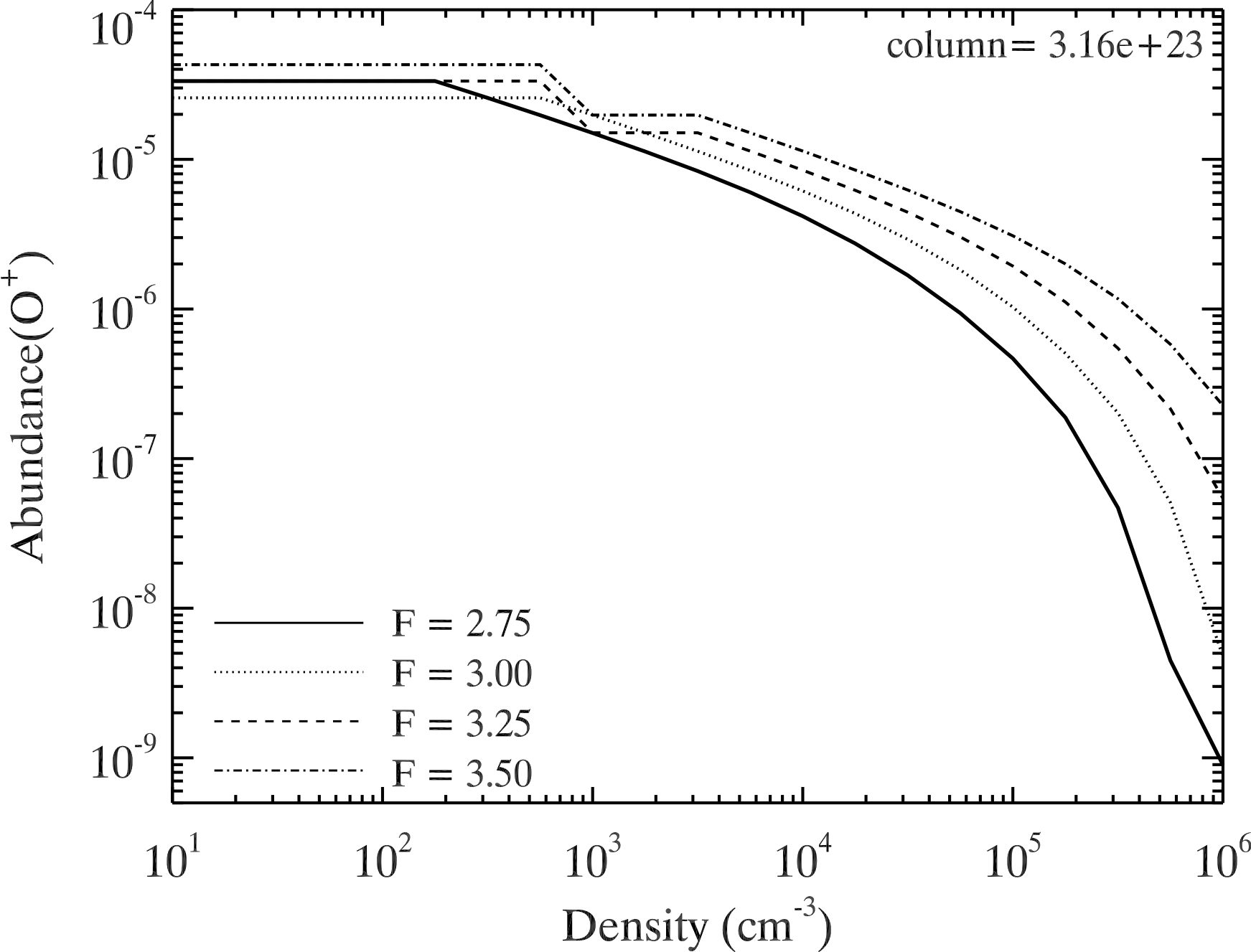}\\
\includegraphics[angle=0,width=7cm]{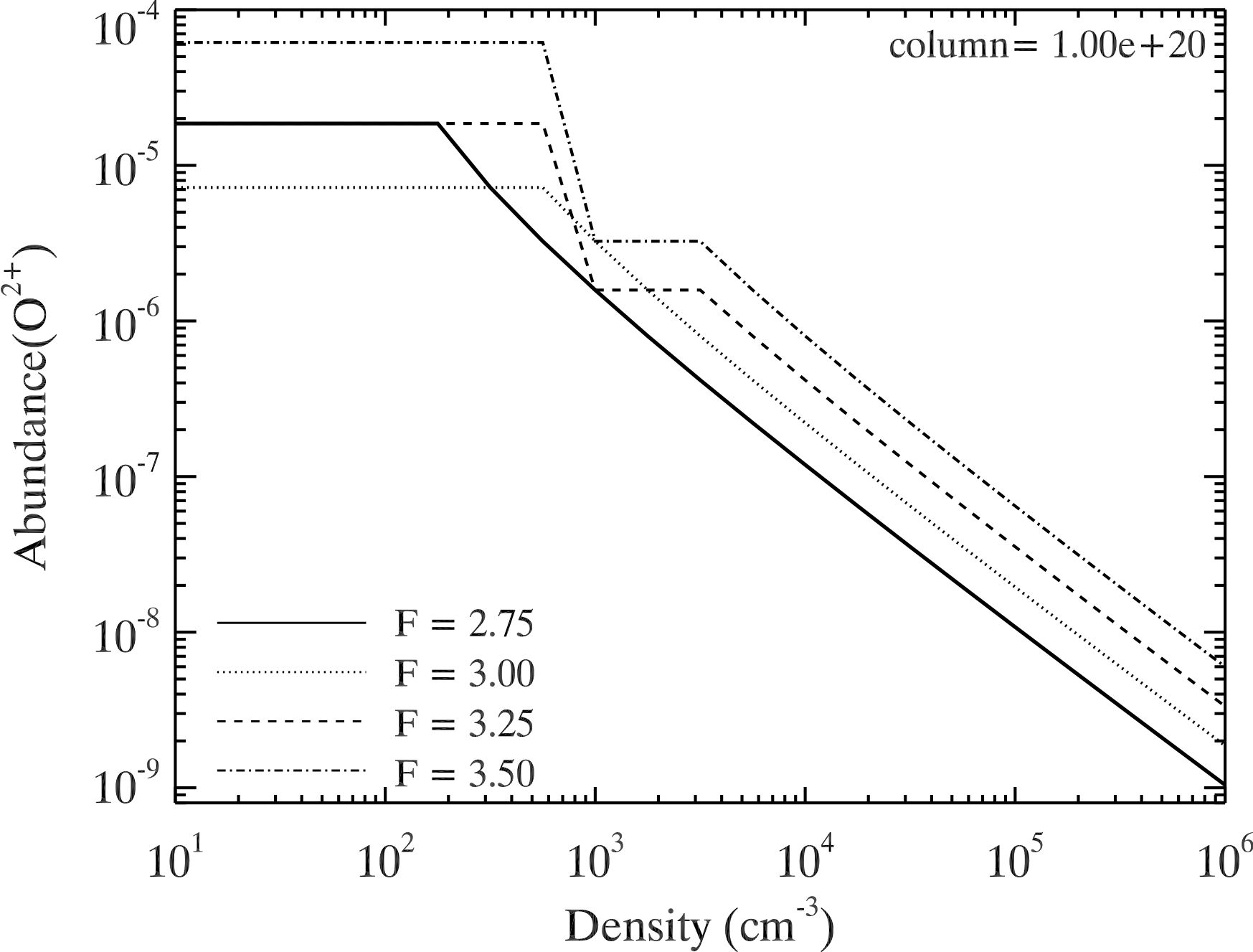}
\vspace{0.05cm}
\includegraphics[angle=0,width=7cm]{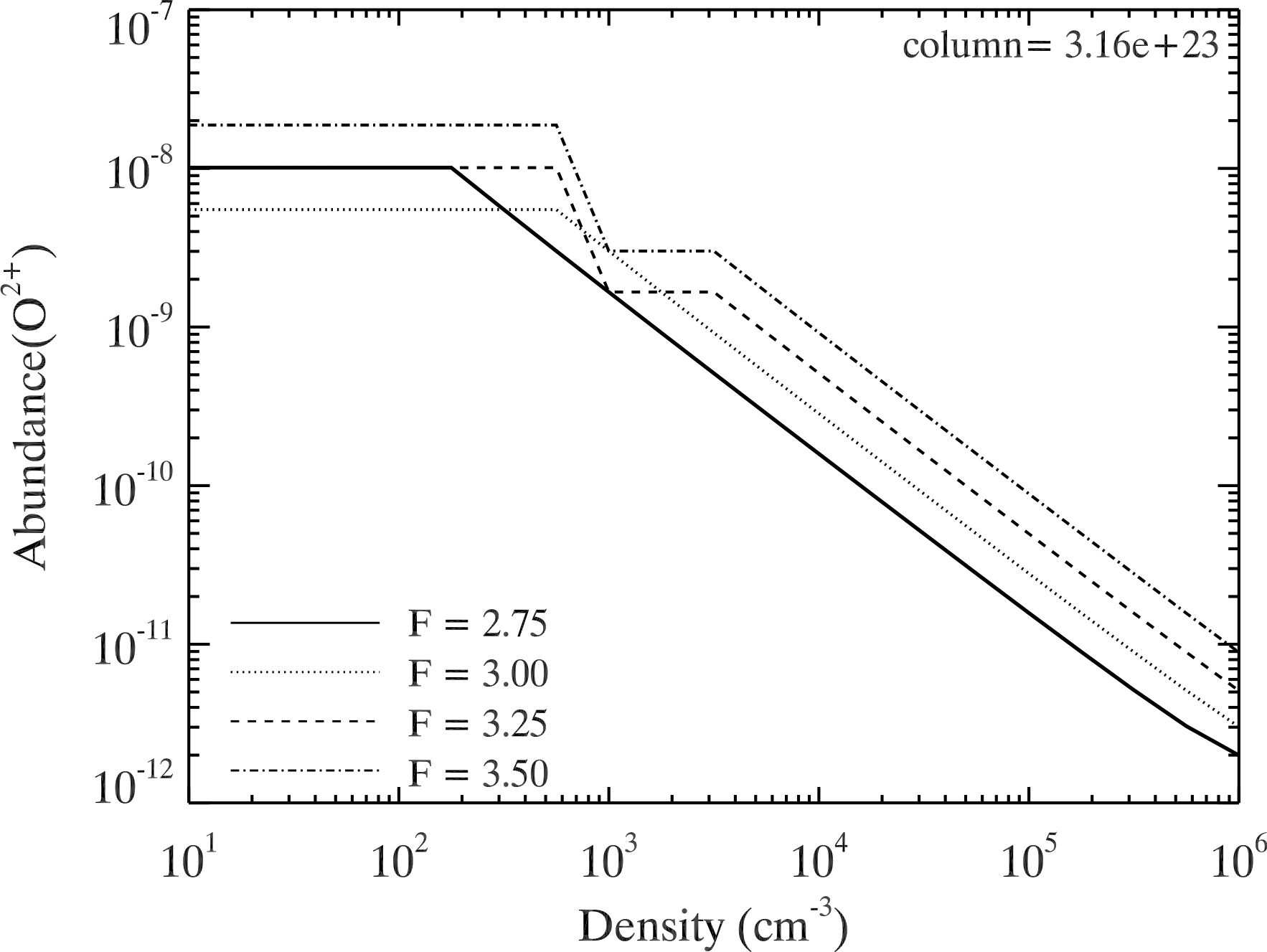}\\
\includegraphics[angle=0,width=7cm]{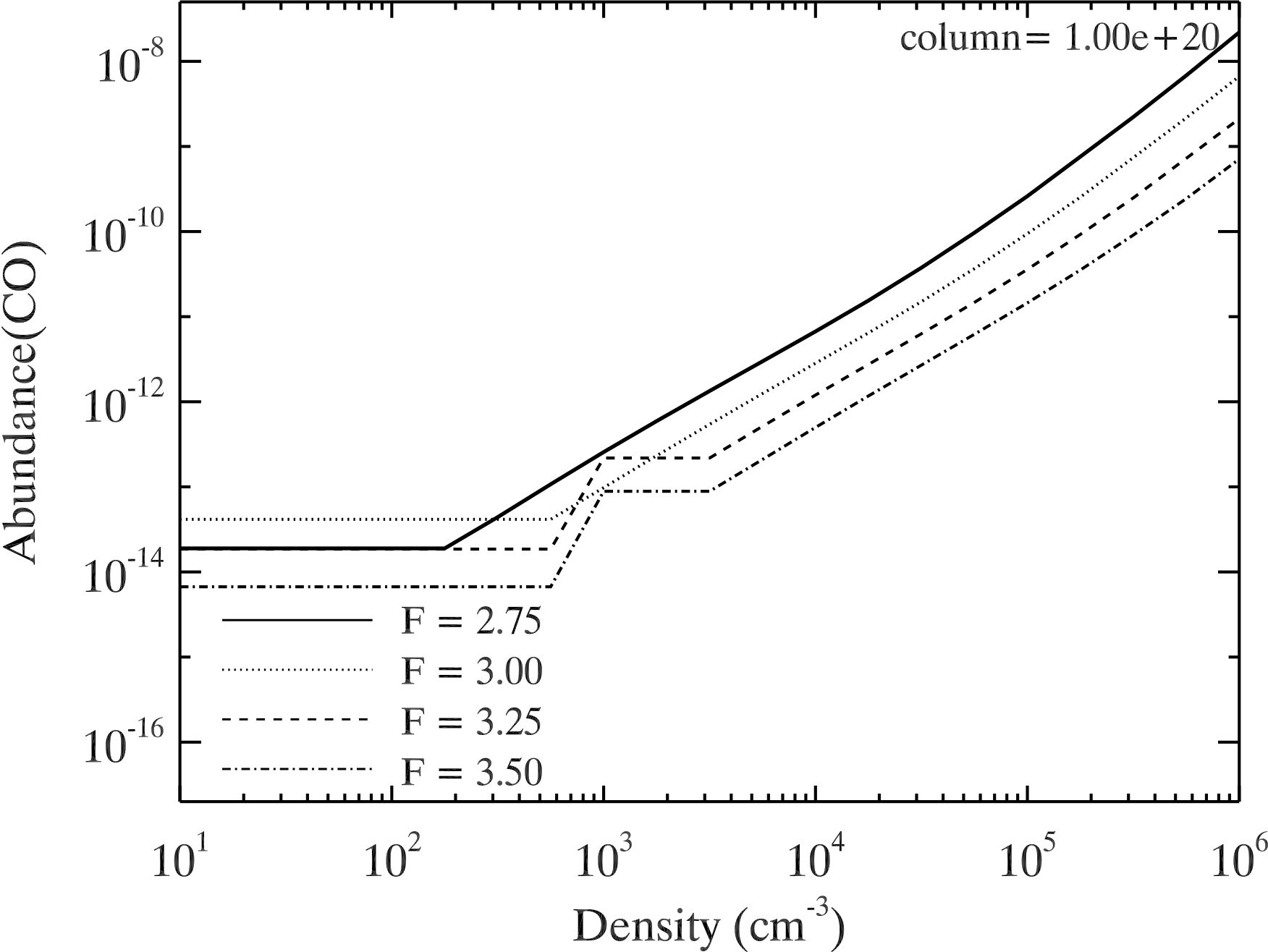}
\vspace{0.05cm}
\includegraphics[angle=0,width=7cm]{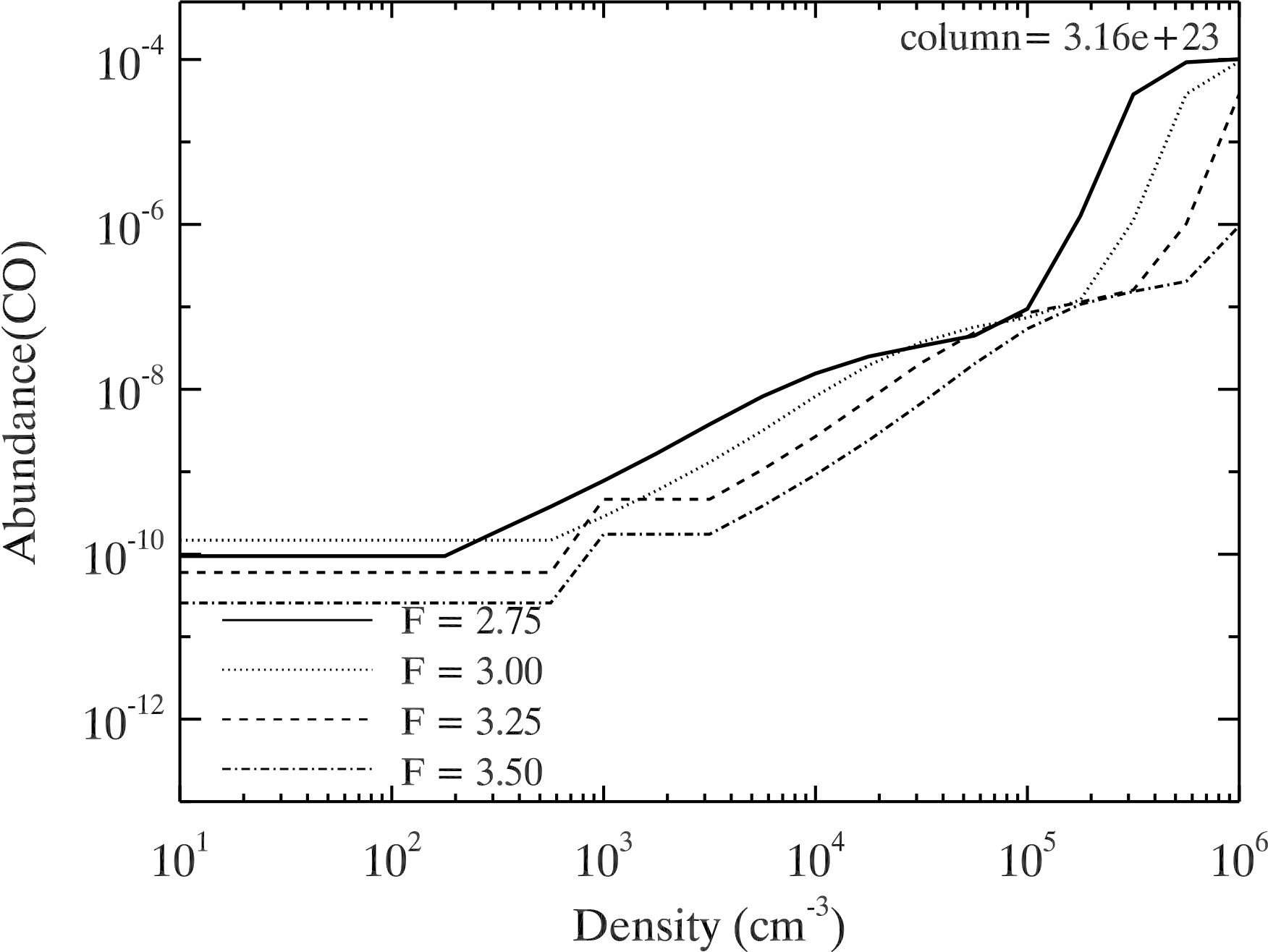}\\
\end{figure*}
\begin{figure*}
\includegraphics[angle=0,width=7cm]{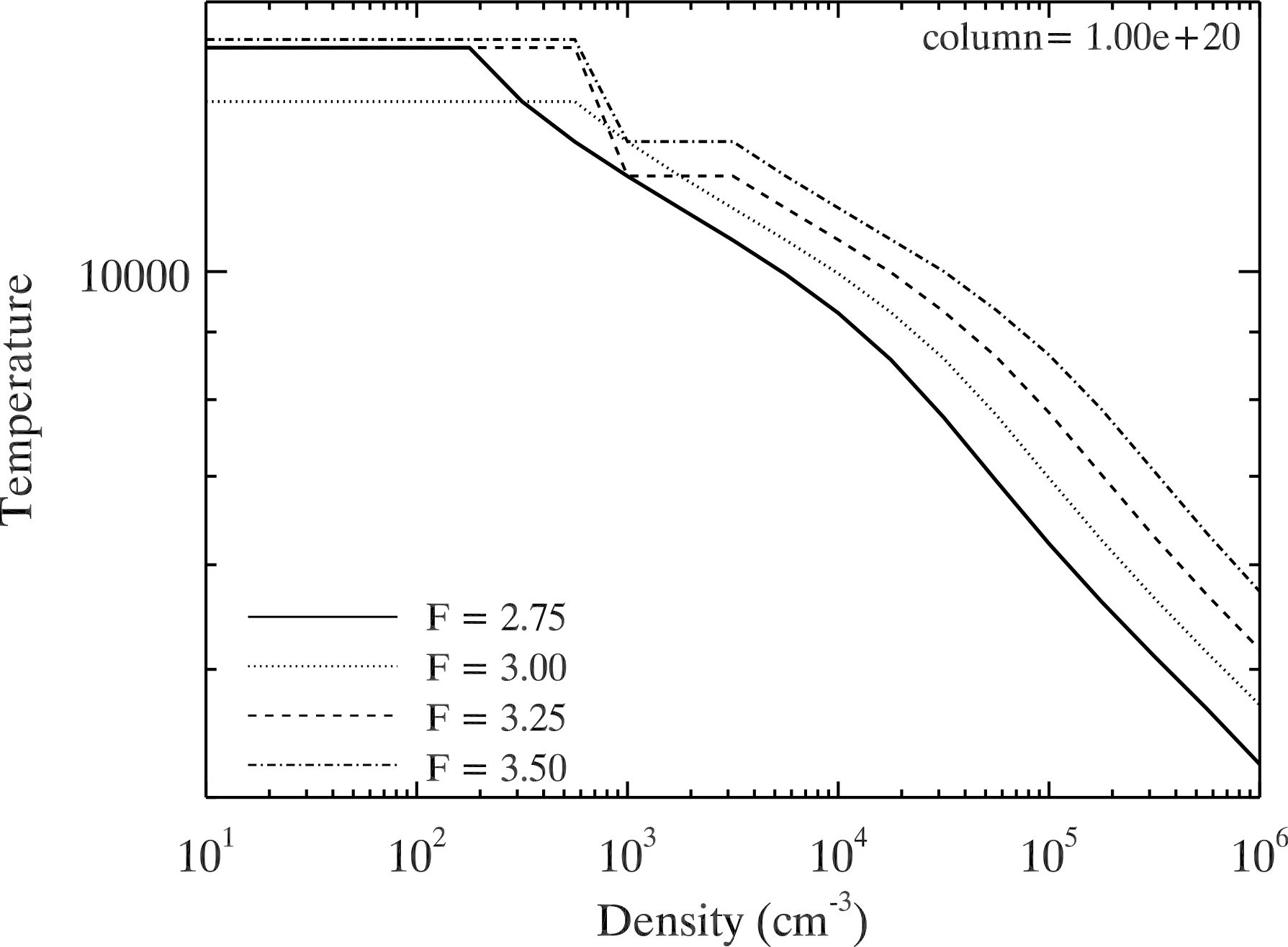}
\vspace{0.05cm}
\includegraphics[angle=0,width=7cm]{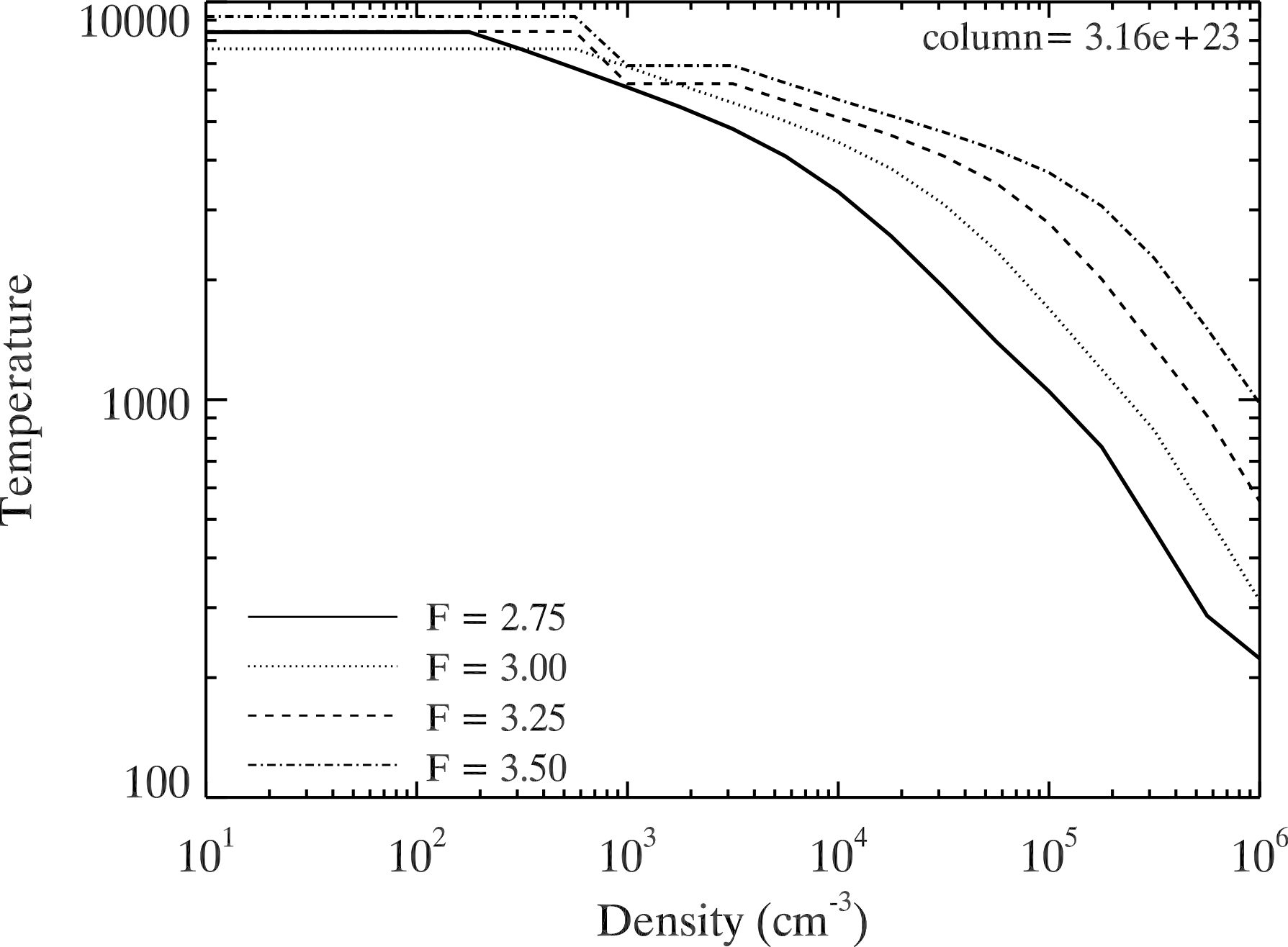}\\
\caption{Abundances of e$^-$, H, H$^+$, H$^-$, H$_2$, H$_2^+$, He, He$^+$, He$^{2+}$, C, C$^+$, C$^{2+}$, O, O$^+$, O$^{2+}$, CO and temperature for column densities of 10$^{20}$ cm$^{-2}$ (left column) and $3.16\times10^{23}$ cm$^{-2}$ (right column) for a X-ray flux of log F = 2.75 - 3.5 erg s$^{-1}$ cm$^{-2}$.} \label{fig17}
\end{figure*}

\bibliographystyle{apj}   
\bibliography{Aykutalp}

\end{document}